\documentclass[11pt]{article}
\usepackage{amssymb}
\usepackage{amsmath}
\usepackage{amscd}
\usepackage{latexsym}
\usepackage{graphics}
\usepackage{color}
\catcode`@=11
\renewcommand{\section}
{\@startsection{section}{1}{0pt}{\medskipamount}{\medskipamount}{\large\bf}}
\makeatletter\renewcommand{\subsection}
{\@startsection{subsection}{2}{\z@}{-3.25ex plus -1ex minus -.2ex}
{1.5ex plus .2ex}{\it }}

\numberwithin{equation}{section}
\catcode`@=12

\def\th{\theta}
\def\a{\alpha}
\def\b{\beta}
\def\g{\gamma}
\def\de{\delta}

\def\vp{\varphi}

\def\vt{\vartheta}
\def\p{\phi}
\def\s{\sigma}

\def\sfrac#1#2{{\textstyle\frac{#1}{#2}}}
\def\m{\mu}
\def\n{\nu}
\def\pa{\partial}

\def\beq{\begin{equation}}
\def\eeq{\end{equation}}
\def\bea{\begin{eqnarray}}
\def\eea{\end{eqnarray}}

\newcommand{\uoo}{{{\rm U}(1){\times}{\rm U}(1)}}

\renewcommand{\time}{{\times}}

\newcommand{\im}{\,\mathrm{i}\,}
\newcommand{\diff}{\mathrm{d}}

\newcommand{\rct}{{\mathbb{R}^4_\theta}}

\newcommand{\rt}{{\mathbb{R}^{2n}_\theta}}

\newcommand{\R}{{\mathbb{R}}}

\newcommand{\C}{{\mathbb{C}}}
\newcommand{\Z}{{\mathbb{Z}}}

\newcommand{\Idd}{\mathbf{1}}
\newcommand{\Hcal}{{\cal H}}
\newcommand{\Ecal}{{\cal E}}

\newcommand{\Pcal}{{\cal P}}
\newcommand{\fh}{\hat{f}}

\newcommand{\xh}{\hat{x}}
\newcommand{\zh}{\hat{z}}
\newcommand{\zbh}{\hat{\bar{z}}}
\newcommand{\yb}{{\bar{y}}}
\newcommand{\zb}{{\bar{z}}}
\newcommand{\ab}{{\bar{a}}}
\newcommand{\bb}{{\bar{b}}}

\newcommand{\ca}{{\cal{A}}}
\newcommand{\cf}{{\cal{F}}}

\newcommand{\cliff}{{{\rm C}\ell}}
\newcommand{\K}{{\rm K}}

\newcommand{\Ka}{{\rm K}^{\rm a}}

\newcommand{\rep}{{\rm R}}

\newcommand{\Lcal}{{\cal L}}
\newcommand{\Tr}{{\rm Tr}}

\newcommand{\tr}{{\rm tr}}

\newcommand{\su}{{{\rm SU}(2)}}
\newcommand{\uo}{{{\rm U}(1)}}
\newcommand{\uk}{{{\rm U}(k)}}
\newcommand{\slc}{{{\rm SL}(2,\C)}}
\newcommand{\slcL}{{{\rm sl}(2,\C)}}
\newcommand{\spin}{{\rm Spin}}
\newcommand{\Pt}{{\rm P}}
\newcommand{\quiver}{{\sf Q}}
\newcommand{\rel}{{\sf R}}

\newcommand{\pathalg}{{\sf A}}

\newcommand{\mphi}{{{\mbf\phi}}}
\newcommand{\mT}{{{\mbf T}^{~}_{(m_1,m_2)}}}
\newcommand{\mcalT}{{{\mbf{\cal T}}^{~}_{(m_1,m_2)}}}
\newcommand{\mA}{{\mbf A}}
\newcommand{\ma}{{\mbf a}}
\newcommand{\mmua}{{{\mbf\mu}_{(m_1,m_2)}^{(1)}}}
\newcommand{\mmub}{{{\mbf\mu}_{(m_1,m_2)}^{(2)}}}
\newcommand{\mF}{{\mbf F}^{(m_1,\,m_2)}}
\newcommand{\mup}{{{\mbf\Upsilon}}}
\newcommand{\mbf}[1]{{\boldsymbol {#1} }}

\def\Dirac{{D\!\!\!\!/\,}} % Dirac operator
\def\spinor{{{\cal S}\;\!\!\!\!\!/\,}}

\def\Hom{{\rm Hom}}
\def\Ext{{\rm Ext}}
\def\End{{\rm End}}

\def\>{\rangle}
\def\<{\langle}
\def\+{\dagger}
\def\={\ =\ }

\begin{document}

\begin{titlepage}
\setcounter{page}{0}
\begin{flushright}
hep-th/0603232\\
ITP--UH--07/06\\
HWM--06--6\\
EMPG--06--03\\
\end{flushright}

\vskip 1.8cm

\begin{center}

{\Large\bf Rank Two Quiver Gauge Theory, Graded Connections
  \\[10pt] and Noncommutative Vortices}

\vspace{15mm}

{\large Olaf Lechtenfeld${}^1$}, \ \ {\large Alexander D. Popov${}^{1,2}$}
\ \ and \ \ {\large Richard J. Szabo${}^3$}
\\[5mm]
\noindent ${}^1${\em Institut f\"ur Theoretische Physik,
Universit\"at Hannover \\
Appelstra\ss{}e 2, 30167 Hannover, Germany }
\\[5mm]
\noindent ${}^2${\em Bogoliubov Laboratory of Theoretical Physics, JINR\\
141980 Dubna, Moscow Region, Russia}
\\[5mm]
\noindent ${}^3${\em Department of Mathematics and Maxwell Institute
  for Mathematical Sciences\\ Heriot-Watt University,
Colin Maclaurin Building, Riccarton, Edinburgh EH14 4AS, U.K.}
\\[5mm]
{Email: {\tt lechtenf, popov @itp.uni-hannover.de , R.J.Szabo@ma.hw.ac.uk}}

\vspace{15mm}

\begin{abstract}
\noindent
We consider equivariant dimensional reduction of Yang-Mills theory on
K\"ahler manifolds of the form $M\time\C P^1\time\C P^1$. This induces 
a rank two quiver gauge theory on $M$ which can be formulated as a
Yang-Mills theory of graded connections on $M$. The reduction of the
Yang-Mills equations on $M\time\C P^1\time\C P^1$ induces quiver gauge 
theory equations on $M$ and quiver vortex equations in the BPS sector. 
When $M$ is the noncommutative space $\R_\theta^{2n}$ both BPS and 
non-BPS solutions are obtained, and interpreted as states of D-branes. 
Using the graded connection formalism, we assign D0-brane charges in 
equivariant K-theory to the quiver vortex configurations. Some 
categorical properties of these quiver brane configurations are also 
described in terms of the corresponding quiver representations.

\end{abstract}

\end{center}
\end{titlepage}

\section{Introduction and summary \label{Intro}}

\noindent
It has become clear in recent years that a proper description of the
nonperturbative vacuum in string theory will require detailed
understanding of the properties of systems of both BPS and non-BPS
brane configurations (see~\cite{Senrev1} for a recent review). The
basic non-BPS system is the unstable brane-antibrane configuration
which corresponds to a pair of vector bundles with a tachyon field
mapping between them. The dynamics of this system can be cast as a
Yang-Mills theory of superconnections~\cite{AIO1}. In some instances
the branes can be realized as instantons of gauge theory in the
appropriate dimensionality~\cite{Douglas1}. Important examples of this are
noncommutative solitons and instantons which find their most natural
physical interpretations in terms of D-branes~\cite{DMR1}. This is
related~\cite{Matsuo1} to the fact that the charges of D-branes are
classified by K-theory~\cite{MM1}. Reviews on noncommutative solitons
and D-branes can be found in~\cite{Haman}, while applications
of BPS soliton solutions in noncommutative (supersymmetric) Yang-Mills
theory to D-brane dynamics are given e.g.~in~\cite{BPSNCD1}.

One way to generate both stable and unstable states of D-branes is by
placing them at singularities of
orbifolds~\cite{quiverG,quivercat}. Regular representation
D-branes then decay into irreducible representation fractional branes
under the action of the discrete orbifold group. The low-energy
dynamics of the D-brane decay is succinctly described by a quiver
gauge theory. Resolving orbifold singularities by non-contractible
cycles blows up the fractional D-branes into higher dimensional branes
wrapping the cycles.
Another way of obtaining quiver gauge theories on a $q$-dimensional
manifold~$M$ is to consider $k$~coincident D($q{+}r$)-branes wrapping
the worldvolume manifold  $X=M\time G/H$ where $G/H$ is an $r$-dimensional
homogeneous space for a Lie group~$G$ with a closed subgroup~$H$.
In the standard interpretation this system of D-branes corresponds to a 
rank~$k$ hermitean vector bundle~$\cal E$ over~$X$ with a connection 
whose dynamics are governed by Yang-Mills gauge theory. For K\"ahler 
manifolds~$X$ the stability of such bundles (BPS conditions) is controlled 
by the Donaldson-Uhlenbeck-Yau (DUY) equations~\cite{DUY1}. 
For $G$-equivariant bundles ${\cal E}\to X$ one finds that Yang-Mills theory 
on~$X$ reduces to a quiver gauge theory on~$M$~\cite{Garcia1}--\cite{A-CG-P2}.

In this paper we will focus on some of these issues in quiver gauge
theories on K\"ahler manifolds $M$ which arise via a quotient by the
natural action of the Lie group $\su\time\su$ on equivariant
Chan-Paton bundles over $M\time\C P^1\time\C P^1$.
Our analysis generalizes previous work on brane-antibrane systems from
reduction on $M\time\C P^1$~\cite{Garcia1,IL,LPS},
and on the generalization to chains of
branes and antibranes arising from $\su$-equivariant dimensional
reduction on $M\time\C P^1$~\cite{A-CG-P3,PS1}. In
particular, we will expand on the formalism introduced in~\cite{PS1}
which merged the low-energy dynamics of brane-antibrane chains with
quiver gauge theory into a Yang-Mills gauge theory of new objects on $M$
termed ``graded connections'', which generalize the usual
superconnections on the worldvolumes of coincident brane-antibrane
pairs. This formalism is particularly well-suited to describe such
physical instances and their novel effects, such as the equivalence
between non-abelian quiver vortices on $M$ and symmetric multi-instantons
on the higher-dimensional space~$M\time\C P^1\time\C P^1$. Moreover,
when $M$ is the noncommutative space $\R_\theta^{2n}$, it enables one to
interpret noncommutative quiver solitons in the present case as states of
D-branes in a straightforward manner, whilst providing a categorical
approach to D-branes which characterizes their moduli beyond their
K-theory charges. These quiver brane configurations require a more
complex description than just that in terms of branes and antibranes,
and we construct a category of D-branes which incorporates both their
locations and their bindings to abelian magnetic monopoles.

The essential new ingredients of the present paper are that our
quivers are of rank two, as opposed to the rank one quivers considered
in~\cite{PS1}, and the necessity of imposing relations on the
quiver. The resulting quiver D-brane configuration is new, and comprises a
two-dimensional lattice of branes and antibranes coupled to
$\uo\time\uo$ Dirac monopole fields with interesting
dynamics formulated through a higher-rank gauge theory of graded
connections. We will also elaborate further on some of the
constructions introduced in~\cite{PS1}.

The outline of this paper is as follows. In Section~\ref{Eqgauge} we
describe general features of the $\su\time\su$-equivariant reduction
of gauge theories on $M\time\C P^1\time\C P^1$ to an arbitrary
K\"ahler manifold $M$, including the special case of the noncommutative
euclidean space $M=\R_\theta^{2n}$. In Section~\ref{QGT} we describe
various features of the induced quiver gauge theory on $M$ and develop
the associated formalism of graded connections in this case. In
Section~\ref{NCinst} we analyse the general structure of quiver gauge
theory on~$M$ and the quiver vortex equations which describe
the BPS sector. We then construct both BPS and non-BPS solutions of
the Yang-Mills equations on the noncommutative space
$\R^{2n}_\theta\time\C P^1\time\C P^1$, describe their induced
quiver representations, and analyse in detail the structure of the
moduli space of noncommutative instantons. Finally, in
Section~\ref{Dcharges} we realize our noncommutative instantons as
configurations of D-branes by computing their topological charges,
by computing their K-theory charges through a noncommutative
equivariant version of the ABS construction, and by realizing them as
objects in the category of quiver representations using some
techniques of homological algebra.

\bigskip

\section{Equivariant gauge theory  \label{Eqgauge}}

\noindent
In this section we will analyse some aspects of
$\su\time\su$-equivariant gauge theory on spaces of the form
$M\time\C P^1\time\C P^1$, where $M$ is a K\"ahler
manifold. After some preliminary definitions, we describe the
equivariant decomposition of generic gauge bundles over $M\time\C
P^1\time\C P^1$, and of their connections and curvatures. We then
write down the corresponding Yang-Mills action functional and explain
the generalization to noncommutative gauge theory. Equivariant
dimensional reduction is described in general in~\cite{A-CG-P1}, while
general aspects of noncommutative field theories are reviewed
in~\cite{Harvey}.

\subsection{The K\"ahler manifold ${M\time\C P^1\time\C P^1}$
  \label{Mdef}}

Let $M$ be a K\"ahler manifold of real dimension $2n$ with local
real coordinates $x=(x^\m )\in\R^{2n}$, where the indices $\m ,\n
,\ldots$ run through $1,\ldots , 2n$. Let $S^2_{(\ell)}\cong\C
P^1_{(\ell)}$, $\ell=1,2$, be two copies of the standard two-sphere of
constant radii $R_\ell$ with coordinates $\vt_\ell\in[0,\pi]$ and
$\vp_\ell\in[0,2\pi]$. We shall consider the product $M\time
\C P_{(1)}^1\time\C P_{(2)}^1$ which is also a K\"ahler manifold with
local complex coordinates $(z^1,\dots,z^n,y_1,y_2)\in\C^{n+2}$ and
their complex conjugates, where
\begin{equation}\label{zz}
z^a\=x^{2a-1}-\im\,x^{2a} \qquad\textrm{and}\qquad
\zb^{\ab}\=x^{2a-1}+\im\,x^{2a} \qquad\textrm{with}\quad
a\=1,\ldots,n
\end{equation}
while
\begin{equation}\label{zn1}
y_\ell\=\frac{\sin\vt_\ell}{1+\cos\vt_\ell}\,\exp{(-\im\vp_\ell)}
\qquad\mbox{and}\qquad
\yb_\ell\=\frac{\sin\vt_\ell}{1+\cos\vt_\ell}\,\exp{(\im\vp_\ell)}
 \qquad\textrm{with}\quad \ell\=1,2 \ .
\end{equation}
In these coordinates the riemannian metric
\beq
\diff s^2=g_{\hat\mu\hat\nu}~\diff x^{\hat\mu}~\diff x^{\hat\nu}
\label{metrichat}\eeq
on $M\time \C P^1_{(1)}\time \C P^1_{(2)}$ takes the form
\bea\label{metric3}
\diff s^2 &=& g_{\m\n}~\diff x^{\m}~\diff x^{\n}
+ R_1^2\,\left(\diff \vt_1^2 + \sin^2\vt_1\ \diff\vp_1^2\right)
+ R_2^2\,\left(\diff \vt_2^2 + \sin^2\vt_2\ \diff\vp_2^2\right)
\nonumber\\[4pt]
&= & 2\,g_{a\bb}\ \diff z^a~\diff \zb^{\bb} +
\frac{4\,R_1^2}{\left(1+y_1\yb_1\right)^2}\ \diff y_1~\diff \yb_1
+ \frac{4\,R_2^2}{\left(1+y_2\yb_2\right)^2}\ \diff y_2~\diff \yb_2\ ,
\eea
where hatted indices $\hat\mu,\hat\nu,\dots$ run over $1,\dots,2n+4$.
The K\"ahler two-form $\Omega $ is given by
\bea\label{kahler}
\Omega&=&\sfrac{1}{2}\,\omega_{\m\n}~\diff x^\m\wedge\diff x^\n +
R^2_1\,\sin\vt_1~\diff\vt_1\wedge\diff\vp_1 + R^2_2\,\sin\vt_2~
\diff\vt_2\wedge\diff\vp_2
\nonumber\\[4pt]
&=&- 2\im g_{a\bb}\ \diff z^a\wedge\diff \zb^{\bb}
- \frac{4\im R_1^2}{\left(1+y_1\yb_1\right)^2}\ \diff y_1\wedge\diff \yb_1
- \frac{4\im R_2^2}{\left(1+y_2\yb_2\right)^2}\ \diff y_2\wedge\diff \yb_2\ .
\eea

\subsection{Equivariant vector bundles \label{eqvecbun}}

Let ${\cal E}\to M\time \C P_{(1)}^1\time \C P_{(2)}^1$ be a
hermitean vector bundle of rank $k$. We wish to impose the condition of
$G$-equivariance on this bundle with the group $G:=\su\time\su$ of
rank~$2$ acting trivially on $M$ and in the standard way on
the homogeneous space $\C P^1\time\C P^1\cong G/H$, where
$H:=\uo\time\uo$ is a maximal torus of $G$. This means that we should
look for representations of the group $G$ inside the $\uk$ structure group
of the bundle $\cal E$, i.e. for $k$-dimensional unitary
representations of $G$. For every pair of positive integers $k_i$ and
$k_\alpha$, up to isomorphism there are unique irreducible
$\su$-modules $\underline{V}_{\,k_i}$
and $\underline{V}_{\,k_\alpha}$ of dimensions $k_i$ and $k_\alpha$,
respectively, and consequently a unique irreducible representation
$\underline{V}_{\,k_{i\alpha}}:=\underline{V}_{\,k_i}\otimes
\underline{V}_{\,k_\alpha}$ of $G$ with dimension
$k_{i\alpha}:=k_i\,k_\alpha$. Thus, for each pair of positive integers
$m_1$ and $m_2$, the module
\beq
\underline{\cal V}\=\bigoplus_{i=0}^{m_1}~\bigoplus_{\a =0}^{m_2}\,
\underline{V}_{\,k_{i\a}} \qquad\textrm{with}\quad
\underline{V}_{\,k_{i\a}} \ \cong \ \C^{k_{i\a}}\qquad\textrm{and}\qquad
\sum_{i=0}^{m_1}~\sum_{\a =0}^{m_2}\,k_{i\a}\=k
\label{genrepSU2Uk}\eeq
gives a representation of $\su\time\su$ inside ${\rm U}(k)$. The
structure group of the bundle $\cal E$ is correspondingly broken as
\beq
{\rm U}(k)~\longrightarrow~\prod_{i=0}^{m_1}~
\prod_{\a =0}^{m_2}\,{\rm U}(k_{i\a}) \ .
\label{gaugebroken}\eeq
As a result, we must construct bundles ${\cal E}\to M\time \C
P_{(1)}^1\time \C P_{(2)}^1$ whose typical fibres $\underline{\cal
  V}$ are complex vector spaces with a direct sum decomposition as in
(\ref{genrepSU2Uk}). We will now describe how this is
done explicitly.

There are natural equivalence functors between the categories of
$G$-equivariant vector bundles over $M\time G/H$ and $H$-equivariant
bundles over $M$, where $H$ acts trivially on $M$~\cite{A-CG-P1}. If $E\to
M$ is an $H$-equivariant bundle, then it defines a $G$-equivariant
bundle ${\cal E}\to M\time\C P^1\time\C P^1$ by induction as
\beq
{\cal E}=G\time_HE \ ,
\label{calEind}\eeq
where the $H$-action on $G\time E$ is given by
$h\cdot(g,e)=(g\,h^{-1},h\cdot e)$ for $h\in H$, $g\in G$ and $e\in
E$. We therefore focus our attention on the structure of
$H$-equivariant bundles $E\to M$. For this, it is more convenient to
work in a holomorphic setting by passing to the universal
complexification $G^{\rm c}:=G\otimes\C=\slc\time\slc$ of the Lie
group $G$. If ${\cal E}\to M\time\C P^1\time\C P^1$ is a
$G$-equivariant vector bundle, then the $G$-action can be extended to
an action of $G^{\rm c}$. Let $K=\Pt\time\Pt$ be the Borel subgroup
of $G^{\rm c}$ with $\Pt$ the group of lower triangular matrices in
$\slc$. Its Levi decomposition is given by $K=U\ltimes H^{\rm c}$,
where $H^{\rm c}:=H\otimes\C=\C^\times\time\C^\times$. A
representation $\underline{V}$ of $K$ is irreducible if and only if
the action of $U$ on $\underline{V}$ is trivial and the restriction
$\underline{V}\,|_{H^{\rm c}}$ is irreducible. It follows that there
is a one-to-one correspondence between irreducible representations of
$K$ and irreducible representations of the Cartan subgroup $H^{\rm
  c}\subset G^{\rm c}$.
The natural map $\C P^1\time\C P^1=G/H\to G^{\rm c}/K$ is a
diffeomorphism of projective varieties. The categorical
equivalence above can then be reformulated as a
one-to-one correspondence between $G^{\rm c}$-equivariant bundles
${\cal E}\to M\time \C P^1\time\C P^1$ and $K$-equivariant bundles
over $M$, with $K$ acting trivially on $M$.

The Lie algebra $\slcL$ is generated by the three Pauli matrices
\beq
\sigma_3\=\begin{pmatrix}1&0\\0&-1\end{pmatrix} \ , \quad
\sigma_+\=\begin{pmatrix}0&1\\0&0\end{pmatrix} \quad\mbox{and}\quad
\sigma_-\=\begin{pmatrix}0&0\\1&0\end{pmatrix}
\label{sl2cmatrices}\eeq
with the commutation relations
\beq
\left[\sigma_3\,,\,\sigma_\pm\right]\=\pm\,2\,\sigma_\pm
\qquad\textrm{and}\qquad \left[\sigma_+\,,\,\sigma_-\right]\=\sigma_3 \ .
\label{sl2cLie}\eeq
The Lie algebra of $U$ is generated by two independent copies of the
element $\sigma_-$, while the Cartan subgroup $H^{\rm c}$ is generated
by two independent copies of the element $\sigma_3$. For each
$p\in\Z$ there is a unique irreducible representation
$\underline{S}_{\,p}\cong\C$ of $\C^\times$ given by $\zeta\cdot
v=\zeta^p\,v$ for $\zeta\in\C^\times$ and $v\in\underline{S}_{\,p}$.
Thus for each pair of integers $p_1,p_2$ there is a unique
irreducible module
$\underline{S}^{(1)}_{\,p_1}\otimes\underline{S}^{(2)}_{\,p_2}\cong\C$
over the subgroup $H^{\rm
  c}=\C^\times_{(1)}\time\C^\times_{(2)}$. Since the manifold $M$
carries the trivial action of the group $H^{\rm c}$, any $K$-equivariant
bundle $E\to M$ admits a finite Whitney sum decomposition into
isotopical components as $E=\bigoplus_{p_1,p_2}\,E_{p_1\,p_2}\otimes
\underline{S}^{(1)}_{\,p_1}\otimes\underline{S}^{(2)}_{\,p_2}$, where
the sum runs over the set of eigenvalues for the $H^{\rm c}$-action on
$E$ and $E_{p_1\,p_2}\to M$ are bundles with the trivial $H^{\rm
  c}$-action. From the commutation relations (\ref{sl2cLie}) it
follows that the $U$-action on $E_{p_1\,p_2}\otimes
\underline{S}^{(1)}_{\,p_1}\otimes\underline{S}^{(2)}_{\,p_2}$
corresponds to independent bundle morphisms $E_{p_1\,p_2}\to
E_{p_1-2\,p_2}$ and $E_{p_1\,p_2}\to E_{p_1\,p_2-2}$, along with the
trivial $\sigma_-$-actions on the irreducible $H^{\rm c}$-modules
$\underline{S}^{(1)}_{\,p_1}\otimes\underline{S}^{(2)}_{\,p_2}$.

After an appropriate twist by an $H^{\rm c}$-module and a relabelling,
the $\sigma_3$-actions are given by the $H^{\rm c}$-equivariant
decomposition
\beq
E=\bigoplus_{i=0}^{m_1}~\bigoplus_{\alpha=0}^{m_2}\,E_{k_{i\alpha}}
\otimes\underline{S}^{(1)}_{\,m_1-2i}\otimes
\underline{S}^{(2)}_{\,m_2-2\alpha} \ ,
\label{Hceqdecomp}\eeq
while the $U$-action is determined through the diagram
\beq
\begin{CD}
E^{}_{k_{m_1\,0}}@>{\phi^{(1)}_{m_1\,0}}>>
E^{}_{k_{m_1-1\,0}}@>{\phi_{m_1-1\,0}^{(1)}}>>\cdots@>
{\phi_{10}^{(1)}}>>E^{}_{k_{00}}\\
@A{\phi^{(2)}_{m_11}}AA@A{\phi^{(2)}_{m_1-1\,1}}AA@.@AA{\phi^{(2)}_{01}}A\\
\vdots@.\vdots@.@.\vdots\\
@A{\phi_{m_1\,m_2-1}^{(2)}}AA@A{\phi^{(2)}_{m_1-1\,m_2-1}}AA@.@AA
{\phi^{(2)}_{0\,m_2-1}}A\\
E^{}_{k_{m_1\:m_2-1}}@>{\phi^{(1)}_{m_1\,m_2-1}}>>
E^{}_{k_{m_1-1\:m_2-1}}@>{\phi_{m_1-1\,m_2-1}^{(1)}}>>
{}~\cdots~@>{\phi^{(1)}_{1\,m_2-1}}>>E^{}_{k_{0\:m_2-1}}\\
@A{\phi^{(2)}_{m_1m_2}}AA@A{\phi^{(2)}_{m_1-1\,m_2}}AA@.@AA
{\phi^{(2)}_{0m_2}}A\\
E^{}_{k_{m_1\:m_2}}@>>{\phi^{(1)}_{m_1m_2}}>
E^{}_{k_{m_1-1\:m_2}}@>>{\phi_{m_1-1\,m_2}^{(1)}}>
{}~\cdots~@>>{\phi^{(1)}_{1\,m_2}}>E^{}_{k_{0\:m_2}}
\end{CD}
\label{bundlediag}\eeq
of holomorphic bundle maps with ${\phi^{(1)}_{m_1+1\;
    \a}}=0=\phi^{(1)}_{0 \a}$ for $\a = 0,1,\ldots ,m_2$ and
${\phi^{(2)}_{i\; m_2+1}}=0=\phi^{(2)}_{i0}$ for $i = 0,1,\ldots
,m_1$. Since the Lie algebra of $U$ is abelian, these maps generate a
{\it commutative} bundle diagram (\ref{bundlediag}), i.e. for each
$i,\alpha$ one has
\beq
\phi^{(1)}_{i+1\:\a}\,\phi^{(2)}_{i+1\:\a+1}=
\phi^{(2)}_{i\:\a+1}\,\phi^{(1)}_{i+1\:\a+1} \ .
\label{phicommrels}\eeq
Finally, we can now consider the underlying $H$-equivariant hermitean
vector bundle and introduce the standard $p_\ell$-monopole line bundles
\beq
\Lcal_{(\ell)}^{p_\ell}=\su\time_\uo\,\underline{S}^{(\ell)}_{\,p_\ell}
\label{monbundles}\eeq
over the homogeneous spaces $\C P^1_{(\ell)}$ for $\ell=1,2$. Then the
original rank~$k$ hermitean vector bundle (\ref{calEind}) over
$M\time\C P^1_{(1)}\time\C P^1_{(2)}$ admits an equivariant
decomposition
\beq
{\cal E}\=\bigoplus_{i=0}^{m_1}~\bigoplus_{\a=0}^{m_2}\,{\cal E}_{i\a}
\qquad\textrm{with}\quad
{\cal E}_{i\a}\=E_{k^{}_{i\a}}\otimes\Lcal^{m_1-2i}_{(1)}\otimes
\Lcal^{m_2-2\a}_{(2)}\ ,
\label{calEansatz}\eeq
where $E_{k_{i\a}}\to M$ is a hermitean vector bundle of rank $k_{i\a}$
with typical fibre the module $\underline{V}_{\,k_{i\a}}$ in
(\ref{genrepSU2Uk}), and ${\cal E}_{i\a}\to M\time\C P^1_{(1)}\time
\C P^1_{(2)}$ is the bundle with fibres
\beq
\bigl({\cal  E}_{i\a}\bigr)_{(x,\,y_1,\,\yb_1,\,y_2,\,\yb_2)}=
\bigl(E_{k_{i\a}}\bigr)_{x}\otimes\bigl(\Lcal_{(1)}^{m_1-2i}
\bigr)_{(y_1,\yb_1)}
\otimes\bigl(\Lcal_{(2)}^{m_2-2\a}\bigr)_{(y_2,\,\yb_2)} \ .
\eeq

\subsection{Equivariant gauge fields \label{invgauge}}

Let $\ca$ be a connection on the hermitean vector bundle $\Ecal\to
M\time\C P^1_{(1)}\time\C P^1_{(2)}$ having the form
$\ca=\ca_{\hat\mu}\,\diff x^{\hat\mu}$ in local coordinates
$(x^{\hat\mu})$ and taking values in the Lie algebra ${\rm u}(k)$. We
will now describe the $G$-equivariant reduction of $\ca$ on $M\time\C
P^1_{(1)}\time\C P^2_{(2)}$. The spherical dependences are completely
determined by the unique $\su$-invariant connections $a_{p_\ell}^{(\ell)}$,
$\ell=1,2$, on the monopole line bundles (\ref{monbundles}) having, in
local complex coordinates on $\C P^1_{(\ell)}$, the forms
\beq\label{f1}
a_{p_\ell}^{(\ell)} = \frac{p_\ell}{2\,\left(1 +y_\ell\yb_\ell
\right)}\ \left(\yb_\ell~
\diff y_\ell -y_\ell~\diff\yb_\ell\right) \ .
\eeq
The curvatures of these connections are
\beq
f_{p_\ell}^{(\ell)}\=\diff a_{p_\ell}^{(\ell)}\= - \frac{p_\ell}
{\left(1 +y_\ell\yb_\ell\right)^2}~
\diff y_\ell\wedge \diff \yb_\ell \ ,
\label{f2}
\eeq
and their topological charges are given by the degrees of the
complex line bundles $\Lcal_{(\ell)}^{p_\ell}\to\C P^1_{(\ell)}$ as
\beq
{\rm deg}~\Lcal^{p_\ell}_{(\ell)} \= \frac{\im}{2\pi}\,
\int_{\C P^1_{(\ell)}}f^{(\ell)}_{p_\ell} \= p_\ell \ .
\label{f3}
\eeq
In the spherical coordinates $(\vt^{}_\ell,\vp^{}_\ell)\in S_{(\ell)}^2$ the
monopole fields can be written as
\beq
a_{p_\ell}^{(\ell)}\=-\frac{\im p_\ell}{2}\,(1-\cos\vt_\ell)~\diff\vp_\ell
\qquad\mbox{and}\qquad
f_{p_\ell}^{(\ell)}\=\diff a_{p_\ell}^{(\ell)}\=-\frac{\im
  p_\ell}{2}\,\sin\vt_\ell~
\diff\vt_\ell\wedge\diff\vp_\ell \ .
\label{am}
\eeq
Related to the monopole fields are the unique, covariantly constant
$\su$-invariant forms of types $(1,0)$ and $(0,1)$ on $\C P^1_{(\ell)}$
given respectively by
\begin{equation}\label{f8}
\beta_\ell\= \frac{\diff y_\ell}{1 +y_\ell\yb_\ell} \qquad \mbox{and}\qquad
\bar{\b}_\ell\= \frac{\diff \yb_\ell}{1 +y_\ell\yb_\ell}\ .
\end{equation}
They take values respectively in the components $\Lcal_{(\ell)}^2$ and
$\Lcal_{(\ell)}^{-2}$ of the complexified cotangent bundle $T^*\C
P^1_{(\ell)}\otimes\C=\Lcal_{(\ell)}^2\oplus\Lcal_{(\ell)}^{-2}$ over $\C
P^1_{(\ell)}$. Note that there is no summation over the index $\ell$
in (\ref{f1})--(\ref{f8}).

With respect to the isotopical decomposition (\ref{calEansatz}), the
twisted ${\rm u}(k)$-valued gauge potential $\ca$ thus splits into
$k^{}_{i\a}\time k^{}_{j\b}$ blocks as
\begin{equation}\label{f4}
\ca\=\left(\ca^{i\a ,\,j\b}\right) \qquad\mbox{with}\quad
\ca^{i\a ,\,j\b}~\in~\mbox{Hom}\bigl(\,\underline{V}^{}_{\,k_{j\b}}\,,\,
\underline{V}^{}_{\,k_{i\a}}\bigr) \ ,
\end{equation}
where
\bea
\ca^{i\a ,\,i\a}&=&A^{i\a}(x)\otimes1\otimes1 
+\, \Idd_{k^{}_{i\a}}\otimes
\left(a^{(1)}_{m_1-2i}(y_1)\otimes1 + 1\otimes a^{(2)}_{m_2-2\a }(y_2)
\right) \ ,
\label{f6a} \\[4pt] \label{f6}
\ca^{i\a,\,i+1\,\a}&=:&\quad\Phi^{(1)}_{i+1\,\a}~\qquad~=~\quad
\phi^{(1)}_{i+1\,\a}(x)\otimes\bar{\b}_1(y_1)\otimes1\  , \\[4pt]
\label{f7}
\ca^{i+1\,\a,\,i\a}&=&-\left(\ca^{i\a,\,i+1\,\a}\right)^\+
\=-\bigl(\phi^{(1)}_{i+1\,\a}(x)\big)^\+
\otimes\beta_1(y_1)\otimes1\  , \\[4pt]
\label{ff6}
\ca^{i\a,\,i\,\a+1}&=:&\quad\Phi^{(2)}_{i\,\a+1}~\qquad~=~\quad
\phi^{(2)}_{i\,\a+1}(x)\otimes1\otimes\bar{\b}_2(y_2)\  , \\[4pt]
\label{ff7}
\ca^{i\,\a+1,\,i\a}&=&-\left(\ca^{i\a,\,i\,\a+1}\right)^\+
\=-\bigl(\phi^{(2)}_{i\,\a+1}(x)\bigr)^\+ \otimes1\otimes\beta_2(y_2)\  .
\eea
All other components $\ca^{i\a ,\,j\b}$ vanish, while the bundle
morphisms $\Phi_{i+1\,\a}^{(1)}\in{\rm Hom}(\Ecal_{i+1\,\a},\Ecal_{i\a})$ and
$\Phi_{i\,\a+1}^{(2)}\in {\rm Hom}(\Ecal_{i\,\a+1},\Ecal_{i\a})$ obey
${\Phi^{(1)}_{m_1+1\;\a}}=0=\Phi^{(1)}_{0 \a}$ for $\a = 0,1,\ldots
,m_2$ and ${\Phi^{(2)}_{i\; m_2+1}}=0=\Phi^{(2)}_{i0}$ for $i = 0,1,\ldots
,m_1$. The gauge potentials $A^{i\a}\in{\rm u}(k_{i\a})$ are connections
on the hermitean vector bundles $E_{k_{i\a}}\to M$, while the
bi-fundamental scalar fields $\phi_{i+1\,\a}^{(1)}$ and $\phi_{i\,\a+1}^{(2)}$
transform in the representations
$\underline{V}_{\,k_{i\a}}\otimes\underline{V}_{\,k_{i+1\,\a}}^\vee$
and
$\underline{V}_{\,k_{i\a}}\otimes\underline{V}_{\,k_{i\,\a+1}}^\vee$
of the subgroups ${\rm U}(k_{i\a})\time{\rm U}(k_{i+1\,\a})$ and
${\rm U}(k_{i\a})\time{\rm U}(k_{i\,\a+1})$ of the original $\uk$ gauge group.

The curvature two-form $\cf =\diff\ca + \ca\wedge\ca$ of the
connection $\ca$ has components $\cf_{\hat\mu\hat\nu}=
\pa_{\hat\mu}\ca_{\hat\nu} - \pa_{\hat\nu}\ca_{\hat\mu} + [\ca_{\hat\mu},
\ca_{\hat\nu}]$ in local coordinates $(x^{\hat\mu})$, where
$\pa_{\hat\mu}:=\pa /\pa x^{\hat\mu}$. It also take values in the Lie
algebra ${\rm u}(k)$, and in local coordinates on $M\time\C
P^1_{(1)}\time\C P^1_{(2)}$ it can be written as
\bea
{\cf}&=&\sfrac{1}{2}\,{\cf}_{\m \n }~\diff x^{\m }\wedge\diff x^{\n } +
{\cf}_{\m  y_1}~\diff x^{\m }\wedge\diff y_1 +
{\cf}_{\m \yb_1}~\diff x^{\m }\wedge\diff\yb_1
+{\cf}_{\m  y_2}~\diff x^{\m }\wedge\diff y_2 \nonumber\\
&&+\,{\cf}_{\m \yb_2}~\diff x^{\m }\wedge\diff\yb_2
+ {\cf}_{y_1\yb_1}~\diff y_1\wedge\diff\yb_1+
{\cf}_{y_2\yb_2}~\diff y_2\wedge\diff\yb_2
+ {\cf}_{y_1 y_2}~\diff y_1\wedge\diff y_2\nonumber\\
&&+\,{\cf}_{\yb_1\yb_2}~\diff \yb_1\wedge\diff\yb_2
+ {\cf}_{y_1\yb_2}~\diff y_1\wedge\diff\yb_2 +
{\cf}_{\yb_1 y_2}~\diff \yb_1\wedge\diff y_2 \ .
\label{curvprod}
\eea
The calculation of the curvature (\ref{curvprod}) for $\ca$ of the
form (\ref{f4})--(\ref{ff7}) yields
\begin{equation}\label{f10}
\cf\=\left(\cf^{i\a ,\, j\b}\right) \qquad\mbox{with}\quad
\cf^{i\a ,\, j\b} \= {\diff}\ca^{i\a ,\, j\b} +
\sum_{l =0}^{m_1}~\sum_{\g =0}^{m_2}\,\ca^{i\a ,\, l\g}\wedge \ca^{l\g ,\, j\b}
\ ,
\end{equation}
where
\bea
\cf^{i\a,\, i\a}&=&F^{i\a}+f^{(1)}_{m_1-2i}
+f^{(2)}_{m_2-2\a} \nonumber\\ &&
+\,\left(\phi^{(1)}_{i+1\, \a}\,\big(\phi^{(1)}_{i+1\, \a}\big)^\+ -
\big(\phi_{i\a}^{(1)}\big)^\+\, \phi_{i\a}^{(1)} \right)
\left(\beta_1\wedge{\bar\beta}_1\right) \nonumber\\ &&
+\,\left(\phi^{(2)}_{i\, \a+1}\,
\big(\phi^{(2)}_{i\, \a+1}\big)^\+ - \big(\phi_{i\a}^{(2)}
\big)^\+\, \phi_{i\a}^{(2)} \right)
\left(\beta_2\wedge{\bar\beta}_2\right) \ , \label{f11} \\[4pt]
\label{f12}
\cf^{i\a,\:i+1\,\a}&=&D \phi^{(1)}_{i+1\:\a}\wedge{\bar\b}_1  \ , \\[4pt]
\label{f13}
\cf^{i+1\,\a,\:i\a}&=&-\left( \cf^{i\a,\:i+1\,\a}\right)^\+
\=-\bigl(D \phi^{(1)}_{i+1\,\a}\bigr)^\+
\wedge \beta_1 \ , \\[4pt]
\label{ff12}
\cf^{i\a,\,i\,\a+1}&=&D \phi^{(2)}_{i\,\a+1}\wedge{\bar\b}_2 \ , \\[4pt]
\label{ff13}
\cf^{i\,\a+1,\:i\a}&=&-\left( \cf^{i\a,\:i\,\a+1}\right)^\+
\=-\bigl(D \phi^{(2)}_{i\,\a+1}\bigr)^\+
\wedge \beta_2 \ , \\[4pt] \label{fc1}
\cf^{i\a,\: i+1\,\a+1}&=&\left(\phi^{(1)}_{i+1\, \a}\,\phi^{(2)}_{i+1\, \a+1}-
\p_{i\,\a+1}^{(2)}\, \phi_{i+1\,\a+1}^{(1)} \right)
\,{\bar\b}_1\wedge{\bar\beta}_2\ ,\\[4pt] \label{fc2}
\cf^{i+1\,\a+1,\: i\,\a}&=&- \big(\cf^{i\a,\: i+1\,\a+1}\big)^\+ \= -
\left(\phi^{(1)}_{i+1\, \a}\,\phi^{(2)}_{i+1\,
    \a+1}-\p_{i\,\a+1}^{(2)}\,\phi_{i+1\,\a+1}^{(1)} \right)^\+\,
{\b}_1\wedge{\beta}_2\ ,\\[4pt] \label{fc3}
\cf^{i\,\a+1,\: i+1\,\a}&=&\left(\big(\phi^{(2)}_{i\, \a+1}
\big)^\+\,\phi^{(1)}_{i+1\, \a}-
\p_{i+1\,\a+1}^{(1)}\,\big(\phi_{i+1\,\a+1}^{(2)}\big)^\+ \right)
\,{\bar\b}_1\wedge{\beta}_2\ ,\\[4pt] \label{fc4}
\cf^{i+1\,\a,\: i\,\a+1}&=&- \big(\cf^{i\,\a+1,\: i+1\,\a}\big)^\+ \=
\left(\big(\phi^{(1)}_{i+1\, \a}
\big)^\+\,\phi^{(2)}_{i\, \a+1}-\p_{i+1\,\a+1}^{(2)}\,
\big(\phi_{i+1\,\a+1}^{(1)}\big)^\+ \right)\,
{\bar\b}_2\wedge{\beta}_1 \label{f14}
\eea
with all other components vanishing. We have suppressed the tensor
product structure pertaining to $M\time \C P^1\time\C P^1$ in
(\ref{f11})--(\ref{f14}). Here $F^{i\a}:=\diff A^{i\a} +
A^{i\a}\wedge A^{i\a}=\frac12\,F_{\mu\nu}^{i\a}~\diff x^\mu\wedge\diff
x^\nu$ are the curvatures of the bundles $E_{k_{i\a}}\to M$, and we
have introduced the bi-fundamental covariant derivatives
\bea
D \phi^{(1)}_{i+1\,\a}&:=&
\diff \phi^{(1)}_{i+1\,\a} + A^{i\a}\,\phi^{(1)}_{i+1\,\a} -
\phi^{(1)}_{i+1\,\a}\,A^{i+1\,\a} \ , \\[4pt]
D \phi^{(2)}_{i\,\a+1}&:=&
\diff \phi^{(2)}_{i\,\a+1} + A^{i\a}\,\phi^{(2)}_{i\,\a+1} -
\phi^{(2)}_{i\,\a+1}\,A^{i\,\a+1} \ .
\label{covderivs}\eea

{}From (\ref{f11})--(\ref{f14}) we find the non-vanishing field
strength components
\bea\label{f15}
\cf^{i\a,\,i\a}_{{\mu} {\nu} }&=&F^{i\a}_{{\mu} {\nu} } \ , \\[4pt]
\label{f16}
\cf^{i\a,\:i+1\,\a}_{{\mu} \yb_1}&=&\frac{1}{1 +y_1{\yb_1}}\,
D_{{\mu} } \phi^{(1)}_{i+1\,\a}~=~-
\bigl(\cf^{i+1\,\a,\:i\a}_{{\mu} y_1}\bigr)^\+ \ , \\[4pt]\label{fc16}
\cf^{i\a,\:i\,\a+1}_{{\mu} \yb_2}&=&\frac{1}{1 +y_2{\yb_2}}\,
D_{{\mu} } \phi^{(2)}_{i\,\a+1}~=~-
\bigl(\cf^{i\,\a+1,\:i\a}_{{\mu} y_2}\bigr)^\+ \ , \\[4pt]\label{fc17}
\cf^{i\a,\,i\a}_{y_1\yb_1}&=& - \frac{1}{(1 +y_1{\yb}_1)^2}\,
\left((m_1-2i)~\Idd_{k_{i\a}}+\big(\phi^{(1)}_{i\a}
\big)^\+\,\phi^{(1)}_{i\a} -
\phi^{(1)}_{i+1\,\a}\,\big(\phi^{(1)}_{i+1\,\a}\big)^\+\right)
\ ,\\[4pt]\label{fc18}
\cf^{i\a,\,i\a}_{y_2\yb_2}&=& - \frac{1}{(1 +y_2{\yb}_2)^2}\,
\left((m_2-2\a)~\Idd_{k_{i\a}}+\big(\phi^{(2)}_{i\a}\big)^\+\,
\phi^{(2)}_{i\a} -
\phi^{(2)}_{i\,\a+1}\,\big(\phi^{(2)}_{i\,\a+1}\big)^\+\right)
\eea
and
\bea
\cf^{i\a,\:i+1\,\a+1}_{\yb_1\yb_2}&=&
\frac{\phi^{(1)}_{i+1\,\a}\,\phi^{(2)}_{i+1\,\a+1} -
\phi^{(2)}_{i\,\a+1}\,\phi^{(1)}_{i+1\,\a+1}}
{(1 +y_1{\yb}_1)\,(1 +y_2{\yb}_2)}\=-
\big(\cf^{i+1\,\a+1,\: i\a}_{y_1y_2}\big)^\+
\ , \label{Fyy} \\[4pt]
\cf^{i\,\a+1,\:i+1\,\a}_{y_1\yb_2}&=&
\frac{\big(\phi^{(2)}_{i\,\a+1}\big)^\+\,\phi^{(1)}_{i+1\,\a}-
\phi^{(1)}_{i+1\,\a+1}\,\big(\phi^{(2)}_{i+1\,\a+1}\big)^\+}
{(1 +y_1{\yb}_1)\,(1 +y_2{\yb}_2)}
\=-\big(\cf^{i+1\,\a,\: i\,\a+1}_{\yb_1y_2}\big)^\+ \ .
\label{Fyyb}\eea
Note that at this stage we do not generally require the imposition of
the holomorphic constraints (\ref{phicommrels}) in this ansatz,
which ensure that the bundle diagram (\ref{bundlediag})
commutes. Later on we will see that they arise as a {\it dynamical}
constraint for BPS solutions of the Yang-Mills equations on $M\time\C
P^1_{(1)}\time\C P^1_{(2)}$ that force the vanishing of the
cross-components (\ref{Fyy}) of the field strength tensor between the
two copies of the sphere. In fact, our particular ansatz in the
noncommutative gauge theory will automatically satisfy this condition,
as well as the analogous ones which force the cross-components
(\ref{Fyyb}) to vanish.

\subsection{The Yang-Mills functional\label{YMfunct}}

Let us now consider the equivariant reduction of the Yang-Mills
lagrangian
\bea
L^{~}_{\rm YM}&:=&-\sfrac{1}{4}\,\sqrt{\hat g}~\tr^{~}_{k\times k}\
\cf_{\hat{\m}\hat{\n}}\,\cf^{\hat{\m} \hat{\n}} \nonumber\\[4pt]
&=&-\sfrac{1}{4}\,\sqrt{\hat g}~\tr^{~}_{k\times k}
\Bigl\{\cf_{\mu \nu }\,\cf^{\mu \nu }+
g^{\mu \nu }\, g^{y_1\yb_1}\,
\left(\cf_{\mu y_1}\,\cf_{\nu \yb_1}+\cf_{\mu  \yb_1}\,\cf_{\nu  y_1}
\right)\Bigr.\nonumber\\
&& +\,g^{\mu \nu }\, g^{y_2\yb_2}\,
\left(\cf_{\mu y_2}\,\cf_{\nu \yb_2}+\cf_{\mu  \yb_2}\,\cf_{\nu  y_2}
\right)-2\,\left(g^{y_1\yb_1}\,\cf_{y_1\yb_1}\right)^2-
2\,\left(g^{y_2\yb_2}\,\cf_{y_2\yb_2}\right)^2\nonumber\\
&& +\,2\, \Bigl.g^{y_1\yb_1}\, g^{y_2\yb_2}\,
\left(\cf_{\yb_1\yb_2}\,\cf_{y_1 y_2}+
\cf_{y_1 y_2}\,\cf_{\yb_1\yb_2} + \cf_{\yb_1 y_2}\,\cf_{y_1 \yb_2}+
\cf_{y_1 \yb_2}\,\cf_{\yb_1 y_2}
\right)\Bigr\}\ ,
\label{lagrprod}
\eea
where $\hat g=\det (g_{\hat\mu\hat\nu})=g~g^{}_{\C P^1_{(1)}}~g^{}_{\C
  P^1_{(2)}}$ with $g=\det(g_{\mu\nu})$ and
\beq
{\sqrt {g^{}_{\C P^1_{(\ell)}}}} \=
\frac{2\,R^2_\ell}{\left(1+y_\ell\bar y_\ell
\right)^2}\=\left(g^{y_\ell\yb_\ell}\right)^{-1} \ .
\eeq
For the ansatz of the Section~\ref{invgauge} above we substitute
(\ref{f15})--(\ref{Fyyb}). After integration over the spherical
factors $\C P^1_{(1)}\time\C P^1_{(2)}$, the dimensional reduction of
the corresponding Yang-Mills action functional is given by
\bea
S^{~}_{\rm YM}&:=&\int_{M\time {\C P^1_{(1)}}\time {\C P^1_{(2)}}}
{\diff^{2n+4}}x~L_{\rm YM}^{~}
\nonumber\\[4pt]
&=&\pi\,R_1^2\,R_2^2\, \int_M\diff^{2n}x ~\sqrt{g }~
\sum_{i=0}^{m_1}~\sum_{\a=0}^{m_2}\,\tr^{~}_{k_{i\a}\time k_{i\a}}\left[
\bigl(F_{{\mu} {\nu} }^{i\a}\bigr)^\+\,F^{i\a\,{\mu} {\nu} }\right.
\label{SYMred} \\
&& +\, \frac{1}{R_1^2}\,\bigl(D_{{\mu} } \phi_{i+1\,\a}^{(1)}\bigr)\,
\bigl(D^{{\mu} }\phi_{i+1\,\a}^{(1)}\bigr)^\+
+ \frac{1}{R_1^2}\,\bigl(D_{{\mu} } \phi_{i\a}^{(1)}\bigr)^\+\,
\bigl(D^{{\mu} }\phi_{i\a}^{(1)}\bigr)
\nonumber\\
&& +\, \frac{1}{R_2^2}\,\bigl(D_{{\mu} } \phi_{i\,\a+1}^{(2)}\bigr)\,
\bigl(D^{{\mu} }\phi_{i\,\a+1}^{(2)}\bigr)^\+
+ \frac{1}{R_2^2}\,\bigl(D_{{\mu} } \phi_{i\a}^{(2)}\bigr)^\+\,
\bigl(D^{{\mu} }\phi_{i\a}^{(2)}\bigr)
\nonumber\\
&& +\,
\frac{1}{2\,R_1^4}\,\left((m_1-2i)~\Idd_{k_{i\a}}
+\big(\phi_{i\a}^{(1)}\big)^\+\,\phi^{(1)}_{i\a} -
\phi^{(1)}_{i+1\, \a}\,\big(\phi^{(1)}_{i+1\,\a}\big)^\+\right)^2
\nonumber\\
&& +\,
\frac{1}{2\,R_2^4}\,\left((m_2-2\a)~\Idd_{k_{i\a}}
+\big(\phi_{i\a}^{(2)}\big)^\+\,\phi^{(2)}_{i\a} -
\phi^{(2)}_{i\, \a+1}\,\big(\phi^{(2)}_{i\,\a+1}\big)^\+\right)^2
\nonumber\\
&& +\,
\frac{1}{2\,R_1^2\,R_2^2}\left(\phi_{i+1\,\a}^{(1)}\,\phi^{(2)}_{i+1\:\a+1} -
\phi^{(2)}_{i\, \a+1}\,\phi^{(1)}_{i+1\,\a+1}\right)\,
\left(\p_{i+1\,\a}^{(1)}\,\phi^{(2)}_{i+1\,\a+1} -
\phi^{(2)}_{i\, \a+1}\,\phi^{(1)}_{i+1\,\a+1}\right)^\+
\nonumber\\
&& +\,
\frac{1}{2\,R_1^2\,R_2^2}\left(\phi_{i\,\a-1}^{(1)}\,\phi^{(2)}_{i\a} -
\phi^{(2)}_{i-1\, \a}\,\phi^{(1)}_{i\a}\right)^\+\,
\left(\p_{i\,\a-1}^{(1)}\,\phi^{(2)}_{i\a} -
\phi^{(2)}_{i-1\, \a}\,\phi^{(1)}_{i\a}\right)
\nonumber\\
&& +\,
\frac{1}{2\,R_1^2\,R_2^2}\left(\big(\p_{i\a}^{(2)}\big)^\+\,
\phi^{(1)}_{i+1\:\a-1} -
\phi^{(1)}_{i+1\, \a}\,\big(\phi^{(2)}_{i+1\,\a}\big)^\+\right)\,
\left(\big(\p_{i\a}^{(2)}\big)^\+\,\phi^{(1)}_{i+1\:\a-1} -
\phi^{(1)}_{i+1\, \a}\,\big(\phi^{(2)}_{i+1\,\a}\big)^\+
\right)^\+
\nonumber\\
&& +\,
\frac{1}{2\,R_1^2\,R_2^2}\left(\big(\p_{i-1\,\a+1}^{(2)}
\big)^\+\,\phi^{(1)}_{i\a}-
\phi^{(1)}_{i\,\a+1}\,\big(\phi^{(2)}_{i\,\a+1}\big)^\+\right)^\+\,
\left(\big(\p_{i-1\,\a+1}^{(2)}\big)^\+\,\phi^{(1)}_{i\a}-
\phi^{(1)}_{i\,\a+1}\,\big(\phi^{(2)}_{i\,\a+1}\big)^\+
\right)\Bigl.\Bigr] \ . \nonumber
\eea

All individual terms in (\ref{SYMred}) are $k_{i\a}\time k_{i\a}$ matrices.
Recall that $\phi^{(1)}_{i+1\,\a}$ are $k^{}_{i\a}\time k^{}_{i+1\,\a}$
matrices,
$\phi^{(2)}_{i\,\a+1}$ are $k^{}_{i\a}\time k^{}_{i\,\a+1}$ matrices and
$A^{i\a}_{\m }$ are $k^{}_{i\a}\time k^{}_{i\a}$ matrices. The
action (\ref{SYMred}) is non-negative, and it can be regarded as
an energy functional for static fields on $\R^{0,1}\time M$ in the
temporal gauge.

\subsection{Noncommutative gauge theory \label{NGT}}

When we come to construct explicit solutions of the Yang-Mills
equations we will specialize to the K\"ahler manifold
$M=\R^{2n}$ with metric tensor $g_{\m\n}=\de_{\m\n}$ and pass to
a noncommutative deformation $\R^{2n}\to\R^{2n}_\th$. The
spherical factors $\C P^1_{(\ell)}$, $\ell=1,2$, will always remain
commutative spaces. The noncommutative space~$\R^{2n}_\th$ is
defined by declaring its coordinate functions
$\xh^1,\ldots,\xh^{2n}$ to obey the Heisenberg algebra relations
\begin{equation}
[ \xh^\mu\,,\,\xh^\nu\,] = \im\th^{\mu\nu}
\label{Heisenalg}\end{equation}
with a constant real antisymmetric tensor~$\th^{\mu\nu}$ of maximal rank~$n$.
Via an orthogonal transformation of the coordinates, the matrix
$\theta=(\th^{\m\n})$ can be rotated into its canonical block-diagonal
form with non-vanishing components
\begin{equation}\label{tha}
\th^{{2a-1}\ {2a}} \= -\th^{{2a}\ {2a-1}} \ =:\ \th^a
\end{equation}
for $a=1,\dots,n$. We will assume for definiteness that all
$\th^a>0$. The noncommutative version of the complex coordinates
(\ref{zz}) has the non-vanishing commutators
\begin{equation}\label{zzb}
\big[\zh^a\,,\,\zbh^{\bb}\,\big] \= -2\,\de^{a\bb}\,\th^a \
=:\ \th^{a\bb} \= -\th^{\bb a}\ < \ 0 \ .
\end{equation}
Taking the product of $\R^{2n}_\th$ with the commutative spheres $\C
P^1_{(1)}\time\C P^1_{(2)}$ means extending the noncommutativity
matrix $\th$ by vanishing entries along the four new directions.

The algebra (\ref{Heisenalg}) can be represented on the Fock
space~$\Hcal$ which may be realized as the linear span
\begin{equation}
\Hcal=\bigoplus_{r_1,\dots,r_n=0}^\infty\,\C|r_1,\ldots,r_n\> \ ,
\label{Fockspace}\eeq
where the orthonormal basis states
\beq
|r_1,\ldots,r_n\>=\prod_{a=1}^{n}\,\left(2\,
\th^a\,r_a!\right)^{-1/2}\,(\zh^{a})^{r_a}|0,\ldots,0\>
\end{equation}
are connected by the action of creation and annihilation operators
subject to the commutation relations
\begin{equation}
\Bigl[\,\frac{\zbh^{\bb}}{\sqrt{2\,\th^b}}\ ,\ \frac{\zh^a}{\sqrt{2\,\th^a}}\,
\Bigr] = \de^{a\bb} \ .
\end{equation}
In the Weyl operator realization $f\mapsto\fh$ which maps Schwartz
functions $f$ on $\R^{2n}$ into compact operators $\fh$ on $\Hcal$,
coordinate derivatives are given by inner derivations of the
noncommutative algebra according to
\begin{equation}\label{pazzf}
\widehat{\pa_{z^a} f}\=\th_{a\bb}\,\big[\zbh^{\bb} \,,\, \fh\,\big]
\ =:\ \pa_{\zh^a} \fh
\qquad\textrm{and}\qquad
\widehat{\pa_{\zb^{\ab}} f}\=\th_{\ab b}\,\big[\zh^b \,,\, \fh\,\big]
\ =:\ \pa_{\zbh^{\,\ab}} \fh\ ,
\end{equation}
where $\th_{a\bb}$ is defined via $\th_{b\bar{c}}\,\th^{\bar{c}a}=\de^a_b$
so that $\th_{a\bb}=-\th_{\bb a}=\frac{\de_{a\bb}}{2\,\th^a}$.
On the other hand, integrals are given by traces over the Fock space
$\Hcal$ as
\begin{equation}\label{intNC}
\int_{\R^{2n}} \diff^{2n} x~f(x)=
{\rm Pf}(2\pi\,\th)~\textrm{Tr}^{~}_\Hcal~\fh\ .
\end{equation}

Vector bundles $E\to\R^{2n}$ whose typical
fibres are complex vector spaces $\underline{V}$ are replaced by the
corresponding (trivial) projective modules $\underline{V}\otimes\Hcal$ over
$\R^{2n}_\theta$. The field strength components along $\rt$
in~(\ref{curvprod}) read $\hat{\cf}_{\m\n}=\pa_{\xh^{\m}}\hat{\ca}_{\n}-
\pa_{\xh^{\n}}\hat{\ca}_{\m}+[\hat{\ca}_{\m},\hat{\ca}_{\n}]$,
where $\hat{\ca}_{\m}$ are simultaneously valued in ${\rm u}(k)$ and
in ${\rm End}(\Hcal)$. To avoid a cluttered notation, we will omit the
hats over operators, so that all equations will have the same form as
previously but considered as equations in ${\rm
  End}(\,\underline{V}\otimes\Hcal)$. The main advantage of this
prescription will arise from the fact that, unlike $\R^{2n}$, the
noncommutative space $\R_\theta^{2n}$ has a non-trivial K-theory which
allows for gauge field configurations of non-trivial topological
charge while retaining the simple geometry of flat contractible
space.

\bigskip

\section{Quiver gauge theory and graded connections\label{QGT}}

\noindent
In this section we will exploit the fact that the $G$-equivariant
reduction carried out in the previous section has a natural
interpretation as the representation of a particular class of quivers
in the category of vector bundles over the K\"ahler manifold $M$,
i.e. as a quiver bundle over $M$~\cite{A-CG-P1,A-CG-P2,GK1}. The most
natural notion of gauge field on a quiver bundle is provided by that
of a {\it graded connection} as introduced in~\cite{PS1}. After
describing some general aspects of the quivers related to our
analysis, we will rewrite the equivariant decomposition of the gauge
fields of the previous section in terms of graded connections on the pertinent
quivers. Besides its mathematical elegance, the main advantage of this
representation is that it will make the physical interpretations of
our field configurations completely transparent later on. Treatments
of the theory of quivers can be found in~\cite{Quiverbooks}.

\subsection{The ${\rm A}_{m_1+1}\oplus{\rm A}_{m_2+1}$ quiver  and its
  representations\label{CPquiver}}

A quiver is an oriented graph, i.e. a set of vertices together with a
set of arrows between the vertices. For a given pair of positive
integers $m_1,m_2$, it is clear that the bundle diagram
(\ref{bundlediag}) can be naturally associated to a quiver
$\quiver_{(m_1,m_2)}$. The nodes of this quiver are labelled by
monopole charges giving the vertex set
$\quiver_{(m_1,m_2)}^{(0)}=\{(v^{(1)}_i,v^{(2)}_\a)=(m_1-2i,m_2-2\a)~|~0\leq
i\leq m_1\,,\,0\leq\a\leq m_2\}$. The arrow set is given by
$\quiver_{(m_1,m_2)}^{(1)}=\{\zeta^{(\ell)}_{i\a}~|~\ell=1,2\,,\,0\leq i\leq
m_1\,,\,0\leq\a\leq m_2\}$ with
$\zeta_{i+1\,\a}^{(1)}:(v_{i+1}^{(1)},v_\a^{(2)})
\mapsto(v_{i}^{(1)},v_\a^{(2)})$
and
$\zeta_{i\,\a+1}^{(2)}:(v_i^{(1)},v_{\a+1}^{(2)})
\mapsto(v_{i}^{(1)},v_{\a}^{(2)})$.
A path in $\quiver_{(m_1,m_2)}$ is a sequence of arrows in
$\quiver_{(m_1,m_2)}^{(1)}$ which compose. If the head of
$\zeta_{i\a}^{(\ell)}$ is the same node as the tail of
$\zeta_{i'\a'}^{(\ell'\,)}$, then we may produce a path
$\zeta_{i'\a'}^{(\ell'\,)}\,\zeta_{i\a}^{(\ell)}$ consisting of
$\zeta_{i\a}^{(\ell)}$ followed by $\zeta_{i'\a'}^{(\ell'\,)}$. To each vertex
$(m_1-2i,m_2-2\a)$ we associate the trivial path $e_{i\a}$ of
length~$0$. Each arrow $\zeta_{i\a}^{(\ell)}$ itself may be associated to
a path of length~$1$. A relation $r$ of the quiver is a formal finite
sum of paths. From (\ref{phicommrels}) it follows that the set
$\rel_{(m_1,m_2)}$ of relations of $\quiver_{(m_1,m_2)}$ are given by
$r_{i\a}=\zeta^{(1)}_{i+1\:\a}\,\zeta^{(2)}_{i+1\:\a+1}-
\zeta^{(2)}_{i\:\a+1}\,\zeta^{(1)}_{i+1\:\a+1}$ for $0\leq i\leq m_1$,
$0\leq \a\leq m_2$.

If we set $M={\rm point}$ in the construction of
Section~\ref{eqvecbun}, then we obtain a representation
$\underline{\cal V}$ of the quiver $\quiver_{(m_1,m_2)}$ obtained by
placing the $G$-modules $\underline{V}_{\,k_{i\a}}$ in
(\ref{genrepSU2Uk}) at the vertices $(m_1-2i,m_2-2\a)$. Recalling that
the nodes of the quiver arose as the set of weights for the action of
the Borel subgroup $K$ on the bundle $E\to M$, we obtain natural
equivalence functors between the categories of holomorphic
representations of $K$ and indecomposable representations of the quiver with
relations $(\quiver_{(m_1,m_2)}\,,\,\rel_{(m_1,m_2)})$, and also with the
category of holomorphic homogeneous vector bundles over $\C
P^1\time\C P^1\cong G^{\rm c}/K$. In particular, there is a
one-to-one correspondence between $G$-equivariant vector bundles over
$\C P^1\time\C P^1$ and commutative diagrams on the quiver
$\quiver_{(m_1,m_2)}$. In the case of a generic K\"ahler manifold $M$,
any $G$-equivariant bundle over $M\time\C P^1\time\C P^1$ defines a
quiver representation obtained by placing the vector bundles
$E_{k_{i\a}}\to M$ at the vertices $(m_1-2i,m_2-2\a)$, as in
(\ref{bundlediag}). It follows that there is a one-to-one
correspondence between such bundles and indecomposable
$(\quiver_{(m_1,m_2)}\,,\,\rel_{(m_1,m_2)})$-bundles over $M$. Neither the
holomorphicity of the quiver representation nor the relations need
generically hold for the decomposition of gauge fields given in
Section~\ref{invgauge}, but instead will arise as a dynamical effect
from a specific choice of ansatz. Note that when one passes to the
corresponding noncommutative gauge theory, one is faced with
infinite-dimensional quiver representations $\underline{\cal
  V}\otimes\Hcal$, and one of the goals of our later constructions
will be to find appropriate truncations to finite-dimensional modules
over $\quiver_{(m_1,m_2)}$.

To aid in the construction of quiver representations, one defines the
path algebra $\pathalg_{(m_1,m_2)}$ of $\quiver_{(m_1,m_2)}$ to be the
vector space over $\C$ generated by all paths, together with the
multiplication given by concatenation of paths. If two paths do not
compose then their product is defined to be~$0$. The trivial paths are
idempotents, $e_{i\a}^2=e^{~}_{i\a}$, and thereby define a collection of
projectors on the finite-dimensional free algebra
$\pathalg_{(m_1,m_2)}$. Imposing relations on the quiver then amounts
to taking the quotient of $\pathalg_{(m_1,m_2)}$ by the ideal
generated by the $r_{i\a}$. Given a representation $\underline{\cal
  V}$ of the algebra $\pathalg_{(m_1,m_2)}$, we can form the vector spaces
$\underline{V}_{\,k_{i\a}}=\underline{\cal V}\cdot
e_{i\a}\cong\C^{k_{i\a}}$. The elements of $\pathalg_{(m_1,m_2)}$
corresponding to arrows in $\quiver_{(m_1,m_2)}$ yield linear
maps between the $\underline{V}_{\,k_{i\a}}$ which have to satisfy the
relations
$r_{i\a}=0$. It follows that representations of the path algebra
$\pathalg_{(m_1,m_2)}/\rel_{(m_1,m_2)}$ are equivalent to quiver
representations of
$(\quiver_{(m_1,m_2)}\,,\,\rel_{(m_1,m_2)})$~\cite{Quiverbooks}. Such
a representation is specified by giving the ordered collection of
positive integers $\vec k=\vec k_{\underline{\cal
    V}}:=(k_{i\a})_{0\leq i\leq m_1,0\leq\a\leq m_2}$, called the
dimension vector of the quiver representation, at the vertices of
$\quiver_{(m_1,m_2)}$.

A useful set of quiver representations $\underline{\cal P}_{\,i\a}$ is
defined for each vertex of $\quiver_{(m_1,m_2)}$ by $\underline{\cal
  P}_{\,i\a}:=e_{i\a}\cdot\pathalg_{(m_1,m_2)}$, which is the subspace of
$\pathalg_{(m_1,m_2)}$ generated by all paths starting at the node
$(m_1-2i,m_2-2\a)$. Multiplying on the right by elements of the
path algebra $\pathalg_{(m_1,m_2)}$ makes $\underline{\cal P}_{\,i\a}$
into a right $\pathalg_{(m_1,m_2)}$-module and hence a quiver
representation. This path algebra representation has many special
properties. The collection of all modules $\underline{\cal
  P}_{\,i\a}$, $0\leq i\leq m_1$, $0\leq\a\leq m_2$ are exactly the
set of all indecomposable projective representations of the quiver
$\quiver_{(m_1,m_2)}$, with the natural isomorphism
\beq
\pathalg_{(m_1,m_2)}=\bigoplus_{i=0}^{m_1}~\bigoplus_{\a=0}^{m_2}\,
\underline{\cal P}_{\,i\a}
\label{pathalgPiso}\eeq
as right $\pathalg_{(m_1,m_2)}$-modules. Furthermore, for any quiver
representation (\ref{genrepSU2Uk}) there is a natural isomorphism
\beq
\Hom\bigl(\,\underline{\cal P}_{\,i\a}\,,\,\underline{\cal V}\,
\bigr)=\underline{V}_{\,k_{i\a}} \ ,
\label{HomPViso}\eeq
and in particular
\beq
\Hom\bigl(\,\underline{\cal P}_{\,j\b}\,,\,\underline{\cal P}_{\,i\a}
\bigr)\=\bigl(\,\underline{\cal P}_{\,i\a}\bigr)_{j\b}\=
e_{i\a}\cdot\pathalg_{(m_1,m_2)}\cdot e_{j\b}
\label{HomPPiso}\eeq
is the vector space spanned by all paths from vertex
$(v_i^{(1)},v_\a^{(2)})$ to vertex $(v_j^{(1)},v_\b^{(2)})$. Imposing
the relations $r_{i\a}$ identifies all such paths and one has
$(\,\underline{\cal P}_{\,i\a})_{j\b}\cong\C$ for the corresponding
quiver representation of
$(\quiver_{(m_1,m_2)}\,,\,\rel_{(m_1,m_2)})$.

A morphism $\,\underline{ f}\,:\underline{\cal
  V}\to\underline{{\cal V}'}$ of two quiver representations is given
by linear maps $ f_{i\a}:\underline{{V}}_{\,k_{i\a}}\to\underline{{
    V}'}_{\,k_{i\a}'}$ for each vertex such that
$\phi_{i+1\,\a}^{\prime\,(1)}\, f_{i\a}=
 f_{i+1\,\a}\,\phi_{i+1\,\a}^{(1)}$
and
$\phi_{i\,\a+1}^{\prime\,(2)}\, f_{i\a}=
 f_{i\,\a+1}\,\phi_{i\,\a+1}^{(2)}$.
This notion defines the abelian category of quiver representations
(or equivalently of right $\pathalg_{(m_1,m_2)}$-modules).
If all linear maps $ f_{i\a}$ are invertible, then
$\,\underline{ f}\,$ is called an isomorphism of quiver
representations. Any two isomorphic representations necessarily have
the same dimension vector $\vec k$. This provides a natural notion of
gauge symmetry in quiver gauge theory. We will return to the issue of
equivalence of representations of the quiver $\quiver_{(m_1,m_2)}$
in Section~\ref{modsp}.

\subsection{Matrix presentation of equivariant gauge
  fields\label{Matrixpres}}

A convenient way of combining the reductions of equivariant gauge fields
is through the formalism of graded connections introduced
in~\cite{PS1}. The first step in this procedure is to rewrite the
decompositions of Section~\ref{invgauge} in a particular matrix form
that reflects the representations of the path algebra given in
(\ref{pathalgPiso})--(\ref{HomPPiso}). The basic idea is that, given
the isomorphisms $(\,\underline{\cal P}_{\,i\a})_{j\b}\cong\C$, one can
identify (\ref{pathalgPiso}) with an algebra of upper triangular
complex matrices. For this, let us write the rank~$k$ equivariant
bundle $E\to M$ in the $\Z_{m_1+1}\time\Z_{m_2+1}$-graded form
\beq\label{E}
E~:=~\bigoplus_{i=0}^{m_1}~\bigoplus_{\a=0}^{m_2}\,E_{k_{i\a}}\=
\bigoplus_{\a=0}^{m_2}\,E_{(m_1)\a} \qquad\mbox{with}\quad
E_{(m_1)\a}~:=~\bigoplus_{i=0}^{m_1}\,E_{k_{i\a}} \ .
\eeq
The algebra $\Omega_\sharp(M,E)$ of differential forms on the manifold~$M$
with values in the bundle $E$ has a total $\Z_{m_1+1}\time\Z_{m_2+1}$
grading defined by combining the grading in (\ref{E}) with the
$\Z$-grading by form degree. Similarly, the
$\Z_{m_1+1}\time\Z_{m_2+1}$ grading of the endomorphism bundle
\beq
\End(E)\=\bigoplus_{i,j=0}^{m_1}~\bigoplus_{\a,\b=0}^{m_2}\,
\Hom(E_{k_{i\a}},E_{k_{j\b}})\=\bigoplus_{\a=0}^{m_2}\,
\End(E_{(m_1)\a})~\oplus~\bigoplus_{\stackrel{\scriptstyle\a,\b=0}
{\a\neq\b}}^{m_2}\,\Hom(E_{(m_1)\a},E_{(m_1)\b})
\label{EndEgrad}\eeq
induces a total $\Z_{m_1+1}\time\Z_{m_2+1}$ grading on the
endomorphism algebra $\Omega_\sharp(M,\End~E)$.

A graded connection on $E$ is a derivation on $\Omega_\sharp(M,E)$ which shifts
the total $\Z_{m_1+1}\time\Z_{m_2+1}$ grading by~$1$, and is thus an
element of the degree~$1$ subspace of $\Omega_\sharp(M,\End~E)$. For a
given module (\ref{genrepSU2Uk}) over the quiver
$\quiver_{(m_1,m_2)}$, the zero-form components in this subspace
represent the arrows of $\quiver_{(m_1,m_2)}$ and are defined by appropriately
assembling the Higgs fields of the equivariant gauge potentials into
off-diagonal operators in (\ref{EndEgrad}) acting on
the decomposition in (\ref{E}). To this end we introduce square
matrices of morphisms acting on the bundles $E_{(m_1)\a}$ through
\beq
\mphi^{(1)}_{(m_1)\a}~:=~\begin{pmatrix}~0&\p^{(1)}_{1\a}&0&\dots&0\\
{}~0&0&\p^{(1)}_{2\a}&\ddots&\vdots\\
{}~\vdots&\vdots&\ddots&\ddots&0\\~0&0&\dots&0&\p^{(1)}_{m_1\a}\\
{}~0&0&\dots&0&0\end{pmatrix}\qquad \mbox{with}\quad \a \=0,1,\dots,m_2
\label{mphi1alph}\eeq
and assemble them into a $k\time k$ matrix with respect to the
grading (\ref{E}) and (\ref{EndEgrad}) as
\beq
\mphi^{(1)}_{(m_1,\, m_2)}:=\begin{pmatrix}\mphi^{(1)}_{(m_1)0}&0&
0&\dots&0\\0&\mphi^{(1)}_{(m_1)1}&0&\dots&0\\
0&0&\mphi^{(1)}_{(m_1)2}&\ddots&\vdots\\
\vdots&\vdots&\ddots&\ddots&0\\
0&0&\dots&0&\mphi^{(1)}_{(m_1)m_2}\end{pmatrix} \ .
\label{mphi1m}\eeq
Remembering that $\p^{(2)}_{i\: m_2+1}:=0~~\forall i=0,1,\dots,m_1$,
we similarly define matrices of morphisms on $E_{(m_1)\,\a+1}$ through
\beq
\mphi^{(2)}_{(m_1)\,\a+1}~:=~\begin{pmatrix}
\p^{(2)}_{0\,\a+1}&0&0&\dots&0\\
0&\p^{(2)}_{1\,\a+1}&0&\dots&0\\
0&0&\p^{(2)}_{2\,\a+1}&\ddots&\vdots\\
\vdots&\vdots&\ddots&\ddots&0\\
0&0&\dots&0&\p^{(2)}_{m_1\,\a+1}\end{pmatrix}
\qquad \mbox{with}\quad \a \=0,1,\dots,m_2
\label{mphi2alph}\eeq
and assemble them into a $k\time k$ matrix acting on (\ref{E}) as
\beq
\mphi^{(2)}_{(m_1,\, m_2)}:=\begin{pmatrix}~0&\mphi^{(2)}_{(m_1)1}
&0&\dots&0\\~0&0&\mphi^{(2)}_{(m_1)2}&\ddots&\vdots\\
{}~\vdots&\vdots&\ddots&\ddots&0\\~0&0&\dots&0&\mphi^{(2)}_{(m_1)m_2}\\
{}~0&0&\dots&0&0\end{pmatrix} \ .
\label{mphi2m}\eeq

The finite dimensionality of the path algebra (\ref{pathalgPiso})
corresponds to the generic nilpotency properties
\bea
\mphi^{(1)}_{(m_1,\, m_2)},\left(\mphi^{(1)}_{(m_1,\, m_2)}
\right)^2,\dots,\left(\mphi^{(1)}_{(m_1,\, m_2)}\right)^{m_1}&\ne& 0
\qquad\mbox{but}\qquad
\left(\mphi^{(1)}_{(m_1,\, m_2)}\right)^{m_1+1}\= 0\ ,\nonumber\\[4pt]
\mphi^{(2)}_{(m_1,\, m_2)},\left(\mphi^{(2)}_{(m_1,\, m_2)}\right)^2,
\dots,\left(\mphi^{(2)}_{(m_1,\, m_2)}\right)^{m_2}&\ne& 0
\qquad\mbox{but}\qquad
\left(\mphi^{(2)}_{(m_1,\, m_2)}\right)^{m_2+1}\= 0\ .
\label{nilconds}\eea
The holomorphic relations (\ref{phicommrels}) now take the simple
algebraic form of commutativity of the matrices (\ref{mphi1m}) and
(\ref{mphi2m}) as
\beq
\left[\mphi^{(1)}_{(m_1,\, m_2)}\,,\,\mphi^{(2)}_{(m_1,\, m_2)}
\right]=0 \ .
\label{mphicommrels}\eeq
Although a very natural requirement, the condition
(\ref{mphicommrels}) is not necessary for the present formulation and
the relations $\rel_{(m_1,m_2)}$ of the quiver $\quiver_{(m_1,m_2)}$
will only play a prominent role in the subsequent sections.

The one-form components of the graded connection represent the
vertices of $\quiver_{(m_1,m_2)}$ and correspond to diagonal
operators in the decomposition (\ref{EndEgrad}). They can be written
using the canonical orthogonal projections $\Pi_{i\a}: E\to
E_{k_{i\a}}$ of rank~$1$ obeying
\beq
\Pi_{i\a}\,\Pi_{j\b}=\delta_{ij}\,\delta_{\a\b}~\Pi_{i\a}
\label{Piortho}\eeq
which may be represented, with respect to the decomposition
(\ref{E}), by the diagonal matrices
\beq
\Pi_{i\a}=
\bigl(\delta_{ij}\;\delta_{il}\;\de_{\a\b}\;\de_{\a\g}
\bigr)^{j,l=0,1,\dots,m_1}_{\b,\g=0,1,\dots,m_2} \ .
\eeq
The gauge potentials living at the vertices of the quiver may then be
assembled into the $k\time k$ matrix
\beq\label{mA}
\mA^{(m_1,\, m_2)}:=\sum_{i=0}^{m_1}~\sum_{\a=0}^{m_2}\,
A^{i\a}\otimes \Pi_{i\a}\ .
\eeq

To rewrite the equivariant decomposition of the components of the
gauge potentials on the bundle ${\cal E}\to M\time\C
P^1_{(1)}\time\C P^1_{(2)}$, we assemble the monopole connections
into the matrices
\bea\label{ma1}
\ma^{(m_1)}&:=&\sum_{i=0}^{m_1}\,a_{m_1-2i}^{(1)}\otimes\Pi_i
\qquad\mbox{with}\quad
\Pi_i~:=~\bigoplus_{\a=0}^{m_2}\,\Pi_{i\a}\ ,\\[4pt]
\ma^{(m_2)}&:=&\sum_{\a=0}^{m_2}\,a_{m_2-2\a}^{(2)}\otimes\Pi_\a
\qquad\mbox{with}\quad
\Pi_\a~:=~\bigoplus_{i=0}^{m_1}\,\Pi_{i\a}
\label{ma2}
\eea
and the monopole charges labelling the vertices of
$\quiver_{(m_1,m_2)}$ into the matrices
\bea
\mup^{(1)}_{(m_1,\, m_2)}&:=&\sum_{i=0}^{m_1}\,(m_1-2i)\,\Pi_i\ ,\\[4pt]
\mup^{(2)}_{(m_1,\, m_2)}&:=&\sum_{\a=0}^{m_2}\,(m_2-2\a)\,\Pi_\a\ .
\eea
Then the ansatz (\ref{f4})--(\ref{ff7}) can be rewritten in terms of
the matrix operators (\ref{mphi1alph})--(\ref{mphi2m}) and
(\ref{mA})--(\ref{ma2}) as
\bea
\ca_\mu&=&\bigl(\mA^{(m_1,\, m_2)}\bigr)_\mu\otimes1\otimes1 \ ,
\label{calAgradedmu}\\[4pt]
\ca_{y_1}&=&\Idd_k\otimes\bigl(\ma^{(m_1)}\bigr)_{y_1}\otimes1-
\bigl(\mphi^{(1)}_{(m_1,\, m_2)}\bigr)^\+
\otimes (\beta_1)_{y_1}\otimes1 \ , \label{calAy1}\\[4pt]
\ca_{y_2}&=&\Idd_k\otimes1\otimes\bigl(\ma^{(m_2)}\bigr)_{y_2}-
\bigl(\mphi^{(2)}_{(m_1,\, m_2)}\bigr)^\+
\otimes1\otimes (\beta_2)_{y_2} \ , \label{calAy2}\\[4pt]
\ca_{\yb_1}&=&\Idd_k\otimes\bigl(\ma^{(m_1)}\bigr)_{\yb_1}\otimes1
+\mphi^{(1)}_{(m_1,\, m_2)}\otimes
(\bar\beta_1)_{\yb_1}\otimes1 \ , \label{calAyb1}\\[4pt]
\ca_{\yb_2}&=&\Idd_k\otimes1\otimes\bigl(\ma^{(m_2)}\bigr)_{\yb_2}+
\mphi^{(2)}_{(m_1,\, m_2)}\otimes1\otimes(\bar\b_2)_{\yb_2} \ .
\label{calAgraded}\eea
As we will see in Section~\ref{gradedconn}, the scalar potential in
(\ref{SYMred}) can be rewritten entirely in terms of the natural
algebraic operators $\mup^{(1)}_{(m_1,\, m_2)} -
\big[ \mphi^{(1)}_{(m_1,\, m_2)}, (\mphi^{(1)}_{(m_1,\, m_2)})^\+\big]$,
$\mup^{(2)}_{(m_1,\, m_2)} - \big[\mphi^{(2)}_{(m_1,\, m_2)},
(\mphi^{(2)}_{(m_1,\, m_2)})^\+\big]$, $\big[\mphi^{(1)}_{(m_1,\,
  m_2)}, \mphi^{(2)}_{(m_1,\, m_2)}\big]$ and $\big[
\mphi^{(1)}_{(m_1,\, m_2)}, (\mphi^{(2)}_{(m_1,\, m_2)})^\+\big]$ on
the quiver $\quiver_{(m_1,m_2)}$.

\subsection{Examples\label{MatrixEx}}

To help understand the forms of the matrix presentations introduced
above, it is instructive to look at some explicit examples of
$(\quiver_{(m_1,m_2)}\,,\,\rel_{(m_1,m_2)})$-bundles over $M$ before
proceeding further with more of the general formalism.

\bigskip

\noindent
{$\mbf{(m_1,m_2)=(m,0)}$.  } In this case the vertical arrows
$\zeta_{i\a}^{(2)}$ of the quiver $\quiver_{(m,0)}$ are all~$0$ and
the quiver bundle (\ref{bundlediag}) collapses to the holomorphic {\it
  chain}~\cite{A-CG-P3}
\beq
\begin{CD}
E^{}_{k_{m\,0}}~@>{\phi^{(1)}_{m\,0}}>>
{}~E^{}_{k_{m-1\,0}}~@>{\phi_{m-1\,0}^{(1)}}>>~\cdots~@>
{\phi_{10}^{(1)}}>>~E^{}_{k_{00}}
\end{CD}
\label{holchain}\eeq
considered in~\cite{PS1}. The quiver $\quiver_{(m,0)}$ is called the ${\rm
  A}_{m+1}$-quiver. The set of relations
$\rel_{(m,0)}$ is empty and the non-vanishing Higgs fields are
assembled into the zero-form graded connection component
\beq
\mphi^{(1)}_{(m,0)}\=\mphi^{(1)}_{(m)0}\=
\begin{pmatrix}~0&\p^{(1)}_{10}&0&\dots&0\\
{}~0&0&\p^{(1)}_{20}&\ddots&\vdots\\
{}~\vdots&\vdots&\ddots&\ddots&0\\~0&0&\dots&0&\p^{(1)}_{m0}\\
{}~0&0&\dots&0&0\end{pmatrix} \qquad\mbox{on}\quad
E\=E_{(m)0}\=\bigoplus_{i=0}^{m}\,E_{k_{i0}} \ .
\label{mphim0}\eeq
The simplest case $m=1$ gives a holomorphic {\it
  triple}~\cite{Garcia1} and corresponds to the more standard
superconnections, having $(\mphi^{(1)}_{(1,0)})^2=0$, which
characterize the low-energy field content on brane-antibrane systems
with the tachyon field $\phi^{(1)}_{10}$ between the branes and
antibranes~\cite{AIO1,LPS}. A completely analogous characterization holds
for the charge configuration $(m_1,m_2)=(0,m)$. As we will discuss
further in the subsequent sections, for generic $m_1,m_2$ the set of
relations $\rel_{(m_1,m_2)}$, making the vector space $(\,\underline{\cal
  P}_{\,i\a})_{j\b}$ one-dimensional, implies that the quiver
$\quiver_{(m_1,m_2)}$ can always be naturally mapped (e.g. via a
lexicographic ordering) onto an ${\rm A}_{m+1}$-quiver. This will
become evident from the other examples considered below, and will have
important physical ramifications later on.

\bigskip

\noindent
{$\mbf{(m_1, m_2)=(1,1)}$.  } In this case the quiver bundle
truncates to a square
\beq
\begin{CD}
{E}^{}_{k_{10}}@>{\p^{(1)}_{10}}>>{E}^{}_{k_{00}}\\
@A{\p^{(2)}_{11}}AA@AA{\p^{(2)}_{01}}A\\
{E}^{}_{k_{11}}@>>{\p^{(1)}_{11}}>{E}^{}_{k_{01}}
\end{CD}
\eeq
and uniqueness of the bundle morphism on $E_{k_{11}}\to E_{k_{00}}$
(or of the corresponding path in the path algebra $\pathalg_{(1,1)}$)
yields the single holomorphic relation
\beq
\p^{(2)}_{01}\,\p^{(1)}_{11}=\p^{(1)}_{10}\,\p^{(2)}_{11} \ .
\label{singleholrel}\eeq
The equivariant graded connection admits the matrix presentation
\beq
{\ca} =
\begin{pmatrix}
\ca^{{00,00}}&\p^{(1)}_{10}&\p^{(2)}_{01}&0\\[4pt]
-\bigl(\p^{(1)}_{10}\bigr)^\+& \ca^{{10,10}}&0&\p^{(2)}_{11}\\[4pt]
-\bigl(\p^{(2)}_{11}\bigr)^\+&0& \ca^{{01,01}}&\p^{(1)}_{11}\\[4pt]
0&-\bigl(\p^{(2)}_{11}\bigr)^\+&-\bigl(\p^{(1)}_{11}\bigr)^\+& \ca^{{11,11}}
\end{pmatrix} \ .
\eeq

\bigskip

\noindent
{$\mbf{(m_1, m_2)=(2,1)}$. \ } The quiver bundle over $M$ associated
to $\quiver_{(2,1)}$ is given by
\beq
\begin{CD}
{E}^{}_{k_{20}}@>{\p^{(1)}_{20}}>>{E}^{}_{k_{10}}@>{\p^{(1)}_{10}}>>
{E}^{}_{k_{00}}\\
@A{\p^{(2)}_{21}}AA@A{\p^{(2)}_{11}}AA@AA{\p^{(2)}_{01}}A\\
{E}^{}_{k_{21}}@>>{\p^{(1)}_{21}}>{E}^{}_{k_{11}}@>>{\p^{(1)}_{11}}>
{E}^{}_{k_{01}}
\end{CD}
\eeq
with the pair of holomorphic relations
\beq
\p^{(2)}_{11}\,\p^{(1)}_{21}\=\p^{(1)}_{20}\,\p^{(2)}_{21} \qquad
\mbox{and}\qquad
\p^{(2)}_{01}\,\p^{(1)}_{11}\=\p^{(1)}_{10}\,\p^{(2)}_{11} \ .
\label{pairholrels}\eeq
The graded connection zero-form components
\beq
\mphi^{(1)}_{(2,1)}~:=~\begin{pmatrix}~0&\p^{(1)}_{10}&0&0&0&0\\
{}~0&0&\p^{(1)}_{20}&0&0&0\\
{}~0&0&0&0&0&0\\
{}~0&0&0&0&\p^{(1)}_{11}&0\\
{}~0&0&0&0&0&\p^{(1)}_{21}\\
{}~0&0&0&0&0&0\end{pmatrix}\qquad\mbox{and}
\qquad
\mphi^{(2)}_{(2,1)}~:=~\begin{pmatrix}
{}~0&0&0&\p^{(2)}_{01}&0&0\\
{}~0&0&0&0&\p^{(2)}_{11}&0\\
{}~0&0&0&0&0&\p^{(2)}_{21}\\
{}~0&0&0&0&0&0\\
{}~0&0&0&0&0&0\\
{}~0&0&0&0&0&0\end{pmatrix}
\eeq
satisfy the nilpotent relations
\beq
\bigl(\mphi^{(1)}_{(2,1)}\bigr)^2~\ne~ 0\ ,\quad \bigl
(\mphi^{(1)}_{(2,1)}\bigr)^3 \= 0\quad\mbox{and}\quad
\bigl(\mphi^{(2)}_{(2,1)}\bigr)^2\= 0\ .
\eeq
It is straightforward to check that the holomorphic relations
(\ref{pairholrels}) follow from the commutativity condition
(\ref{mphicommrels}) in this case.

\bigskip

\noindent
{$\mbf{(m_1, m_2)=(2,2)}$. \ } Finally, the
$(\quiver_{(2,2)}\,,\,\rel_{(2,2)})$-bundle is given by
\beq
\begin{CD}
{E}^{}_{k_{20}}@>{\p^{(1)}_{20}}>>{E}^{}_{k_{10}}@>{\p^{(1)}_{10}}>>
{E}^{}_{k_{00}}\\
@A{\p^{(2)}_{21}}AA@A{\p^{(2)}_{11}}AA@AA{\p^{(2)}_{01}}A\\
{E}^{}_{k_{21}}@>>{\p^{(1)}_{21}}>{E}^{}_{k_{11}}@>>{\p^{(1)}_{11}}>
{E}^{}_{k_{01}}\\
@A{\p^{(2)}_{22}}AA@A{\p^{(2)}_{12}}AA@AA{\p^{(2)}_{02}}A\\
{E}^{}_{k_{22}}@>>{\p^{(1)}_{22}}>{E}^{}_{k_{12}}@>>{\p^{(1)}_{12}}>
{E}^{}_{k_{02}}
\end{CD}
\eeq
with
\beq
\mphi^{(1)}_{(2,2)}\oplus \mphi^{(2)}_{(2,2)}=
\begin{pmatrix}
{}~0&\p^{(1)}_{10}&0&\p^{(2)}_{01}&0&0&0&0&0\\
{}~0&0&\p^{(1)}_{20}&0&\p^{(2)}_{11}&0&0&0&0\\
{}~0&0&0&0&0&\p^{(2)}_{21}&0&0&0\\
{}~0&0&0&0&\p^{(1)}_{11}&0&\p^{(2)}_{02}&0&0\\
{}~0&0&0&0&0&\p^{(1)}_{21}&0&\p^{(2)}_{12}&0\\
{}~0&0&0&0&0&0&0&0&\p^{(2)}_{22}\\
{}~0&0&0&0&0&0&0&\p^{(1)}_{12}&0\\
{}~0&0&0&0&0&0&0&0&\p^{(1)}_{22}\\
{}~0&0&0&0&0&0&0&0&0
\end{pmatrix}
\eeq
satisfying
\beq
\bigl(\mphi^{(\ell)}_{(2,2)}\bigr)^2~\ne~ 0\qquad\mbox{and}\qquad
\bigl(\mphi^{(\ell)}_{(2,2)}\bigr)^3 \= 0\ \qquad\mbox{for}\quad
\ell\=1,2 \ .
\eeq

\subsection{Graded connections on
  $\quiver_{(m_1,m_2)}$\label{gradedconn}}

We would now like to write the graded connections as intrinsic objects
to the quiver bundle (\ref{bundlediag}) over $M$, without explicit
reference to their origin as connections on the equivariant gauge
bundle ${\cal E}\to M\time\C P^1_{(1)}\time\C P^1_{(2)}$. For this, we will
introduce a more direct dimensional reduction of the gauge potential
$\ca$. The construction exploits the usual canonical isomorphism
between the complexified exterior algebra bundle over $M\time\C
P^1_{(1)}\time\C P^1_{(2)}$ and the corresponding graded Clifford
algebra bundle, which sends the exterior product into completely
antisymmetrized Clifford multiplication and the local cotangent basis $\diff
x^{\hat\mu}$ onto the Clifford algebra generators $\Gamma^{\hat\mu}$
obeying the anticommutation relations
\beq
\Gamma^{\hat\mu}\,\Gamma^{\hat\nu}+\Gamma^{\hat\nu}\,\Gamma^{\hat\mu}\=
-2\,g^{\hat\mu\hat\nu}~\Idd_{2^{n+2}}
\qquad\mbox{with}\quad \hat\mu,\hat\nu\=1,\dots,2n+4 \ .
\label{2n2Cliffalg}\eeq
The gamma-matrices in (\ref{2n2Cliffalg}) may be decomposed as
\beq\label{gamma2n2decomp}
\bigl\{\Gamma^{\hat\mu}\bigr\}\=\bigl\{\Gamma^\mu,\;\Gamma^{y_1},\;
\Gamma^{\yb_1},\;\Gamma^{y_2},\;\Gamma^{\yb_2}\bigr\}
\eeq
where
\beq
\Gamma^\mu=\gamma^\mu\otimes\Idd_2\otimes\Idd_2 \ ,
\label{Gammamug}\eeq
and $\gamma^\mu=-(\gamma^\mu)^\dag$ are the $2^n\time 2^n$ matrices
which locally generate the Clifford algebra bundle over $M$ and which
obey the anticommutation relations
\beq
\gamma^\mu\,\gamma^\nu+\gamma^\nu\,\gamma^\mu\=-2\,g^{\mu\nu}~\Idd_{2^n}
\qquad\mbox{with}\quad \mu,\nu\=1,\dots,2n \ .
\label{2nCliffalg}\eeq
The spherical components are given by
\bea
\Gamma^{y_1}&=&\gamma\otimes\gamma^{y_1}\otimes\Idd_2 \ , \quad
\Gamma^{\yb_1}~=~\gamma\otimes\gamma^{\yb_1}\otimes\Idd_2 \ ,
\\[4pt]
\Gamma^{y_2}&=&\gamma\otimes\s_3\otimes\gamma^{y_2} \ , \quad
\Gamma^{\yb_2}~=~\gamma\otimes\s_3\otimes\gamma^{\yb_2} \ ,
\eea
where
\beq
\g^{y_\ell}\=-\frac1{R_\ell}\,\left(1+y_\ell\yb_\ell\right)\,\s_+
\qquad\mbox{and}\qquad\g^{\yb_\ell}\=\frac1{R_\ell}\,\left(1+y_\ell\yb_\ell
\right)\,\s_-
\label{CP1Cliffalg}\eeq
are the Clifford algebra generators over $\C P^1_{(\ell)}$ for
$\ell=1,2$, with the constant $\slcL$ generators given by
(\ref{sl2cmatrices},\ref{sl2cLie}). The chirality operator over
$M$ is
\beq
\gamma\=\frac{\im^n}{(2n)!~\sqrt{g}}~\epsilon_{\mu_1\cdots\mu_{2n}}\,
\gamma^{\mu_1}\cdots\gamma^{\mu_{2n}} \qquad\mbox{with}\qquad
(\gamma)^2\=\Idd_{2^n} \quad\mbox{and}\quad
\gamma\,\gamma^\mu\=-\gamma^\mu\,\gamma \ .
\label{chiralityop}\eeq

With this set-up we may now write the equivariant gauge potential
given by (\ref{f4})--(\ref{ff7}) as the graded connection
\bea
\hat\ca&:=&\Gamma^{\hat\mu}\,\ca_{\hat\mu}\=
\Gamma^{\mu}\,\ca_{\mu} + \Gamma^{y_1}\,\ca_{y_1} +
\Gamma^{\yb_1}\,\ca_{\yb_1} + \Gamma^{y_2}\,
\ca_{y_2} + \Gamma^{\yb_2}\,\ca_{\yb_2}
 \nonumber\\[4pt]
&=&\g^\m\,\bigl(\mA^{(m_1,\, m_2)}\bigr)_\mu\otimes\Idd_2\otimes\Idd_2+
\frac1{R_1}\,\bigl(\mphi^{(1)}_{(m_1,\, m_2)}\bigr)
\,\g\otimes\s_-\otimes\Idd_2
+\frac1{R_1}\,\bigl(\mphi^{(1)}_{(m_1,\, m_2)}\bigr)^\+
\,\g\otimes\s_+\otimes\Idd_2
\nonumber\\
&& +\,\frac1{R_2}\,\bigl(\mphi^{(2)}_{(m_1,\, m_2)}\bigr)
\,\g\otimes\s_3\otimes\s_- + \frac1{R_2}\,
\bigl(\mphi^{(2)}_{(m_1,\, m_2)}\bigr)^\+
\,\g\otimes\s_3\otimes\s_+\nonumber\\
&& +\,\g\otimes\left(\g^{\yb_1}\,
\bigl(\ma^{(m_1)}\bigr)_{\yb_1}+\g^{y_1}\,
\bigl(\ma^{(m_1)}\bigr)_{y_1}\right)\otimes\Idd_2
+\g\otimes\s_3\otimes\left(\g^{\yb_2}\,
\bigl(\ma^{(m_2)}\bigr)_{\yb_2}+\g^{y_2}\,
\bigl(\ma^{(m_2)}\bigr)_{y_2}\right) \ , \nonumber\\ &&
\label{calAgammas}\eea
where
\beq
\g^{\yb_\ell}\,\bigl(\ma^{(m_\ell)}\bigr)_{\yb_\ell}+\g^{y_\ell}\,
\bigl(\ma^{(m_\ell)}\bigr)_{y_\ell}
\=\frac1{R_\ell}\,\left(1+y_\ell\yb_\ell\right)\,
\left(\bigl(\ma^{(m_\ell)}\bigr)_{\yb_\ell}\,\s^{}_- -
\bigl(\ma^{(m_\ell)}\bigr)_{y_\ell}\,\s^{}_+\right) \qquad
\mbox{for}\quad \ell\=1,2 \ .
\eeq
As desired, the zero-form components in (\ref{calAgammas})
involving $\mphi^{(\ell)}_{(m_1,m_2)}$ are independent of the
coordinates $(y_\ell,\yb_\ell)\in\C P^1_{(\ell)}$ and they anticommute
with the one-form components involving $\mA^{(m_1,m_2)}$ due to their
couplings with the chirality operator (\ref{chiralityop}). From
(\ref{f15})--(\ref{Fyyb}) the curvature of the graded connection
(\ref{calAgammas}) is found to be
\bea
\hat\cf&:=&\mbox{$\frac14$}\,\bigl[\Gamma^{\hat\mu}\,,\,
\Gamma^{\hat\nu}\bigr]\,\cf_{\hat\mu\hat\nu}\nonumber\\[4pt]&=&
\mbox{$\frac14$}\,\bigl[\gamma^\mu\,,\,\gamma^\nu\bigr]\,
\bigl(\mF\bigr)_{\mu\nu}\otimes\Idd_2\otimes\Idd_2
\nonumber\\
&&-\,
\frac1{R_1}\,\g\,\bigl(\gamma^\mu\,
D_\mu\mphi^{(1)}_{(m_1,\, m_2)}  \bigr)\otimes\s^{}_-\otimes\Idd_2
+\frac1{R_1}\,\gamma\,\bigl(\gamma^\mu\,
D_\mu\mphi^{(1)}_{(m_1,\, m_2)}
\bigr)^\+\otimes\sigma^{}_+\otimes\Idd_2
\nonumber\\
&&-\,
\frac1{R_2}\,\g\,\bigl(\gamma^\mu\,
D_\mu\mphi^{(2)}_{(m_1,\, m_2)}  \bigr)\otimes\s^{}_3\otimes\s^{}_-
+\frac1{R_2}\,\gamma\,\bigl(\gamma^\mu\,
D_\mu\mphi^{(2)}_{(m_1,\, m_2)} \bigr)^\+\otimes\sigma^{}_3
\otimes\s^{}_+\nonumber\\
&& +\,
\frac1{2\,R_1^2}\,
\left(\mup^{(1)}_{(m_1,\, m_2)} -\left[
\mphi^{(1)}_{(m_1,\, m_2)}\,,\, \bigl(\mphi^{(1)}_{(m_1,\, m_2)} \bigr)^\+
\right]\right)\,\Idd^{}_{2^n}\otimes\s^{}_3\otimes\Idd_2\nonumber\\
&& +\,
\frac1{2\,R_2^2}\,
\left(\mup^{(2)}_{(m_1,\, m_2)} -\left[
\mphi^{(2)}_{(m_1,\, m_2)}\,,\, \bigl(\mphi^{(2)}_{(m_1,\, m_2)} \bigr)^\+
\right]\right)\,\Idd^{}_{2^n}\otimes\Idd_2\otimes\s^{}_3\nonumber\\
&& +\,
\frac1{R_1\,R_2}\,\left[\mphi^{(1)}_{(m_1, m_2)}\,,\,
\mphi^{(2)}_{(m_1,\, m_2)}\right]\,
\Idd^{}_{2^n}\otimes\s^{}_-\otimes\s^{}_-\nonumber\\
&& +\,
\frac1{R_1\,R_2}\,\left[\mphi^{(1)}_{(m_1,\, m_2)}\,,\,
\mphi^{(2)}_{(m_1,\, m_2)}\right]^\+\,
\Idd^{}_{2^n}\otimes\s^{}_+\otimes\s^{}_+
\nonumber\\
&& +\,
\frac1{R_1\,R_2}\,\left[\mphi^{(1)}_{(m_1,\, m_2)}\;,\;
\bigl(\mphi^{(2)}_{(m_1,\, m_2)}\bigr)^\+\right]\,
\Idd^{}_{2^n}\otimes\s^{}_-\otimes\s^{}_+\nonumber\\
&& +\,
\frac1{R_1\,R_2}\,\left[\mphi^{(1)}_{(m_1,\, m_2)}\;,\;
\bigl(\mphi^{(2)}_{(m_1,\, m_2)}\bigr)^\+\right]^\+\,
\Idd^{}_{2^n}\otimes\s^{}_+\otimes\s^{}_-
\label{gradedcurv}\eea
where $\mF:=\diff\mA^{(m_1,\, m_2)}+\mA^{(m_1,\,
  m_2)}\wedge\mA^{(m_1,\, m_2)}=\frac12\,\bigl(\mF\bigl)_{\mu\nu}~\diff
x^\mu\wedge\diff x^\nu$.

The graded curvature (\ref{gradedcurv}) is completely independent of
the spherical coordinates. Using (\ref{gradedcurv}) and standard
gamma-matrix trace formulas~\cite{PS1}, it is possible to recast
the dimensionally reduced Yang-Mills action functional (\ref{SYMred})
in the compact form
\beq
S^{~}_{\rm YM}=\frac{\pi^2\,R_1^2\,R_2^2}{2^n}\,\int_{M}\diff^{2n}x~
\sqrt{g}~\tr_{k\time k}^{~}~\Tr^{~}_{\C^{2^{n+2}}}~\hat\cf^2 \ ,
\label{EFgraded}\eeq
where the trace $\Tr^{~}_{\C^{2^{n+2}}}$ is taken over the
representation space of (\ref{2n2Cliffalg}) and may be thought of as
an ``integral'' over the Clifford algebra. Thus
the entire equivariant gauge theory on $M\time\C P^1_{(1)}\time\C
P^1_{(2)}$ may be elegantly rewritten as an {\it ordinary\/} Yang-Mills
gauge theory of {\it graded connections\/} on the corresponding
{\it quiver bundle\/} over~$M$.

\bigskip

\section{Noncommutative instantons and quiver vortices\label{NCinst}}

\noindent
We will now proceed to the construction of explicit equivariant instanton
solutions. We will build both BPS and non-BPS configurations of the
Yang-Mills equations on the noncommutative space
$\R_\theta^{2n}\time\C P^1\time\C P^1$. We then describe some general
properties of the moduli space of noncommutative instantons in this instance.

\subsection{BPS equations\label{BPSeqs}}

The equations of motion which follow from varying the Yang-Mills
lagrangian (\ref{lagrprod}) on the K\"ahler manifold $M\time\C
P^1\time\C P^1$ are given by
\begin{equation}\label{YM}
\frac{1}{\sqrt{\hat g}}\, \pa_{\hat\m}\bigl(\sqrt{\hat g}~
\cf^{\hat{\m}\hat{\n}}\bigr) +
\bigl[\ca_{\hat\m}\,,\, \cf^{\hat\m\hat\n}\bigr]=0 \ .
\end{equation}
The BPS configurations which satisfy (\ref{YM}) are provided by
solutions of the DUY equations~\cite{DUY1}
\begin{equation}\label{DUY}
*\Omega\wedge {\cf}\ =\ 0 \qquad\textrm{and}\qquad
{\cf}^{2,0}\=0\=\cf^{0,2}\ ,
\end{equation}
where $*$ is the Hodge duality operator and
$\cf =\cf^{2,0}+\cf^{1,1}+\cf^{0,2}$ is the K\"ahler decomposition of
the gauge field strength. In the local complex coordinates
$(z^a,y_1,y_2)$ these equations take the form
\begin{eqnarray}\label{DUY1}
g^{a\bb}\,{\cf}_{z^a\zb^{\bb}}+ g^{y_1\yb_1}\,{\cf}_{y_1\yb_1}+
g^{y_2\yb_2}\,{\cf}_{y_2\yb_2}&=&0 \ ,
\\[4pt]\label{DUY2}
{\cf}_{\zb^{\ab}\zb^{\bb}}&=&0\ , \\[4pt] {\cf}_{\zb^{\ab}\yb_1}&=&0~~=~~
{\cf}_{\zb^{\ab}\yb_2} \ ,\\[4pt]
\cf_{\yb_1\yb_2}&=&0\ ,\label{DUY3}
\end{eqnarray}
along with their complex conjugates for $a,b=1,\dots,n$.

In terms of the equivariant decomposition (\ref{f15})--(\ref{Fyyb}),
the DUY equations read
\bea
g^{a\bb}\,F^{i\a}_{a{\bb}}&=&\frac{1}{2\,R_1^2}\,\left
[m_1-2i+\bigl(\phi^{(1)}_{i\a}\bigr)^\+\,
\phi^{(1)}_{i\a}-\phi^{(1)}_{i+1\,\a}\,\bigl
(\phi^{(1)}_{i+1\,\a}\bigr)^\+\,\right]\nonumber\\ &\quad+\!\!&
\frac{1}{2\,R_2^2}\,\left[m_2-2\a+\bigl(\phi^{(2)}_{i\a}\bigr)^\+\,
\phi^{(2)}_{i\a}-\phi^{(2)}_{i\,\a+1}\,\bigl
(\phi^{(2)}_{i\,\a+1}\bigr)^\+\right]
\label{f24}\eea
and
\bea
\label{f240} 
F_{\ab\bb}^{i\a}&=&0\ , \\[4pt]
\label{f25}
\pa^{~}_{\bar a}\phi^{(1)}_{i+1\,\a} + A^{i\a}_{\bar a}\,\phi^{(1)}_{i+1\,\a} -
\phi^{(1)}_{i+1\,\a}\,A^{i+1\,\a}_{\bar a}&=&0\ ,\\[4pt]
\label{f25a}
\pa^{~}_{\bar a}\phi^{(2)}_{i\,\a+1} + A^{i\a}_{\bar a}\,\phi^{(2)}_{i\,\a+1} -
\phi^{(2)}_{i\,\a+1}\,A^{i\,\a+1}_{\bar a}&=&0\ ,\\[4pt]
\label{f25b}
\phi^{(1)}_{i+1\,\a}\,\phi^{(2)}_{i+1\,\a+1} -
\phi^{(2)}_{i\,\a+1}\,\p^{(1)}_{i+1\,\a+1}&=&0\ ,
\eea
along with their complex conjugates. Eq.~(\ref{f24}) gives
hermitean conditions on the curvatures of $E_{k_{i\a}}\to M$, while
(\ref{f240}) implies that $E_{k_{i\a}}$ are holomorphic vector bundles
with connections $A^{i\a}$. The conditions (\ref{f25})
and (\ref{f25a}) then mean that the bundle maps on the quiver bundle
(\ref{bundlediag}) are holomorphic. Eq.~(\ref{f25b}) imposes the
relations $\rel_{(m_1,m_2)}$ on the quiver bundle. Note that the
analogous non-holomorphic relations, specified by the vanishing of
(\ref{Fyyb}), do not arise as BPS conditions.

The BPS energies may be computed by noting that the action functional
(\ref{SYMred}) evaluated on equivariant connections $\ca$ of the bundle
${\cal E}\to M\time\C P^1\time\C P^1$ may be written
as~\cite{A-CG-P2}
\beq
S_{\rm YM}^{~}=\frac14\,
\int_{M\time\C P^1\time\C P^1}\diff^{2n+4}x~\sqrt{\hat g}~
\tr^{~}_{k\time k}\bigl(\Omega^{\hat\mu\hat\nu}\,
\cf_{\hat\mu\hat\nu}\bigr)^2-2\pi^2~{\rm Ch}_2({\cal E}) \ ,
\label{SYMhol}\eeq
where
\beq
{\rm Ch}_2({\cal E})=-\frac1{8\pi^2}\,\int_{M\time\C P^1\time\C P^1}
\,\frac{\Omega^n}{n!}\wedge\tr^{~}_{k\time k}\,
\cf\wedge\cf
\label{Ch2def}\eeq
is a Chern-Weil topological invariant of $\cal E$. Eq.~(\ref{SYMhol})
shows that the Yang-Mills action is bounded from below
as $S_{\rm YM}^{~}\geq S^{~}_{\rm BPS}:=-2\pi^2~{\rm
  Ch}_2({\cal E})$, with equality precisely when the DUY equations
(\ref{DUY}) are satisfied. By substituting in (\ref{kahler}) and the
equivariant decomposition (\ref{f15})--(\ref{Fyyb}), after integration
over $\C P^1\time\C P^1$ one finds
\bea
&&S^{~}_{\rm BPS}~=~2\pi^2\,\sum_{i=0}^{m_1}~\sum_{\a=0}^{m_2}\,\Bigl\{
{\rm vol}\,M\,\bigl[(m_1-2i)\,(m_2-2\a)\,k_{i\a}\bigr.\Bigr.
\nonumber\\ &&\qquad+\bigl.4\,\bigl(R_2^2\,(m_1-2i)+R_1^2\,(m_2-2\a)\bigr)~
{\rm deg}~E_{k_{i\a}}\bigr]-64\pi^2\,R_1^2\,R_2^2~{\rm Ch}_2
(E_{k_{i\a}})\nonumber\\ &&\qquad+\,\int_M\!\!\diff^{2n}x~\sqrt g~
\tr^{~}_{k_{i\a}\time k_{i\a}}\left[\bigl(\phi^{(1)}_{i+1\,\a+1}
\bigr)^\dag\,\bigl(\phi^{(2)}_{i\,\a+1}\bigr)^\dag\,\phi^{(1)}_{i+1\,\a}\,
\phi^{(2)}_{i+1\,\a+1}-\bigl(\phi^{(1)}_{i\a}\bigr)^\dag\,
\phi^{(1)}_{i\a}\,\bigl(\phi^{(2)}_{i\,\a+1}
\bigr)^\dag\,\phi^{(2)}_{i\,\a+1}\right.\nonumber\\ &&\qquad+
\Bigl.\left.\bigl(\phi^{(1)}_{i+1\,\a}\bigr)^\dag\,
\phi^{(1)}_{i+1\,\a+1}\,\bigl(\phi^{(2)}_{i+1\,\a+1}\bigr)^\dag\,
\phi^{(2)}_{i\,\a+1}
-\phi^{(1)}_{i+1\,\a}\,\bigl(
\phi^{(1)}_{i+1\,\a}\bigr)^\dag\,\bigl(\phi^{(2)}_{i\a}\bigr)^\dag\,
\phi^{(2)}_{i\a}\right]\Bigr\} \ ,
\label{SBPSgen}\eea
where ${\rm vol}\,M=\int_M\,\omega^n/n!$ is the volume
of the K\"ahler manifold $M$ and
\beq
{\rm deg}~E_{k_{i\a}}=\frac{\im}{{\rm vol}\,M}\,\int_M\,
\frac{\omega^{n-1}}{(n-1)!}\wedge\tr^{~}_{k_{i\a}\time k_{i\a}}\,
F^{i\a}
\label{degreeEdef}\eeq
is the degree of the rank $k_{i\a}$ bundle $E_{k_{i\a}}\to M$.

To cast these equations on the noncommutative space $M=\R_\theta^{2n}$,
we introduce the operators
\begin{equation}\label{X}
X_{a}^{i\a}\ :=\ A_{a}^{i\a} + \th_{a\bb}\,\zb^{\bb}
\qquad\textrm{and}\qquad
X_{{\ab}}^{i\a}\ :=\ A_{{\ab}}^{i\a} + \th^{~}_{\ab b}\,z^b\ .
\end{equation}
In terms of these operators the antiholomorphic bi-fundamental
covariant derivatives take the form
\beq
D_\ab^{~}\phi^{(1)}_{i+1\,\a}\=
X_\ab^{i\a}\,\phi^{(1)}_{i+1\,\a}-\phi^{(1)}_{i+1\,\a}
\,X_\ab^{i+1\,\a} \qquad\mbox{and}\qquad
D_\ab^{~}\phi^{(2)}_{i\,\a+1}\=
X_\ab^{i\a}\,\phi^{(2)}_{i\,\a+1}-\phi^{(2)}_{i\,\a+1}
\,X_\ab^{i\,\a+1}\ ,
\label{bifundX}\eeq
while the components of the field strength tensor become
\begin{equation}
F_{{a}{\bb}}^{i\a}\ =\ \big[X_{a}^{i\a}\,,\,X_{{\bb}}^{i\a}\,\big] + \th_{a\bb}
\ , ~~ F_{{\ab}{\bb}}^{i\a}\ =\ \big[X_{{\ab}}^{i\a}\,,\, X_{\bb}^{i\a}\,\big]
\quad\textrm{and}\quad F_{{a}{b}}^{i\a}\ =\ \big[X_{{a}}^{i\a}\,,\,
X_{b}^{i\a}\,\big]\ .
\label{fieldstrengthX}\end{equation}
The noncommutative DUY equations (without the complex conjugates)  then read
\bea
\delta^{a\bb}\,\Bigl(\big[X_{a}^{i\a}\,,\,X_{{\bb}}^{i\a}\,\big] +
\th_{a\bb}\Bigr)&=&\frac{1}{2\,R_1^2}\,\left[m_1-2i+\big(
\phi^{(1)}_{i\a}\big)^\+\,
\phi^{(1)}_{i\a}-\phi^{(1)}_{i+1\,\a}\,\big(
\phi^{(1)}_{i+1\,\a}\big)^\+\;\right]
\label{ddd1}\\ &\quad+\!\!&
\frac{1}{2\,R_2^2}\,\left[m_2-2\a+\big(\phi^{(2)}_{i\a}\big)^\+\,
\phi^{(2)}_{i\a}-\phi^{(2)}_{i\,\a+1}\,\big(
\phi^{(2)}_{i\,\a+1}\big)^\+\right]
\ ,\nonumber\\[4pt]
\big[X_{\ab}^{i\a}\,,\,X_{{\bb}}^{i\a}\,\big] &=& 0
\ , \label{ddd1a} \\[4pt]
X_\ab^{i\a}\,\phi^{(1)}_{i+1\,\a}-\phi^{(1)}_{i+1\,\a}\,
X_\ab^{i+1\,\a}&=&0 \ , \\[4pt]
X_\ab^{i\a}\,\phi^{(2)}_{i\,\a+1}-\phi^{(2)}_{i\,\a+1}\,
X_\ab^{i\,\a+1}&=&0 \ , \\[4pt]
\phi^{(1)}_{i+1\,\a}\,\phi^{(2)}_{i+1\,\a+1} -
\phi^{(2)}_{i\,\a+1}\,\p^{(1)}_{i+1\,\a+1}&=&0 \ . \label{ddd2}
\eea

\subsection{Examples\label{BPSEx}}

Before proceeding with a more general analysis, we will provide some
illustration of the meaning of the quiver vortex equations
(\ref{f24})--(\ref{f25b}) through special cases and limiting solutions.

\bigskip

\noindent
{\bf Chain vortex equations. \ } Consider a holomorphic chain~(\ref{holchain})
with $(m_1,m_2)=(m,0)$. Its equations, obtainable from
(\ref{f24})--(\ref{f25b}) by taking $\phi^{(2)}_{i\,\a+1}=0$ in the ansatz
for $\ca$ and $\cf$, read
\bea
&& g^{a\bb}\,F^i_{a\bb} \=
\frac{1}{2\,R^2}\,( m-2i+\phi_i^\+\,\phi^{~}_i-\phi^{~}_{i+1}\,\phi_{i+1}^\+ )\ ,\qquad
F^i_{\ab\bb} \= 0 \ , \label{chvort1} \\[4pt]
&& \bar\pa_{\ab}\phi_{i+1}\ +\
A^i_{\ab}\,\phi_{i+1}\ -\ \phi_{i+1}\,A^{i+1}_{\ab} \= 0
\qquad\textrm{for}\quad i=0,1,\ldots,m \ , \label{chvort2}
\eea
where $\phi^{~}_i:=\phi^{(1)}_{i\,0}$, $A^i:=A^{i0}$, $F^i:=F^{i0}$ and $R=R_1$.
Noncommutative chain vortex configurations solving (\ref{chvort1}) and
(\ref{chvort2}) on $M=\rt$ were constructed in~\cite{PS1}.

\bigskip

\noindent
{\bf Holomorphic triples.  } For $m=1$ the holomorphic chain~(\ref{holchain})
reduces to a holomorphic triple~$(\Ecal_1,\Ecal_2,\phi)$~\cite{Garcia1}
described by the equations
\bea
&& g^{a\bb}\,F^0_{a\bb} \=
+\frac{1}{2\,R^2}\,( 1 - \phi\phi^\+ ) \ ,\qquad
F^0_{\ab\bb} \= 0 \ , \label{holtri1} \\[4pt]
&& g^{a\bb}\,F^1_{a\bb} \=
-\frac{1}{2\,R^2}\,( 1 - \phi^\+\phi ) \ ,\qquad
F^1_{\ab\bb} \= 0 \ , \label{holtri2} \\[6pt]
&& \bar\pa_{\ab}\phi\ +\ A^0_{\ab}\,\phi\ -\ \phi\,A^1_{\ab}\= 0\ .
\label{holtri3}
\eea
Solutions of~(\ref{holtri1})--(\ref{holtri3}) for $M=\rt$ and their
D-brane interpretation were presented in~\cite{IL,LPS}.

\bigskip

\noindent
{\bf Four-dimensional case.  } For $\textrm{dim}_\R M=4$, $k_0=k_1=r$ and
$\phi=\Idd_r$, we infer from~(\ref{holtri3}) that $A^0=A^1$, hence both
(\ref{holtri1}) and~(\ref{holtri2}) simplify to the self-dual Yang-Mills
equations on~$M$. In the case of $M=\rct$ their solutions are noncommutative
instantons (see e.g.~\cite{Ncinst1,Ncinst2} and references therein).
In string theory they are interpreted as states of noncommutative D-branes
(see e.g.~\cite{Dbranes} and references therein).
On the other hand, when $k_0=k_1=1$ and $\phi$ is non-constant
eqs.~(\ref{holtri1})--(\ref{holtri3}) reduce to the perturbed abelian 
Seiberg-Witten monopole equations~\cite{Witten}. For $M=\rct$ one encounters 
the noncommutative ${\rm U}_+(1)\time{\rm U}_-(1)$ Seiberg-Witten monopole
equations studied in~\cite{PSW}.

\bigskip

\noindent
{\bf Vortices in two dimensions.  } For $\textrm{dim}_\R M=2$ and $k_0=k_1=1$,
the set (\ref{holtri1})--(\ref{holtri3}) coincides with the standard
vortex equations, whose solutions on $M=\R^2_\theta$ were considered
e.g.~in~\cite{Vortex}.

\bigskip

\noindent
{\bf Quiver Toda equations.  } Let us investigate the equations
(\ref{f24})--(\ref{f25b}) in the limit $R_1,R_2\to\infty$ which decompactifies
the spherical parts of our K\"ahler manifold $M\time\C P^1\time\C P^1$.
With the redefinitions $\phi^{(\ell)}_{i\a}\to R_\ell\,\phi^{(\ell)}_{i\a}$
for $i=0,1,\ldots,m_1$ and $\a=0,1,\ldots,m_2$, the quiver vortex equations then
descend to the {\it quiver Toda equations\/}
\bea
2 g^{a\bb}\,F^{i\a}_{a\bb} \ \=\ 
\big(\phi^{(1)}_{i\a}\big)^\+\,\phi^{(1)}_{i\a}\ -\
\phi^{(1)}_{i+1\,\a}\,\big(\phi^{(1)}_{i+1\,\a}\big)^\+ \!&+&\!
\big(\phi^{(2)}_{i\a}\big)^\+\,\phi^{(2)}_{i\a}\ -\
\phi^{(2)}_{i\,\a+1}\,\big(\phi^{(2)}_{i\,\a+1}\big)^\+ \ ,\\[4pt]
F_{\ab\bb}^{i\a}&=&0\ ,\\[4pt]
\pa^{~}_{\bar a}\phi^{(1)}_{i+1\,\a} + A^{i\a}_{\bar a}\,
\phi^{(1)}_{i+1\,\a} -
\phi^{(1)}_{i+1\,\a}\,A^{i+1\,\a}_{\bar a}&=&0\ ,\\[4pt]
\pa^{~}_{\bar a}\phi^{(2)}_{i\,\a+1} + A^{i\a}_{\bar a}\,
\phi^{(2)}_{i\,\a+1} -
\phi^{(2)}_{i\,\a+1}\,A^{i\,\a+1}_{\bar a}&=&0\ ,\\[4pt]
\phi^{(1)}_{i+1\,\a}\,\phi^{(2)}_{i+1\,\a+1} -
\phi^{(2)}_{i\,\a+1}\,\p^{(1)}_{i+1\,\a+1}&=&0\ .
\eea
In this limit the induced quiver gauge theory on~$M$ is independent of the
additional spherical dimensions. 
In the case $\phi^{(2)}_{i\a}=0\ ~\forall i,\a$
and $\phi^{~}_i:=\phi^{(1)}_{i\,0}$ we arrive at
\beq
2 g^{a\bb}\,F^i_{a\bb} \=
\phi_i^\+\,\phi^{~}_i -\phi^{~}_{i+1}\,\phi_{i+1}^\+\ ,\qquad
F^i_{\ab\bb} \= 0\ ,\qquad
\bar\pa_{\ab}\phi_{i+1}\ +\
A^i_{\ab}\,\phi_{i+1}\ -\ \phi_{i+1}\,A^{i+1}_{\ab} \= 0\ ,
\eeq
which may be called the {\it holomorphic chain Toda equations\/} on the
K\"ahler manifold~$M$.

\bigskip

\noindent
{\bf Symmetric instantons on $\mbf{\C P^1\time\C P^1}$.  } A
somewhat opposite limit to the decompactification limit above comes
from choosing the vacuum solution for generic monopole charges
$(m_1,m_2)$ on $M\time\C P^1\time\C P^1$. Let us set $A^{i\alpha}=0$ in
(\ref{f6a}), $\phi_{i+1\,\alpha}^{(1)}$ and $\phi^{(2)}_{i\,\alpha+1}$
to constant matrices in (\ref{f6})--(\ref{ff7}), and $F^{i\alpha}=0$
in (\ref{f11}). Then the field strength components
(\ref{f12})--(\ref{ff13}) are identically zero, but
(\ref{fc1})--(\ref{fc4}) are generically non-vanishing. The components
(\ref{f15})--(\ref{fc16}) vanish, while (\ref{fc17})--(\ref{Fyyb}) are
non-vanishing and give the components of the gauge fields on $\C
P^1\time\C P^1$. The BPS equations (\ref{f240})--(\ref{f25a}) are
identically satisfied in this case, while eqs.~(\ref{f24}) and
(\ref{f25b}) should be solved with constant matrices
$\phi^{(\ell)}_{i\alpha}$. The simplest choice is square matrices with
$(m_1,m_2)=(1,1)$. The BPS equations (\ref{f24}) and (\ref{f25b}) are
respectively equivalent in this case to the equations
\bea
{\cal F}^{i\alpha,i\alpha}_{y_1 {\bar y}_1} +
{\cal F}^{i\alpha,i\alpha}_{y_2 {\bar y_2}}&=&0 \ ,
\nonumber\\[4pt]
{\cal F}^{i+1\,\alpha+1,i\alpha}_{y_1 y_2}&=&0~=~
{\cal F}^{i\alpha,i+1\,\alpha+1}_{{\bar y}_1 {\bar y}_2} \ .
\label{redvaceqs}\eea
Furthermore, ${\cal F}^{i\,\alpha+1,i+1\,\alpha}_{y_1\bar y_2}$ is
given by (\ref{Fyyb}). The equations (\ref{redvaceqs}) give
$\su\time\su$-equivariant solutions of the self-dual Yang-Mills
equations on $\C P^1\time\C P^1$ which are vacuum BPS solutions of
the original DUY equations. These solutions have non-zero energy, and
the entire structure of these non-abelian instantons on $\C
P^1\time\C P^1$ is reduced to equations for finite-dimensional
matrices from our equivariant fields.

\subsection{Finite energy solutions\label{NCYMsols}}

Let us fix monopole charges $m_1,m_2>0$ and an arbitrary integer
$0<r\le k$. Consider the ansatz
\bea\label{ansatz3}
X_{a}^{i\a} &=& \th_{a\bb}\,T^{~}_{N_{i\a}}\,\zb^{\bb}\,T_{N_{i\a}}^\+
\qquad\mbox{and}\qquad
X_{{\ab}}^{i\a}~=~\th^{~}_{\ab b}\, T^{~}_{N_{i\a}}
\,z^{b}\,T_{N_{i\a}}^\+ \ ,
\\[4pt] \label{ansatz3p1}
\p^{(1)}_{i+1\,\a} &=& \lambda^{(1)}_{i+1\,\a}\,T^{~}_{N_{i\a}}\,
T_{N_{i+1\,\a}}^\+\qquad\mbox{and}\qquad
\big(\p^{(1)}_{i+1\,\a}\big)^\+~=~\bar\lambda^{(1)}_{i+1\,\a}\,
T^{~}_{N_{i+1\,\a}}\,T_{N_{i\a}}^\+ \ ,
\\[4pt] \label{ansatz3p}
\p^{(2)}_{i\,\a+1} &=& \lambda^{(2)}_{i\,\a+1}\,T^{~}_{N_{i\a}}\,
T_{N_{i\,\a+1}}^\+\qquad\mbox{and}\qquad
\big(\p^{(2)}_{i\,\a+1}\big)^\+~=~\bar\lambda^{(2)}_{i\,\a+1}\,
T^{~}_{N_{i\,\a+1}}\,T_{N_{i\a}}^\+ \ ,
\eea
where $\lambda^{(1)}_{i\a}, \lambda^{(2)}_{i\a}\in\C$ are some constants with
$\lambda^{(1)}_{0\a}=0=\lambda^{(1)}_{m_1+1\,\a}$ and
$\lambda^{(2)}_{i0}=0=\lambda^{(2)}_{i\,m_2+1}$ for $i=0,1,\dots,m_1$,
$\a=0,1,\dots,m_2$. Denoting by $\Hcal$ the $n$-oscillator Fock space
as before, the Toeplitz operators
\beq
T^{~}_{N_{i\a}}\ :\quad \C^r\otimes\Hcal~\longrightarrow~
\underline{V}_{\,k_{i\a}}\otimes\Hcal
\eeq
are partial isometries described by {\it rectangular} $k_{i\a}\time r$
matrices (with values in ${\rm End}~\Hcal$) possessing the
properties
\begin{equation}\label{ansatz4}
T_{N_{i\a}}^\+\,T^{~}_{N_{i\a}}\=\Idd_r \qquad\textrm{while}\qquad
T^{~}_{N_{i\a}}\,T_{N_{i\a}}^\+\ =\ \Idd_{k_{i\a}} - P^{~}_{N_{i\a}}\ ,
\end{equation}
where $P^{~}_{N_{i\a}}$ is a hermitean projector of finite rank $N_{i\a}$ on
the Fock space $\underline{V}_{\,k_{i\a}}\otimes\Hcal$ so that
\beq
P_{N_{i\a}}^2\=P^{~}_{N_{i\a}}\=P^\dag_{N_{i\a}} \qquad\mbox{and}\qquad
\Tr^{~}_{\underline{V}_{\,k_{i\a}}\otimes\Hcal}~P^{~}_{N_{i\a}}\=N_{i\a} \ .
\label{PNiTrace}
\eeq
It follows that
\beq
\ker T^{~}_{N_{i\a}}\=\{0\} \qquad\mbox{but}\qquad
\ker T^\dag_{N_{i\a}}\={\rm im}\,P^{~}_{N_{i\a}}~\cong~
\C^{N_{i\a}} \ .
\label{dimkerTNi}
\eeq

For the ansatz (\ref{ansatz3})--(\ref{ansatz3p}) the equations
(\ref{ddd1a})--(\ref{ddd2}) are satisfied along with the
non-holomorphic relations
\beq
\big(\p^{(2)}_{i\a}\big)^\+ \,\p^{(1)}_{i+1\,\a-1} -
\p^{(1)}_{i+1\,\a}\,\big(\p^{(2)}_{i+1\,\a} \big)^\+=0 \ ,
\label{antiholrels}\eeq
or equivalently in terms of graded connections one has the
commutativity condition
\beq
 \left[\mphi^{(1)}_{(m_1,\, m_2)}\,,\,
\bigl(\mphi^{(2)}_{(m_1,\, m_2)}\bigr)^\+\right]=0\ .
\label{mphiantiholcommrels}\eeq
The non-vanishing gauge field strength components are given by
\begin{equation}
F_{a{\bb}}^{i\a} \=  \th_{a\bb}\,P^{~}_{N_{i\a}}\=\frac1{2\,\theta^a}\,
\delta_{a\bb}\,P^{~}_{N_{i\a}} \ .
\label{ansatzfieldstrength}\end{equation}
It follows that our ansatz determines a {\it finite-dimensional}
representation of the quiver with relations
$(\quiver_{(m_1,m_2)}\,,\,\rel_{(m_1,m_2)})$. The projectors
$P^{~}_{N_{i\a}}$ give representations of the trivial path idempotents
$e_{i\a}$ and project the infinite-dimensional Fock module
$\underline{\cal V}\otimes\Hcal$ over the path algebra
$\pathalg_{(m_1,m_2)}$, given by the noncommutative quiver bundle, onto
finite-dimensional vector spaces $P^{~}_{N_{i\a}}\cdot(\,\underline{\cal
  V}\otimes\Hcal)=\ker T^\dag_{N_{i\a}}$. This module will be denoted
as
\beq
\underline{\cal T}:=\bigoplus_{i=0}^{m_1}~\bigoplus_{\a=0}^{m_2}\,
\ker T^\dag_{N_{i\a}}
\label{calTdef}\eeq
with dimension vector
\beq
\vec N~:=~\vec k_{\underline{\cal T}}\=
\big(N_{i\a}\big)^{i=0,1,\dots,m_1}_{\a=0,1,\dots,m_2} \ .
\label{Ndimvec}\eeq
These dimensions correspond to the degrees of the corresponding
noncommutative sub-bundles determined by (\ref{ansatzfieldstrength}).

The noncommutative Yang-Mills action for the ansatz
(\ref{ansatz3})--(\ref{ansatz3p}) can be evaluated by using
(\ref{SYMred}), (\ref{intNC}), (\ref{ddd2}), (\ref{ansatz4}),
(\ref{antiholrels}) and (\ref{ansatzfieldstrength}) to get
\bea
S^{~}_{\rm YM}&=&-\pi\,R_1^2\,R_2^2~{\rm Pf}(2\pi\,\theta)~
\sum_{i=0}^{m_1}~\sum_{\a=0}^{m_2}\,
\Tr^{~}_{\underline{V}_{\,k_{i\a}}\otimes\Hcal}\,\biggl\{~
\tr^{~}_{2n\time 2n}\left(\theta^{-2}\right)~P^{~}_{N_{i\a}}\biggr.
\nonumber\\ &&-\,\frac1{2\,R_1^4}\,\left[(m_1-2i)\,\Idd_{k_{i\a}}+
\left(\big|\lambda^{(1)}_{i\a}\big|^2-\big|\lambda^{(1)}_{i+1\,\a}
\big|^2\right)\,
\big(\Idd_{k_{i\a}}-P^{~}_{N_{i\a}}\big)\right]^2\nonumber\\ &&
-\biggl.\frac1{2\,R_2^4}\,\left[(m_2-2\a)\,\Idd_{k_{i\a}}+
\left(\big|\lambda^{(2)}_{i\a}\big|^2-\big|\lambda^{(2)}_{i\,\a+1}
\big|^2\right)\,
\big(\Idd_{k_{i\a}}-P^{~}_{N_{i\a}}\big)\right]^2~\biggr\} \ .
\label{SNCYMansatz}\eea
Requiring that $S_{\rm YM}^{~}<\infty$ yields a pair of equations
determining the moduli of the complex coefficients
$\lambda^{(1)}_{i\a}$ and $ \lambda^{(2)}_{i\a}$ respectively. Up to a
phase they are thus uniquely fixed, by demanding that the ansatz
(\ref{ansatz3})--(\ref{ansatz3p}) be a finite energy field
configuration, as
\beq
\big|\lambda^{(1)}_{i\a}\big|^2\=i\,(m_1-i+1) \qquad\mbox{and}\qquad
\big|\lambda^{(2)}_{i\a}\big|^2\=\a\,(m_2-\a+1) \ .
\label{lambdafixed}\eeq
The corresponding finite action (\ref{SNCYMansatz}) then reads
\bea
S_{\rm YM}^{~}&=&\pi\,R_1^2\,R_2^2~{\rm Pf}(2\pi\,\theta)
\sum_{i=0}^{\big\lfloor\frac{m_1}2\big\rfloor}~
\sum_{\a=0}^{\big\lfloor\frac{m_2}2\big\rfloor}\,\bigl(N_{i\a}+
N_{m_1-i\,m_2-\a}+N_{m_1-i\,\a}+N_{i\,m_2-\a}\bigr)\nonumber\\ &&
\time\,\left[~\frac{(m_1-2i)^2}{2\,R_1^4}+\frac{(m_2-2\a)^2}{2\,R_2^4}-
\tr^{~}_{2n\time 2n}\left(\theta^{-2}\right)~\right] \ ,
\label{SNCYMfinitegen}\eea
where we have split the sum over nodes of the quiver
$\quiver_{(m_1,m_2)}$ into contributions from Dirac monopoles and
antimonopoles which each have the same Yang-Mills energies on the
spheres $\C P^1_{(1)}$ and $\C P^1_{(2)}$. This splitting will be the
crux later on for the physical interpretation of our instanton
solutions.

Finally, let us check that the Yang-Mills equations on
$\R^{2n}_\theta\time\C P^1_{(1)}\time\C P^1_{(2)}$ are indeed
satisfied by our choice of ansatz. We have
\bea\label{chia}
\ca_a- \th_{a\bb}\,\zb^{\bb} & =&
\sum_{i=0}^{m_1}~ \sum_{\a=0}^{m_2}\, X^{i\a}_a\otimes \Pi^{~}_{i\a}~=~
\th_{a\bb}\,\sum_{i=0}^{m_1}~ \sum_{\a=0}^{m_2}\,
T^{~}_{N_{i\a}}\,\zb^{\bb}\,T_{N_{i\a}}^\+
\otimes\Pi^{~}_{i\a} \ , \\[4pt]
\label{chiab}
\ca_{\ab} - \th_{\ab b}\, z^{b} & =&
\sum_{i=0}^{m_1}~ \sum_{\a=0}^{m_2}\, X^{i\a}_{\ab}\otimes \Pi^{~}_{i\a}~=~
\th_{{\ab} b}\,\sum_{i=0}^{m_1} ~\sum_{\a=0}^{m_2} \,
T^{~}_{N_{i\a}}\,z^{b}\,T_{N_{i\a}}^{\+}
\otimes\Pi^{~}_{i\a} \ ,
\eea
while $\ca_{y_{1}}$, $\ca_{y_{2}}$, $\ca_{\yb_{1}}$ and
$\ca_{\yb_{2}}$ are given by (\ref{calAy1})--(\ref{calAgraded}).
For our ansatz the field strength tensor has components
\bea\label{cfabb}
\cf_{a\bb} &=&\th_{a\bb}\, \sum_{i=0}^{m_1}~ \sum_{\a=0}^{m_2} \,
P^{~}_{N_{i\a}}\otimes
\Pi^{~}_{i\a} \ , \\[4pt] \label{cfvtvp1}
\cf_{y_1\yb_1} &=& \frac{1}{\left(1+y_1\yb_1\right)^2}\,
\sum_{i=0}^{m_1} ~\sum_{\a=0}^{m_2}\,
(m_1-2i)~P^{~}_{N_{i\a}}\otimes\Pi^{~}_{i\a} \ ,
\\[4pt] \label{cfvtvp2}
\cf_{y_2\yb_2} &=&\frac{1}{\left(1+y_2\yb_2\right)^2}\,
\sum_{i=0}^{m_1}~ \sum_{\a=0}^{m_2}\,
(m_2-2\a)~P^{~}_{N_{i\a}}\otimes\Pi^{~}_{i\a} \ ,
\eea
where we have imposed the finite energy conditions
(\ref{lambdafixed}). One can now easily check in the same way as
in~\cite{PS1} that the Yang-Mills equations (\ref{YM}) are satisfied.

\subsection{BPS solutions\label{BPSsolns}}

The configurations described above are generically non-BPS solutions
of the Yang-Mills equations on $\R^{2n}_\theta\time\C
P^1_{(1)}\time\C P^1_{(2)}$. Let us now describe the structure of the
BPS states. Substituting (\ref{ansatz3p1}), (\ref{ansatz3p}) and
(\ref{ansatzfieldstrength}) into the remaining DUY equations
(\ref{ddd1}) and using the finite energy constraints
(\ref{lambdafixed}), one finds the BPS conditions
\beq
\sum_{a=1}^n\,\frac1{\theta^a}=\frac{m_1-2i}{2\,R_1^2}+
\frac{m_2-2\a}{2\,R_2^2}
\label{BPScondn}\eeq
for all $i,\a$ with $N_{i\a}>0$. Generically, these conditions are
incompatible with one another unless only one of the degrees, say
$N_{00}$ for definiteness, is non-zero. Then the solution
(\ref{ansatz3})--(\ref{ansatz3p}) is truncated by setting
$T^{~}_{N_{i\a}}=\Idd_r$ for all $(i,\a)\neq(0,0)$ which correspond to
vacuum gauge potentials $A^{i\a}=0$ with trivial bundle maps
$\phi^{(\ell)}_{i\a}$ acting as multiplication by the complex numbers
$\lambda^{(\ell)}_{i\a}$ satisfying (\ref{lambdafixed}). The BPS
solutions are also restricted to the special class of quiver
representations (\ref{genrepSU2Uk}) having dimension vectors $\vec k$
with $k_{i\a}=r~~\forall(i,\a)\neq(0,0)$ and
$k_{00}+m_1\,m_2\,r=k$. As we will see in Section~\ref{modsp}, these
quiver representations are essentially generic and hence BPS solutions
always exist. The corresponding BPS energy (\ref{SNCYMfinitegen}) is
proportional to the degree $N_{00}$ and corresponds to the topological
invariants displayed in (\ref{SBPSgen}), with the remaining terms
vanishing due to the non-holomorphic relations (\ref{antiholrels}).

Notice that there are special points in the quiver vortex moduli space
where the generic BPS gauge symmetry ${\rm U}(k_{00})\time{\rm
  U}(r)^{m_1\,m_2}$ is enhanced. For example, if $R_1=R_2$ and $p$ is
any fixed integer with $0\leq p\leq\min(m_1,m_2)$, then a BPS solution
with $N_{i\,p-i}>0$ for $i=0,1,\dots,p$ is possible. This solution
corresponds to a holomorphic chain along the diagonal vertices
$(i,\a)$ of the quiver $\quiver_{(m_1,m_2)}$ with $i+\a=p$. The
corresponding BPS energies depend on $p$ and are minimized precisely
at $p=0$.

The BPS solution having $N_{i\a}>0$ may be characterized in quiver
gauge theory as $N_{i\a}$ copies of the simple Schur representation
$\underline{\Lcal}_{\,i\a}$ for each $i=0,1,\dots,m_1$,
$\a=0,1,\dots,m_2$. This is the $\quiver_{(m_1,m_2)}$-module given by
a one-dimensional vector space at vertex $(m_1-2i,m_2-2\a)$ with all
maps equal to~$0$, i.e. the $\pathalg_{(m_1,m_2)}$-module with
$(\,\underline{\Lcal}_{\,i\a})_{j\b}=\delta_{ij}\,\delta_{\a\b}~\C$
and dimension vector $(\vec
k_{\underline{\Lcal}_{\,i\a}})_{j\b}=\delta_{ij}\,\delta_{\a\b}$. The
generic non-BPS configurations give modules $\underline{\cal T}$
which are extensions of the BPS
modules~$(\,\underline{\Lcal}_{\,i\a})^{\oplus N_{i\a}}$~\cite{PS1}
describing noncommutative quiver vortex configurations.

\subsection{Instanton moduli space\label{modsp}}

We will now describe the moduli space of the generic (non-BPS)
solutions that we have obtained. The equations of motion are fixed
first of all by the positive integers $n$ and $k$. The condition of
$G$-equivariance then specifies a quiver representation
(\ref{genrepSU2Uk}) with dimension vector $\vec k$. The Yang-Mills
action (\ref{SNCYMfinitegen}) is independent of $\vec k$, and later on
we will find that in fact no physical quantities depend on the particular
choice of quiver representation. As we now proceed to demonstrate,
this independence is due to the triviality of the moduli space of
$\quiver_{(m_1,m_2)}$-modules.

Let us fix a dimension vector $\vec k$. Then with the identifications
$\underline{V}_{\,k_{i\a}}\cong\C^{k_{i\a}}$ we can regard the module
(\ref{genrepSU2Uk}) as an element in the space of quiver
representations {\it into} $\underline{\cal V}$ given by
\beq
{\rm Rep}\bigl(\quiver_{(m_1,m_2)}\,,\vec k\,\bigr):=
\bigoplus_{i=0}^{m_1}~\bigoplus_{\a=0}^{m_2}\,\left(
{\rm Hom}\big(\C^{k_{i+1\,\a}}\,,\,\C^{k_{i\a}}\big)\oplus{\rm Hom}
\big(\C^{k_{i\,\a+1}}\,,\,\C^{k_{i\a}}\big)\right)
\label{RepQkdef}\eeq
with $k_{m_1+1\,\a}:=0=:k_{i\,m_2+1}$. This is the space of
representations with fixed dimension vector $\vec
k$. The set of representations of $\quiver_{(m_1,m_2)}$ into
$\underline{\cal V}$ satisfying the relations $\rel_{(m_1,m_2)}$ is an
affine variety inside the space (\ref{RepQkdef}).

The gauge group of the corresponding quiver gauge theory is given
by (\ref{gaugebroken}). As in Section~\ref{eqvecbun}, it is useful to
work instead with the complexified gauge group
\beq
{\sf G}(\vec k\,)=\prod_{i=0}^{m_1}~\prod_{\a=0}^{m_2}\,
{\rm GL}(k_{i\a},\C) \ .
\label{Qcomplexgaugegp}\eeq
Suppose that $\underline{\cal V}\,,\,\underline{{\cal V}'}\in{\rm
  Rep}(\quiver_{(m_1,m_2)},\vec k\,)$ and
$\,\underline{ f}\,:\underline{\cal
  V}\to\underline{{\cal V}'}$ is an isomorphism of quiver
representations. Then $\,\underline{ f}\,$ can be naturally
regarded as an element of the gauge group
(\ref{Qcomplexgaugegp}). Conversely, any element
$\,\underline{ f}\,=\{ f_{i\a}\in{\rm GL}(k_{i\a},\C)\}_{0\leq
  i\leq m_1,0\leq\a\leq m_2}\in{\sf G}(\vec k\,)$ acts on
$\underline{\cal V}\in{\rm Rep}(\quiver_{(m_1,m_2)},\vec k\,)$ in the
same fashion. It follows that the gauge group ${\sf G}(\vec k\,)$ acts
on ${\rm Rep}(\quiver_{(m_1,m_2)},\vec k\,)$ and two quiver
representations are isomorphic if and only if they lie in the same
orbit of ${\sf G}(\vec k\,)$. Thus there is a one-to-one
correspondence between ${\sf G}(\vec k\,)$-orbits in ${\rm
  Rep}(\quiver_{(m_1,m_2)},\vec k\,)$ and isomorphism classes of
$\quiver_{(m_1,m_2)}$-modules with dimension vector $\vec k$.

This set defines the moduli space ${\cal M}(\quiver_{(m_1,m_2)},\vec
k\,)$ of quiver representations. It has virtual dimension~\cite{Kac1}
\bea
\dim\left[{\cal M}\big(\quiver_{(m_1,m_2)}\,,\,\vec k\,\big)
\right]^{\rm vir}&=&1+\dim\,
{\rm Rep}\bigl(\quiver_{(m_1,m_2)}\,,\vec k\,\bigr)-\dim\,
{\sf G}\bigl(\vec k\,\bigr)\nonumber\\ &=&
1-\sum_{i=0}^{m_1}~\sum_{\a=0}^{m_2}\,k_{i\a}\,
\big(k_{i\a}-k_{i+1\,\a}-k_{i\,\a+1}\big) \ .
\label{dimcalMvir}\eea
Restricting to representations which satisfy the relations
$\rel_{(m_1,m_2)}$ lowers (\ref{dimcalMvir}) by
$\sum_{i,\a}\,k_{i\a}\,k_{i+1\,\a+1}$. Representations with moduli
space dimension greater than the virtual dimension can arise due to
additional unbroken gauge symmetry, as described in
Section~\ref{BPSsolns}. Schur representations, describing generic BPS
states, are those modules for which the stable dimension equals the
virtual dimension. Rigid representations carry no moduli and have
vanishing virtual dimension. As we now show, it is these latter
$\quiver_{(m_1,m_2)}$-modules that parametrize our noncommutative
quiver vortices.

The scalar subgroup $\C^\times\subset{\sf G}(\vec k\,)$ acts trivially
on ${\rm Rep}(\quiver_{(m_1,m_2)},\vec k\,)$, and we are left with a
free action of the projective gauge group ${\sf PG}(\vec k\,):={\sf
  G}(\vec k\,)/\C^\times$. Since ${\sf PG}(\vec k\,)$ is not compact,
we must use geometric invariant theory to obtain a quotient which is
well-defined as a projective variety~\cite{GIT}. The representation space
$X={\rm Rep}(\quiver_{(m_1,m_2)},\vec k\,)$ is an affine variety. Let
$\C[X]$ denote the ring of polynomial functions on $X$. The ${\sf
  PG}(\vec k\,)$-action on $X$ induces a ${\sf PG}(\vec k\,)$-action
on $\C[X]$ in the usual way by pull-back. Let $\C[X]^{{\sf PG}(\vec
  k\,)}\subset\C[X]$ be the subalgebra of ${\sf PG}(\vec
k\,)$-invariant polynomials. Since the gauge group
(\ref{Qcomplexgaugegp}) is reductive, the graded ring $\C[X]^{{\sf
    PG}(\vec k\,)}$ is finitely generated and by the Gel'fand-Naimark
theorem it can be regarded as the polynomial ring of a complex
projective affine variety $X\,/\!\!/\,{\sf PG}(\vec k\,)$. This
defines the desired moduli space
\beq
{\cal M}\bigl(\quiver_{(m_1,m_2)}\,,\,\vec k\,\bigr)~:=~
{\rm Rep}\bigl(\quiver_{(m_1,m_2)}\,,\,\vec k\,\bigr)\,/\!\!/\,
{\sf PG}(\vec k\,)\={\rm Proj}~\C\bigl[
{\rm Rep}\bigl(\quiver_{(m_1,m_2)}\,,\,\vec k\,\bigr)
\bigr]^{{\sf PG}(\vec k\,)} \ .
\label{calMdef}\eeq

Now since the quiver $\quiver_{(m_1,m_2)}$ has no oriented cycles, we
may lexicographically order its vertex set as
$\quiver_{(m_1,m_2)}^{(0)}=\{1,2,\dots,(m_1+1)\,(m_2+1)\}$ and assume
that the integer label of the tail node of each arrow is smaller than
that of the head node. For $\zeta\in\C^\times$ we define
$\,\underline{ f}\,_\zeta\in{\sf G}(\vec k\,)$ by
$(\,\underline{ f}\,_\zeta)_i=\zeta^i~\Idd_{k_i}\in{\rm
  GL}(k_i,\C)$ for each $i\in\quiver_{(m_1,m_2)}^{(0)}$. Then by
considering the action of $\,\underline{ f}\,_\zeta$ on $X={\rm
  Rep}(\quiver_{(m_1,m_2)},\vec k\,)$ and on
$\C[X]^{{\sf PG}(\vec k\,)}$, one easily deduces that $\C[X]^{{\sf
    PG}(\vec k\,)}\cong\C$. This means that the moduli space
(\ref{calMdef}) is trivial,
\beq
{\cal M}\bigl(\quiver_{(m_1,m_2)}\,,\,\vec k\,\bigr)={\rm point} \ ,
\label{calMtrivial}\eeq
and all quiver representations are gauge equivalent.

Thus the only moduli of our solutions arise from the moduli space of
noncommutative solitons~\cite{GHS1}. They are parametrized by the pair
of monopole charges $(m_1,m_2)$ and by the dimension vector $\vec N$
of the quiver representation (\ref{calTdef}). The above argument again
shows that there are no extra moduli associated with the
$\quiver_{(m_1,m_2)}$-modules $\underline{\cal T}$. For each $i,\a$ we let
$b_{l_{i\a}}=(b_{l_{i\a}}^a)$, $l_{i\a}=1,\dots,N_{i\a}$ be the
holomorphic components of fixed points
in $\C^n$, and let $|b_{l_{i\a}}\rangle$ be the corresponding coherent
states in the $n$-oscillator Fock space $\Hcal$. For the projector
$P_{N_{i\a}}$ in the solution of Section~\ref{NCYMsols} we may take the
orthogonal projection of $\Hcal$ onto the linear span
$\bigoplus_{l_{i\a}=1}^{N_{i\a}}\,\C|b_{l_{i\a}}\rangle$. Modulo the standard
action of the noncommutative gauge group ${\rm U}(\Hcal)\cong{\rm
  U}(\infty)$, the moduli space of these projectors can be described
as an ideal $\cal I$ of the ring of polynomials
$\C[\zb^1,\dots,\zb^n]$ in the noncommutative coordinates acting on
the vacuum state $|0,\dots,0\rangle$. The zero set of $\cal I$ gives
the locations of the instantons in $\C^n$ and the codimension of $\cal
I$ in $\C[\zb^1,\dots,\zb^n]$ is the number $N_{i\a}$ of
instantons. The moduli space of partial isometries $T^{~}_{N_{i\a}}$
thereby coincides with the Hilbert scheme ${\rm Hilb}^{N_{i\a}}(\C^n)$
of $N_{i\a}$ points in $\C^n$~\cite{GHS1}, and thus the total moduli
space of the solutions constructed in Section~\ref{NCYMsols} is
\beq
{\cal M}^n_{(m_1,m_2)}(\vec N\,)=\prod_{i=0}^{m_1}~\prod_{\a=0}^{m_2}\,
{\rm Hilb}^{N_{i\a}}(\C^n) \ .
\label{totmodsp}\eeq
The quiver representation (\ref{calTdef}) thereby specifies the
supports of the noncommutative quiver vortices in $\R^{2n}$. Explicit forms
for the Toeplitz operators $T^{~}_{N_{i\a}}$ corresponding to
specific points in (\ref{totmodsp}) may be constructed exactly as
in~\cite{PS1} by using the noncommutative ABS construction. We will
return to this point in the next section.

\bigskip

\section{D-brane realizations\label{Dcharges}}

\noindent
In this final section we will elucidate the physical interpretation of
our solutions as particular configurations of branes and antibranes in
Type~IIA superstring theory. We will first compute, in the original
gauge theory on $\R_\theta^{2n}\time\C P^1\time\C P^1$, the
topological charges of the multi-instanton solutions constructed in
Section~\ref{NCYMsols}. This will make clear the D-brane
interpretation which we describe in detail. We then present two
independent checks of the proposed identification. Firstly, we work out the
K-theory charges associated to the noncommutative quiver vortices. Secondly,
we compute the topological charge in the quiver gauge theory arising
after dimensional reduction to $\R_\theta^{2n}$. While formally similar
to the construction of~\cite{PS1} in the case of holomorphic
chains, the new feature of the higher rank quiver is that all of these
computations of D-brane charges agree {\it only} when one imposes the
appropriate relations derived earlier. The ensuing calculations
thereby also provide a nice physical realization of the quiver with
relations $(\quiver_{(m_1,m_2)}\,,\,\rel_{(m_1,m_2)})$. Details of the
homological algebra techniques used in this section may be found
in~\cite{Quiverbooks,Hom}.

\subsection{Topological charges\label{Topcharge}}

Let us compute the topological charge of the configurations
(\ref{ansatz3})--(\ref{PNiTrace}). The non-vanishing components of the
field strength tensor along $\R_\theta^{2n}$ are given by
\beq
{\cf}^{~}_{2a-1\, 2a} \= 2\im\, {\cf}^{~}_{a\ab} \= -\frac{\im}{\th^a}\,
\sum_{i=0}^{m_1}~\sum_{\a=0}^{m_2}\,
P^{~}_{N_{i\a}}\otimes\Pi^{~}_{i\a} \ ,
\label{calFalongR}\eeq
while the non-vanishing spherical components can be written in terms
of angular coordinates on $S_{(1)}^2\time S^2_{(2)}$ as
\bea
{\cf}^{~}_{\vt_1\vp_1} &= &-\im\,\frac{\sin\vt_1}{2}\,
\sum_{i=0}^{m_1}~\sum_{\a=0}^{m_2} \,(m_1-2i)~P^{~}_{N_{i\a}}
\otimes\Pi^{~}_{i\a}\ ,\\[4pt]
{\cf}^{~}_{\vt_2\vp_2} &= &-\im\,\frac{\sin\vt_2}{2}\,
\sum_{i=0}^{m_1}~\sum_{\a=0}^{m_2} \,(m_2-2\a)~P^{~}_{N_{i\a}}
\otimes\Pi^{~}_{i\a} \ .
\eea
This gives
\bea
&& {\cf}^{~}_{12}\,{\cf}^{~}_{34}\cdots{\cf}^{~}_{2n-1\, 2n}\,
{\cf}^{~}_{\vt_1\vp_1}\,{\cf}^{~}_{\vt_2\vp_2}\nonumber\\[4pt] &&
\qquad =~(-\im)^{n}\,\frac{\sin\vt_1\,\sin\vt_2}{4~{\rm Pf}(
\th)}\,\Bigl(\,\sum_{i=0}^{m_1}~\sum_{\a=0}^{m_2}\,
P^{~}_{N_{i\a}}\otimes\Pi^{~}_{i\a} \Bigr)^n\nonumber\\
&& \qquad\qquad \times\,
\Bigl(\,\sum_{j_1=0}^{m_1}~ \sum_{\g_1=0}^{m_2}\,(m_1-2j_1)~
P^{~}_{N_{j_1\g_1}}\otimes\Pi^{~}_{j_1\g_1}\Bigr)\,
\Bigl(\,\sum_{j_2=0}^{m_1}~ \sum_{\g_2=0}^{m_2}\,(m_2-2j_2)~
P^{~}_{N_{j_2\g_2}}\otimes\Pi^{~}_{j_2\g_2}\Bigr)
\nonumber\\[4pt] && \qquad =~
(-\im)^{n}\,\frac{\sin\vt_1\,\sin\vt_2}{4~{\rm Pf}(
\th)}\,\sum_{i=0}^{m_1} ~\sum_{\a=0}^{m_2} \,(m_1-2i)\,
(m_2-2\a)~P^{~}_{N_{i\a}}\otimes\Pi^{~}_{i\a} \ .
\eea
The instanton charge is then given by the $(n+2)$-th Chern number
\beq
Q:=\frac{1}{(n+2)!}\ \Bigl(\frac{\im}{2\pi}\Bigr)^{n+2}~
{\rm Pf}(2\pi\,\th)~\int_{S_{(1)}^2\time S_{(2)}^2}\,
\Tr^{~}_{\underline{\cal V}\otimes{\cal H}}~
\underbrace{{\cf}\wedge\cdots\wedge {\cf}}_{n+2} \ .
\label{topchargedef}\eeq
The calculation now proceeds exactly as in~\cite{PS1} and one finds
\beq
Q=\sum_{i=0}^{m_1}~ \sum_{\a=0}^{m_2}\,(m_1-2i)\,(m_2-2\a)\,N_{i\a} \ .
\label{Qsame}\eeq
For the BPS configurations described in Section~\ref{BPSsolns} the
energy functional (\ref{SNCYMfinitegen}) is proportional to the
topological charge (\ref{Qsame}), as expected for a BPS instanton
solution.

As we did in (\ref{SNCYMfinitegen}), let us rewrite (\ref{Qsame}) in
the form
\beq
Q=\sum_{i=0}^{\big\lfloor\frac{m_1}2\big\rfloor}~
\sum_{\a=0}^{\big\lfloor\frac{m_2}2\big\rfloor}\,
(m_1-2i)\,(m_2-2\a)\,\Bigl[\big(N_{i\,\a}+N_{m_1-i\,m_2-\a}
\big)-\big(N_{m_1-i\,\a}+N_{i\,m_2-\a}\big)\Bigr] \ .
\label{Qsugg}\eeq
This formula suggests that one should regard the nodes of the quiver
bundle (\ref{bundlediag}) which live in the upper right and lower left
quadrants as branes (with positive charges), and those in the upper
left and lower right quadrants as antibranes (with negative
charges). The branes and antibranes are realized as a quiver vortex
configuration on $\R^{2n}_\theta$ of D0-branes in a system of
$k=\sum_{i,\a}\,k_{i\a} \ $ D$(2n)$-branes. The twisting of the
Chan-Paton bundles by the Dirac multi-monopole bundles over the $\C
P^1$ factors is crucial in this construction. This system is
equivalent to a configuration of spherical D2-branes, wrapping
$\C P^1_{(\ell)}$ for $\ell=1,2$, inside a system of D$(2n+4)$-branes on
$\R_\theta^{2n}\time\C P^1_{(1)}\time\C P^1_{(2)}$. The monopole
flux through each $\C P^1$ factor stabilizes the D2-branes. After
equivariant dimensional reduction, the D$(2n)$-branes which
carry negative magnetic flux on their worldvolume have opposite
orientation with respect to those which carry positive magnetic flux,
and are thus antibranes. The bi-fundamental scalar fields
$\phi^{(\ell)}_{i\a}$ correspond to massless open string excitations
between nearest neighbour D-branes on the quiver
$\quiver_{(m_1,m_2)}$. The relations $\rel_{(m_1,m_2)}$ of the quiver,
given by (\ref{phicommrels}), imply that there is a unique Higgs
excitation marginally binding any given pair of D-branes. As will become
apparent in Section~\ref{Kcharge}, only those brane-antibrane pairs
whose total monopole charge vanishes are actually unstable and possess
tachyonic excitations causing them to annihilate to the vacuum. Other
pairs are stabilized by the non-trivial monopole bundles over the two
$\C P^1$ factors which act as a source of flux stabilization. This
interpretation is consistent with the form of the energy
(\ref{SNCYMfinitegen}) of our solutions, and the stability of the
brane configuration is consistent with the structure of BPS solutions
found in Section~\ref{BPSsolns}. In the remainder of this section we
will justify and expand on these statements.

\subsection{Symmetric spinors\label{Symspinors}}

The standard explicit realization of the basic partial isometry
operators $T_{N_{i\a}}^{~}$ describing the noncommutative
multi-instanton solutions is provided by a $G$-equivariant version of
the (noncommutative) Atiyah-Bott-Shapiro (ABS) construction of tachyon
field configurations~\cite{PS1}, where $G=\su\time\su$. Let us now
describe some general aspects of this construction. We begin with the
equivariant excision theorem~\cite{Segal1} which computes the
$G$-equivariant K-theory of the space $M\time\C P_{(1)}^1\time\C
P^1_{(2)}$ through the isomorphism
\beq
\K_G\bigl(M\time\C P_{(1)}^1\time\C P^1_{(2)}\bigr)
\=\K_G(G\time_HM)\=\K_H(M) \ .
\label{excisiongen}\eeq
Since the closed subgroup $H=\uoo\subset G$ acts trivially on $M$,
from the K\"unneth theorem we arrive at
\beq
\K_G\bigl(M\time\C P_{(1)}^1\time\C P^1_{(2)}\bigr)=
\K(M)\otimes\rep_\uo^{(1)}\otimes\rep_\uo^{(2)} \ ,
\label{excisionslc}\eeq
where $\rep_\uo$ is the representation ring of the group
$\uo$. Setting $M={\rm point}$ in this isomorphism and using
(\ref{monbundles}), we may describe this representation ring as the
formal Laurent polynomial ring $\rep_H=\K_G(\C P_{(1)}^1\time\C
P^1_{(2)})=\Z[\Lcal_{(1)}^{~},\Lcal^\vee_{(1)}]\otimes
\Z[\Lcal_{(2)}^{~},\Lcal^\vee_{(2)}]$. Then (\ref{excisionslc}) is
just the generalization of the isomorphism described in
Section~\ref{eqvecbun} to the case of virtual bundles.

In the case of main interest, $M=\R^{2n}$, we can make the above
isomorphism very explicit. Let $\rep_{\spin_H(2n)}$ be the
Grothendieck group of isomorphism classes of finite-dimensional
$\Z_2$-graded $H\time\cliff_{2n}$-modules, where
$\cliff_{2n}:=\cliff(\R^{2n})$ denotes the Clifford algebra of the
vector space $\R^{2n}$ with the canonical inner product $\delta_{\mu\nu}$.
Extending the standard ABS construction~\cite{ABS1}, we may then compute
the $H$-equivariant K-theory $\K_H(\R^{2n})$ with $H$ acting trivially
on $\R^{2n}$ and commuting with the Clifford action. Any such
$H\time\cliff_{2n}$-module is a direct sum of products of an
$H$-module and a spinor module, and hence
\beq
\rep_{\spin_H(2n)}=
\rep_{\spin(2n)}\otimes\rep_\uo^{(1)}\otimes\rep_\uo^{(2)} \ .
\label{spinortrivdecomp}\eeq
The first factor can be treated by the standard ABS construction and
yields the ordinary K-theory group $\K(\R^{2n})$.
Therefore, our equivariant K-theory group reduces to
\beq
\K_H\bigl(\R^{2n}\bigr)=\K\bigl(\R^{2n}\bigr)
\otimes\rep_\uo^{(1)}\otimes\rep_\uo^{(2)}\ .
\label{eqABSiso}\eeq
In the present context of the equivariant ABS construction, this
isomorphism may be described in terms of the isotopical decomposition
of the spinor module
\beq
\underline{\Delta}_{\,2n}~:=~\underline{\Delta}\,\bigl(\R^{2n}
\bigr)\=\bigoplus_{i=0}^{m_1}~\bigoplus_{\a=0}^{m_2}\,
\Delta_{i\a}\otimes\underline{S}^{(1)}_{\,m_1-2i}\otimes
\underline{S}_{\,m_2-2\a}^{(2)}
\label{spinmoddecomp}\eeq
obtained by restricting $\underline{\Delta}_{\,2n}$ to representations
of $\uoo\subset\spin(2n)\subset\cliff_{2n}$. Let
$\iota:H\hookrightarrow G$ be the inclusion map. It induces a
restriction map from representations of $G$ to representations of $H$,
and hence a homomorphism of representation rings
\beq
\iota^*\,:\,\rep_G~\longrightarrow~\rep_H \ .
\label{represtrmap}\eeq
The $\Delta_{i\a}$'s in (\ref{spinmoddecomp}) are then the
corresponding multiplicity spaces
\beq
\Delta_{i\a}={\rm Hom}^{~}_H\bigl(\iota^*\underline{\Delta}_{\,2n}\,,\,
\underline{S}^{(1)}_{\,m_1-2i}\otimes\underline{S}_{\,m_2-2\a}^{(2)}
\bigr) \ .
\label{multspaces}\eeq

To compute the spaces (\ref{multspaces}) explicitly, consider the
homomorphism of representation rings
\beq
\iota_*\,:\,\rep_H~\longrightarrow~\rep_G
\label{repindmap}\eeq
induced by the induction map from representations of $H$ to
representations of $G$. On generators it is given by the space of
sections
\beq
\iota_*\bigl(\,\underline{S}^{(1)}_{\,p_1}\otimes
\underline{S}_{\,p_2}^{(2)}\bigr)=\Gamma\bigl(\Lcal_{(1)}^{p_1}
\otimes\Lcal_{(2)}^{p_2}\bigr)
\label{spacesections}\eeq
of the homogeneous line bundle
$\Lcal_{(1)}^{p_1}\otimes\Lcal_{(2)}^{p_2}=G\time_H\big(\,
\underline{S}^{(1)}_{\,p_1}\otimes\underline{S}_{\,p_2}^{(2)}\big)$ over
the base space $G/H\cong\C P_{(1)}^1\time\C P^1_{(2)}$, with
$G$-action induced by the standard action on the base. By Frobenius
reciprocity we have $\dim\,{\rm
  Hom}^{~}_G(\,\underline{V}\,,\,\iota_*\underline{W}\,)=\dim\,{\rm
  Hom}^{~}_H(\iota^*\underline{V}\,,\,\underline{W}\,)$ for
$\underline{V}$ a representation of $G$ and $\underline{W}$ a
representation of $H$. As a consequence we can identify the
multiplicity spaces (\ref{multspaces}) as
\beq
\Delta_{i\a}={\rm Hom}^{~}_G\Bigl(\,\underline{\Delta}_{\,2n}\,,\,
\Gamma\big(\Lcal_{(1)}^{m_1-2i}\otimes\Lcal_{(2)}^{m_2-2\a}\big)\Bigr) \ .
\label{multspHomG}\eeq

We may now calculate the isotopical decomposition
(\ref{spinmoddecomp}) by using (\ref{multspHomG}) to construct the
$\su\time\su$-invariant dimensional reduction of spinors from
$\R^{2n}\time\C P^1\time\C P^1$ to $\R^{2n}$. To this end,
we introduce the twisted Dirac operator on $\R^{2n}\time\C
P^1\time\C P^1$ using the graded connection formalism of
Section~\ref{gradedconn} to write the
$\Z_{m_1+1}\time\Z_{m_2+1}$-graded Clifford connection
\bea
\hat\Dirac := \Gamma^{\hat\mu}\,D_{\hat\mu}\=
\gamma^\mu\,D_\mu\otimes\Idd_2\otimes\Idd_2 \!\!&+&\!\!
\bigl(\mphi^{(1)}_{(m_1,\,m_2)}\bigr)\,\g \otimes \g^{\yb_1}\,\b_{\yb_1}
\otimes\Idd_2-
\bigl(\mphi^{(1)}_{(m_1,\,m_2)}\bigr)^\+\,\g \otimes \g^{y_1}\,\b_{y_1}
\otimes\Idd_2
\quad \nonumber \\ &+&\!\!
\bigl(\mphi^{(2)}_{(m_1,\,m_2)}\bigr)\,\g \otimes\Idd_2
\otimes \g^{\yb_2}\,\b_{\yb_2}-
\bigl(\mphi^{(2)}_{(m_1,\,m_2)}\bigr)^\+\,\g \otimes\Idd_2
\otimes \g^{y_2}\,\b_{y_2}
\nonumber\\  &+&\!\!
\g \otimes \Dirac_{\C P^1}^{(1)} \otimes \Idd_2+
\g \otimes \Idd_2 \otimes \Dirac_{\C P^1}^{(2)}
\label{Diracgradeddef}\eea
where
\beq
\Dirac_{\C P^1}^{(\ell)}\ :=\ \g^{y_\ell}\,\Bigl(
\partial_{y_\ell}+\omega_{y_\ell}+\bigl(\ma^{(m_\ell)}\bigr)_{y_\ell}
\Bigr)\ +\ \g^{\yb_\ell}\,\Bigl(
\partial_{\yb_\ell}+\omega_{\yb_\ell}+\bigl(\ma^{(m_\ell)}\bigr)_{\yb_\ell}
\Bigr)
\label{DiracS2def}\eeq
with $\ell=1,2$, and $\omega_y,\omega_\yb$ are the components of the
Levi-Civita spin connection on the tangent bundle of $\C P^1$. The
operator~(\ref{Diracgradeddef}) acts on sections $\Psi$ of the twisted
spinor bundle
\beq
\spinor=
\bigoplus_{i=0}^{m_1}~\bigoplus_{\a=0}^{m_2}\,\left(E_{k_{i\a}}\otimes
\underline{\Delta}_{\,2n}\right)\otimes
\begin{pmatrix}\Lcal_{(1)}^{m_1-2i+1}\\ \Lcal_{(1)}^{m_1-2i-1}
\end{pmatrix}\otimes
\begin{pmatrix}\Lcal_{(2)}^{m_2-2\a+1}\\ \Lcal_{(2)}^{m_2-2\a-1}
\end{pmatrix}
\label{spinortotgen}\eeq
over $\R^{2n}\time\C P^1\time\C P^1$, where
$\Lcal^{p+1}\oplus\Lcal^{p-1}$ are the twisted spinor bundles of
rank~$2$ over the sphere $\C P^1$. We are therefore interested in the
product of the spinor module
$\underline{\Delta}_{\,2n}\otimes\underline{\Delta}\,(\C
P^1)\otimes\underline{\Delta}\,(\C P^1)$ with the fundamental
representation (\ref{genrepSU2Uk}) of the gauge group ${\rm U}(k)$
broken as in (\ref{gaugebroken}).

The symmetric fermions on $\R^{2n}$ that we are interested in
correspond to $\su\time\su$-invariant spinors on $\R^{2n}\time\C
P^1\time\C P^1$. They belong to the kernels $\ker(\Dirac_{\C
  P^1}^{(1)})\otimes\ker(\Dirac_{\C P^1}^{(2)})$ of the two Dirac
operators~(\ref{DiracS2def}) on $\C P^1$. By using (\ref{ma1}),
(\ref{ma2}) and (\ref{CP1Cliffalg}) one can write chiral
decompositions of the Dirac operators (\ref{DiracS2def}) acting on
(\ref{spinortotgen}) in the form
\beq
\Dirac^{(1)}_{\C P^1}\=
\bigoplus_{i=0}^{m_1}\,\begin{pmatrix}0&\Dirac^{(1)\,+}_{m_1-2i}\\
\Dirac^{(1)\,-}_{m_1-2i}&0 \end{pmatrix} \qquad\mbox{and}\qquad
\Dirac^{(2)}_{\C P^1}\=
\bigoplus_{\a=0}^{m_2}\,\begin{pmatrix}0&\Dirac^{(2)\,+}_{m_2-2\a}\\
\Dirac^{(2)\,-}_{m_2-2\a}&0 \end{pmatrix} \ ,
\label{DiracS2decomp}\eeq
where
\bea
\Dirac^{(1)\,+}_{m_1-2i}&=&-\frac1{R_1^2}\,\left[(R_1^2+y_1\yb_1)\,
\partial_{y_1}+\mbox{$\frac12$}\,(m_1-2i+1)\,\yb_1\right] \ , \\[4pt]
\Dirac^{(1)\,-}_{m_1-2i}&=&{\phantom{-}}\frac1{R_1^2}\,\left[(R_1^2+y_1\yb_1)\,
\partial_{\yb_1}-\mbox{$\frac12$}\,(m_1-2i+1)\,y_1\right]
\label{Dirac1pm}\eea
and analogously for $\Dirac^{(2)\,\pm}_{m_2-2\a}$. The non-trivial
kernels are naturally isomorphic to irreducible
$\su$-modules~\cite{PS1} given by
\bea
\ker\Dirac^{(\ell)\,+}_p\=\{0\}
&\quad\mbox{and}\quad& \ker\Dirac^{(\ell)\,-}_p\=\underline{V}_{\,|p|}
\quad\mbox{for}\quad p<0 \ , \nonumber\\[4pt]
\ker\Dirac^{(\ell)\,+}_p\=\underline{V}_{\,p}
&\quad\mbox{and}\quad& \ker\Dirac^{(\ell)\,-}_p
\=\{0\} \quad\mbox{for}\quad p>0 \ ,
\label{kersumod}\eea
with $p=m_1-2i$ for $\ell=1$ and $p=m_2-2\a$ for $\ell=2$. Thus the
chirality gradings are by the signs of the corresponding magnetic
charges.

It follows that the $\su\time\su$-equivariant reduction of the
twisted spinor representation of $\cliff(\R^{2n}\time\C P^1\time\C P^1)$
decomposes as a $\Z_2\time\Z_2$-graded bundle giving
\beq
\underline{\Delta}_{\,\underline{\cal V}}{}^{\su\time\su}
=\underline{\Delta}_{\,2n}\otimes\bigl(\,\underline{
\Delta}_{\,\underline{\cal V}}^{++}~\oplus~\underline{\Delta}_{\,
\underline{\cal V}}^{+-}~\oplus~\underline{\Delta}_{\,\underline{\cal V}}^{-+}
\,\oplus~\underline{\Delta}_{\,\underline{\cal V}}^{--}\bigr) \ ,
\label{twistedspingrad}\eeq
where
\beq
\begin{aligned}
&\underline{\Delta}_{\,\underline{\cal V}}^{++}\=
\bigoplus_{i=0}^{m_1^-}\ \ \bigoplus_{\a=0}^{m_2^-}\,\
\underline{\Delta}_{\,i\a}\qquad\text{and}\qquad
\underline{\Delta}_{\,\underline{\cal V}}^{+-}\=
\bigoplus_{i=0}^{m_1^-}\ \bigoplus_{\a=m_2^+}^{m_2}\
\underline{\Delta}_{\,i\a} \ ,\\[4pt]
&\underline{\Delta}_{\,\underline{\cal V}}^{-+}\=\
\bigoplus_{i=m_1^+}^{m_1}\;\ \bigoplus_{\a=0}^{m_2^-}\,\
\underline{\Delta}_{\,i\a}\qquad\text{and}\qquad
\underline{\Delta}_{\,\underline{\cal V}}^{--}\=\
\bigoplus_{i=m_1^+}^{m_1}~\bigoplus_{\a=m_2^+}^{m_2}\
\underline{\Delta}_{\,i\a}
\end{aligned}
\label{twistedspinpm}\eeq
with
\beq
\underline{\Delta}_{\,i\a} \= \underline{V}_{\,k_{i\a}}
\otimes\underline{V}_{\,|m_1-2i|} \otimes\underline{V}_{\,|m_2-2\a|}
\qquad\text{and}\qquad
m_\ell^\pm\=\bigl\lfloor\mbox{$\frac{m_\ell\pm1}2$}\bigr\rfloor \ .
\label{Deltaiadef}\eeq
The reduction (\ref{twistedspingrad}) is valid for $m_1\,m_2$ odd,
which we henceforth assume for brevity. When $m_1\,m_2$ is even, one
should also couple eigenspaces of spinor harmonics in the appropriate
manner~\cite{PS1}.

The chirality bi-grading in (\ref{twistedspingrad}) is by the signs of
the magnetic charges. The multiplicative $\Z_2$-grading induced by
this $\Z_2\time\Z_2$-grading coincides with the grading into
brane-antibrane pairs infered from (\ref{Qsugg}). The corresponding
actions of the two Clifford multiplications
\beq
\mu^{(1)}_{\underline{\cal V}}\,:\,
\underline{\Delta}_{\,\underline{\cal V}}^{-\,\bullet}~
\longrightarrow~\underline{\Delta}_{\,\underline{\cal V}}^{+\,\bullet}
\qquad\text{and}\qquad
\mu^{(2)}_{\underline{\cal V}}\,:\,
\underline{\Delta}_{\,\underline{\cal V}}^{\bullet\,-}~
\longrightarrow~\underline{\Delta}_{\,\underline{\cal V}}^{\bullet\,+}
\label{CliffmultV}\eeq
are uniquely fixed on isotopical components in the same manner as
in~\cite{PS1}. They give the tachyon fields which are maps between
branes of equal and opposite charge.

The equivalence between D-brane charges on $M\time\C P^1\time\C P^1$
and on $M$ asserted by the isomorphism (\ref{excisionslc}) can now be
understood heuristically through equivariant dimensional reduction as
follows. The graded Clifford connection (\ref{Diracgradeddef}) defines
a class $[\hat\Dirac]$ in the analytic K-homology group $\K^{\rm
  a}(M\time\C P^1\time\C P^1)$. Corresponding to $[\hat\Dirac]$, we
may define a fermionic action functional on the space of sections
$\Psi$ of the bundle (\ref{spinortotgen}) by
\beq
S_{\rm D}^{~}:=\int_{M\time\C P^1\time\C P^1}\,\diff^{2n+4}x~
\sqrt g~\Psi^\dag\,\hat\Dirac\Psi \ .
\label{SDdef}\eeq
Let us evaluate (\ref{SDdef}) on symmetric spinors given by
\beq
\Psi\=\bigoplus_{i=0}^{m_1}~\bigoplus_{\a=0}^{m_2}\,\Psi_{i\a}
\qquad\mbox{with}\qquad \Psi_{i\a}\=\begin{pmatrix}
\psi^{(1)\,+}_{(m_1-2i)}\\\psi^{(1)\,-}_{(m_1-2i)}\end{pmatrix}
\otimes\begin{pmatrix}\psi^{(2)\,+}_{(m_2-2\a)}\\
\psi^{(2)\,-}_{(m_2-2\a)}\end{pmatrix}
\label{symmspindecomp}\eeq
with respect to the decomposition (\ref{spinortotgen}), where
$\psi_{(p)}^{(\ell)\,\pm}$ are sections of $\Lcal^{p\pm1}$ and
$\Psi_{i\a}$ takes values in
$\underline{\Delta}_{\,2n}\otimes\underline{V}_{\,k_{i\a}}$ with
coefficient functions on $M$. After integration over $\C P^1\time\C
P^1$, one easily computes analogously to~\cite{PS1} that the action
functional (\ref{SDdef}) on symmetric spinors becomes
\bea
&& S_{\rm D}^{~}\=16\pi^2\,R_1^2\,R_2^2~\int_M\,\diff^{2n}x
\label{SDred}\\ &&
\times\,\Biggl[~\sum_{i=0}^{m_1^-}~\sum_{\a=0}^{m_2^-}~
\sum_{k_1=0}^{m_1-2i-1}~\sum_{k_2=0}^{m_2-2\a-1}\,\bigl(
\psi^{(1)\,-}_{(m_1-2i)\,k_1}\,\psi^{(2)\,-}_{(m_2-2\a)\,k_2}
\bigr)^\dag\,\Dirac\bigl(\psi^{(1)\,-}_{(m_1-2i)\,k_1}\,
\psi^{(2)\,-}_{(m_2-2\a)\,k_2}\bigr)\Biggr.\nonumber\\ &&
 +\,\sum_{i=0}^{m_1^-}~\sum_{\a=m_2^+}^{m_2}~
\sum_{k_1=0}^{m_1-2i-1}~\sum_{k_2=0}^{|m_2-2\a|-1}\,\bigl(
\psi^{(1)\,-}_{(m_1-2i)\,k_1}\,\psi^{(2)\,+}_{(m_2-2\a)\,k_2}
\bigr)^\dag\,\Dirac\bigl(\psi^{(1)\,-}_{(m_1-2i)\,k_1}\,
\psi^{(2)\,+}_{(m_2-2\a)\,k_2}\bigr) \nonumber\\ &&
+\,\sum_{i=m_1^+}^{m_1}~\sum_{\a=0}^{m_2^-}~
\sum_{k_1=0}^{|m_1-2i|-1}~\sum_{k_2=0}^{m_2-2\a-1}\,\bigl(
\psi^{(1)\,+}_{(m_1-2i)\,k_1}\,\psi^{(2)\,-}_{(m_2-2\a)\,k_2}
\bigr)^\dag\,\Dirac\bigl(\psi^{(1)\,+}_{(m_1-2i)\,k_1}\,
\psi^{(2)\,-}_{(m_2-2\a)\,k_2}\bigr) \nonumber\\ &&
+\,\Biggl.\sum_{i=m_1^+}^{m_1}~\sum_{\a=m_2^+}^{m_2}~
\sum_{k_1=0}^{|m_1-2i|-1}~\sum_{k_2=0}^{|m_2-2\a|-1}\,\bigl(
\psi^{(1)\,+}_{(m_1-2i)\,k_1}\,\psi^{(2)\,+}_{(m_2-2\a)\,k_2}
\bigr)^\dag\,\Dirac\bigl(\psi^{(1)\,+}_{(m_1-2i)\,k_1}\,
\psi^{(2)\,+}_{(m_2-2\a)\,k_2}\bigr)~\Biggr] \ , \nonumber
\eea
where $\Dirac:=\gamma^\mu\,D_\mu$ and the component functions
$\psi^{(\ell)\,\pm}_{(p)\,k}(x)$ on $M$ with $k=0,1,\dots,|p|-1$ form
the irreducible representation $\underline{V}_{\,|p|}\cong\C^{|p|}$ of
the group $\su$. The action functional (\ref{SDred}) corresponds to a
K-homology class $[\Dirac]$ in $\Ka(M)$ twisted by appropriate
monopole contributions and $\su\time\su$-modules. We shall now
proceed to describe this class more precisely.

\subsection{K-theory charges\label{Kcharge}}

Consider a holomorphic chain as in~(\ref{holchain}) and suppose
that it is a complex at the same time. Let us set
$E_+=\bigoplus_{i\,{\rm even}}\,E_{k_{i0}}$ and
$E_-=\bigoplus_{i\,{\rm odd}}\,E_{k_{i0}}$, and define
\beq \label{newphi}
\Phi\ :=\ \left.\left[
\mphi^{(1)}_{(m,0)}+\big(\mphi^{(1)}_{(m,0)}\big)^\dag
\right]\right\vert_{E_-} \ .
\eeq
With respect to this grading, the graded connection (\ref{newphi})
is an odd map $\Phi:E_-\to E_+$. Hence, the triple
$\big[E_-,E_+;\Phi\big]$
represents the K-theory class of a brane-antibrane system with tachyon
field~$\Phi$~\cite{Fold}.
The same construction would carry through for a higher-rank
quiver bundle of the form~(\ref{bundlediag}) if the latter was also a
bi-complex, i.e.~if both the horizontal and vertical arrows
defined complexes. In this case the commutativity conditions
(\ref{mphicommrels}) and~(\ref{mphiantiholcommrels}) would allow us
to lexicographically map the lattice onto a chain, and hence
make contact with the above well-known K-theory construction.

However, for generic monopole numbers $m_1$ and $m_2$ the quiver
bundle (\ref{bundlediag}) does not have the requisite feature of a
bi-complex due to the nilpotency properties
(\ref{nilconds}). Following the interpretation of
Section~\ref{Topcharge} above, we need to fold the holomorphic lattice
into maps between branes and antibranes~\cite{PS1,Fold}. This accomplished
by decomposing the quiver module (\ref{genrepSU2Uk}) with respect to the
multiplicative $\Z_2$-grading induced by the $\Z_2\time\Z_2$-grading
defined by the signs of the monopole charges $m_1-2i$ and $m_2-2\a$ at
each vertex of $\quiver_{(m_1,m_2)}$. As a $\Z_2$-graded vector space
we have
\beq
\underline{\cal V}\=\underline{\cal V}_{\,+}\oplus
\underline{\cal V}_{\,-} \qquad\mbox{with}\qquad
\underline{\cal V}_{\,+}\=\underline{\cal V}_{\,++}\oplus
\underline{\cal V}_{\,--} \quad\mbox{and}\quad
\underline{\cal V}_{\,-}\=\underline{\cal V}_{\,-+}\oplus
\underline{\cal V}_{\,+-} \ ,
\label{calVZ2grading}\eeq
where the bi-graded components are given analogously to
(\ref{twistedspinpm}) as
\beq
\begin{aligned}
&\underline{\cal
  V}_{\,++}\=\bigoplus_{i=0}^{m_1^-}~\bigoplus_{\a=0}^{m_2^-}
\,\underline{V}_{\,k_{i\a}}\qquad\text{and}\qquad
\underline{\cal V}_{\,--}\=\bigoplus_{i=m_1^+}^{m_1}~
\bigoplus_{\a=m_2^+}^{m_2}\,
\underline{V}_{\,k_{i\a}} \ , \\[4pt]
&\underline{\cal V}_{\,+-}\=\bigoplus_{i=0}^{m_1^-}\
\bigoplus_{\a=m_2^+}^{m_2}
\,\underline{V}_{\,k_{i\a}}\qquad\text{and}\qquad
\underline{\cal V}_{\,-+}\=\bigoplus_{i=m_1^+}^{m_1}\;\
\bigoplus_{\a=0}^{m_2^-}\,\underline{V}_{\,k_{i\a}} \ .
\end{aligned}
\label{calVZ2gradcomps}\eeq
Using (\ref{mphi1alph})--(\ref{mphicommrels}), we now introduce the
operators
\beq
\mmua~:=~\Bigl(\mphi_{(m_1,m_2)}^{(1)}\Bigr)^{m_1^-}
\qquad\mbox{and}\qquad \mmub~:=~
\Bigl(\mphi_{(m_1,m_2)}^{(2)}\Bigr)^{m_2^-}
\label{mmudef}\eeq
constructed from the finite-energy Yang-Mills solutions of
Section~\ref{NCYMsols}. With respect to the $\Z_2\time\Z_2$-grading
in (\ref{calVZ2grading}), they are odd maps
\bea
\mmua\,:\,\underline{\cal V}_{\,-\,\bullet}\otimes{\cal H}&\longrightarrow&
\underline{\cal V}_{\,+\,\bullet}\otimes{\cal H}
\qquad\mbox{with}\qquad\Bigl(\mmua\Bigr)^2\=0 \ , \nonumber\\[4pt]
\mmub\,:\,\underline{\cal V}_{\,\bullet\,-}\otimes{\cal H}&\longrightarrow&
\underline{\cal V}_{\,\bullet\,+}\otimes{\cal H}
\qquad\mbox{with}\qquad\Bigl(\mmub\Bigr)^2\=0
\label{mmuoddmap}\eea
which together form the requisite bi-complex of noncommutative tachyon
fields between branes and antibranes.

Let $\mmua{}^{~}_{i\a}$ and $\mmub{}^{~}_{i\a}$ denote the
restrictions of the operators (\ref{mmudef}) to the isotopical
component $\underline{V}_{\,k_{i\a}}$. These operators can be written
in terms of bundle morphisms as
\beq
\mmua{}^{~}_{i\a}\=\phi^{(1)}_{i-m_1^-\,\a}\cdots
\phi^{(1)}_{i\a}\qquad\mbox{and}\qquad\mmub{}^{~}_{i\a}\=
\phi^{(2)}_{i\,\a-m_2^-}\cdots\phi^{(2)}_{i\a} \ ,
\label{mmuiamorph}\eeq
where it is understood that $\phi^{(1)}_{i\a}=0=\phi^{(2)}_{i\a}$ if
$i<0$ or $\a<0$. From (\ref{ansatz3p1}) and (\ref{ansatz3p}) it
follows that the pair of operators (\ref{mmuiamorph}) are respectively
proportional to the Toeplitz operators
\beq
T_{i\a}^{(1)}~:=~T^{~}_{N_{i-m_1^--1\,\a}}\,T^\dag_{N_{i\a}}
\qquad\mbox{and}\qquad
T^{(2)}_{i\a}~:=~T^{~}_{N_{i\,\a-m_2^--1}}\,T^\dag_{N_{i\a}} \ .
\eeq
The tachyon fields (\ref{mmudef}) are thus holomorphic maps between
branes of equal and opposite magnetic charges,
\bea
\mmua{}^{~}_{i\a}\,:\,\underline{V}_{\,k_{i\a}}\otimes{\cal H}
&\longrightarrow&
\underline{V}_{\,k_{i-m_1^--1\,\a}}\otimes\Hcal \ , \nonumber\\[4pt]
\mmub{}^{~}_{i\a}\,:\,\underline{V}_{\,k_{i\a}}\otimes{\cal H}
&\longrightarrow&\underline{V}_{\,k_{i\,\a-m_2^--1}}
\otimes\Hcal \ ,
\label{mmuiabranemap}\eea
with the implicit understanding that $\underline{V}_{\,k_{i\a}}=\{0\}$
when $i<0$ or $\a<0$. Furthermore, from (\ref{dimkerTNi}) it follows
that when the operators (\ref{mmuiamorph}) are non-vanishing their
kernels and cokernels are the finite dimensional vector spaces given
by
\bea
\ker\left(\mmua{}^{~}_{i\a}\right)\={\rm im}\,P^{~}_{N_{i\a}}
&\quad\mbox{and}\quad&
\ker\left(\mmua{}^{~}_{i\a}\right)^\dag\={\rm im}\,P^{~}_{N_{i-m_1^--1\,\a}}
 \ , \nonumber\\[4pt]
\ker\left(\mmub{}^{~}_{i\a}\right)\={\rm im}\,P^{~}_{N_{i\a}}
&\quad\mbox{and}\quad&
\ker\left(\mmub{}^{~}_{i\a}\right)^\dag\={\rm im}\,P^{~}_{N_{i\,\a-m_2^--1}}
\label{dimkermmuia}\eea
with $N_{i\a}:=0$ for $i<0$ or $\a<0$.

The operators $\mmua$ and $\mmub$ are $k\time k$ matrices whose sum
can be written as
\beq
\mmua\oplus\mmub=\begin{pmatrix}~0&\mmua{}^{~}_{-+}&\mmub{}^{~}_{+-}&0\\
~0&0&0&\mmub{}^{~}_{--}\\~0&0&0&\mmua{}^{~}_{--}\\~0&0&0&0
\end{pmatrix}
\label{mu12sum}\eeq
on $\underline{\cal V}\otimes{\cal H}$ with $\underline{\cal V}=\underline{\cal
  V}_{\,++}\oplus\underline{\cal V}_{\,-+}\oplus\underline{\cal
  V}_{\,+-}\oplus\underline{\cal V}_{\,--}$, where
$\mmua{}^{~}_{-\pm}:=\mmua|_{\underline{\cal
    V}_{\,-\pm}\otimes{\cal H}}$ and
$\mmub{}^{~}_{\pm-}:=\mmub|_{\underline{\cal
    V}_{\,\pm-}\otimes{\cal H}}$. This matrix presentation
corresponds to the bundle diagram
\beq
\begin{CD}
\underline{\cal V}_{\,-+}\otimes{\cal H}@>{\mmua{}^{~}_{-+}}>>
\underline{\cal V}_{\,++}\otimes{\cal H}\\
@A{\mmub{}^{~}_{--}}AA@AA{\mmub{}^{~}_{+-}}A\\
\underline{\cal V}_{\,--}\otimes{\cal H}@>>{\mmua{}^{~}_{--}}>
\underline{\cal V}_{\,+-}\otimes{\cal H} \ .
\end{CD}
\label{mu12diag}\eeq
Via an appropriate change of basis of the Hilbert space
$\underline{\cal V}\otimes{\cal H}$, from (\ref{mu12diag}) it follows
that the operator (\ref{mu12sum}) can be rewritten as
\beq
\mT:=\begin{pmatrix}~0&0&\mmua{}^{~}_{-+}&\mmub{}^{~}_{+-}\\
~0&0&\big(\mmub{}^{~}_{--}\big)^\dag&\big(\mmua{}^{~}_{--}
\big)^\dag\\~0&0&0&0\\~0&0&0&0\end{pmatrix}
\label{mTdef}\eeq
on $\underline{\cal V}\otimes{\cal H}$ with $\underline{\cal
  V}=\underline{\cal V}_{\,++}\oplus\underline{\cal V}_{\,--}\oplus
\underline{\cal V}_{\,-+}\oplus\underline{\cal V}_{\,+-}$.

The important ingredients in this construction are the holomorphic
relations $\rel_{(m_1,m_2)}$ of the quiver $\quiver_{(m_1,m_2)}$ which
enable us to commute the graded connections as in
(\ref{mphicommrels}), along with the non-holomorphic relations
(\ref{mphiantiholcommrels}). Together they imply that, with respect to
the $\Z_2$-grading in (\ref{calVZ2grading}), the operator
(\ref{mTdef}) is an odd map
\beq
\mT\,:\,\underline{\cal V}_{\,-}\otimes\Hcal~\longrightarrow~\underline{
\cal V}_{\,+}\otimes\Hcal \qquad\mbox{with}\qquad \bigl(\mT\bigr)^2\=0
\label{mToddmap}\eeq
and hence it produces the appropriate two-term complex representing
the brane-antibrane system with noncommutative tachyon field
(\ref{mTdef}). Again, when acting on isotopical components the operator
$\mT{}^{~}_{i\a}$ relates a given brane to the two possible antibranes
of equal but opposite charge as
\bea
\left.\mT{}^{~}_{i\a}\right|_{\underline{\cal V}_{\,-+}}
\,:\,\underline{V}_{\,k_{i\a}}\otimes\Hcal&\longrightarrow&
\bigl(\,\underline{V}_{\,k_{i-m_1^--1\,\a}}\otimes\Hcal\bigr)~\oplus~
\bigl(\,\underline{V}_{\,k_{i\,\a+m_2^-+1}}\otimes\Hcal\bigr) \ ,
\nonumber\\[4pt]\left.\mT{}^{~}_{i\a}\right|_{\underline{\cal V}_{\,+-}}
\,:\,\underline{V}_{\,k_{i\a}}\otimes\Hcal&\longrightarrow&\bigl(\,
\underline{V}_{\,k_{i\,\a-m_2^--1}}\otimes\Hcal\bigr)
{}~\oplus~\bigl(\,\underline{V}_{\,k_{i+m_1^-+1\,\a}}
\otimes\Hcal\bigr) \ .
\label{mTiabranemap}\eea
{}From (\ref{dimkermmuia}) it then follows that the operators
(\ref{mTiabranemap}) have kernels and cokernels of finite dimensions
given by
\bea
\dim\ker\Bigl.\left(\mT{}^{~}_{i\a}\right)^\dag\Bigr|_{
\underline{\cal V}_{\,++}}&=&\dim\Bigl[\ker\left(\mmua{}^{~}_{i\a}
\right)\cap\ker\left(\mmub{}^{~}_{i\a}\right)^\dag\Bigr]~=~
N_{i-m_1^--1\,\a-m_2^--1} \ ,
\nonumber\\[4pt]\dim\ker\Bigl.\left(\mT{}^{~}_{i\a}\right)^\dag\Bigr|_{
\underline{\cal V}_{\,--}}&=&\dim\Bigl[\ker\left(\mmua{}^{~}_{i\a}
\right)^\dag\cap\ker\left(\mmub{}^{~}_{i\a}\right)\Bigr]~=~
N_{i\a} \ ,
\nonumber\\[4pt]\dim\ker\Bigl.\left(\mT{}^{~}_{i\a}\right)\Bigr|_{
\underline{\cal V}_{\,-+}}&=&\dim\Bigl[\ker\left(\mmua{}^{~}_{i\a}
\right)^\dag\cap\ker\left(\mmub{}^{~}_{i\a}\right)^\dag\Bigr]~=	~
N_{i\,\a-m_2^--1} \ ,
\nonumber\\[4pt]\dim\ker\Bigl.\left(\mT{}^{~}_{i\a}\right)\Bigr|_{
\underline{\cal V}_{\,+-}}&=&\dim\Bigl[\ker\left(\mmua{}^{~}_{i\a}
\right)\cap\ker\left(\mmub{}^{~}_{i\a}\right)\Bigr]~=~
N_{i-m_1^--1\,\a} \ .
\label{dimkermTia}\eea

To incorporate the twistings by the magnetic monopole bundles, we
use the ABS construction of Section~\ref{Symspinors} above to modify
the tachyon field (\ref{mTdef}) to the operator
\beq
\mcalT~:=~\mT\otimes\Idd\,:\,\underline{\Delta}_{\,\underline{\cal
    V}}^{-}\otimes\Hcal~\longrightarrow~
\underline{\Delta}_{\,\underline{\cal V}}^{+}\otimes\Hcal
\label{mcalTdef}\eeq
where  $\underline{\Delta}_{\,\underline{\cal
    V}}^{+}:=\underline{\Delta}_{\,\underline{\cal
    V}}^{++}\oplus\underline{\Delta}_{\,\underline{\cal V}}^{--}$ and
$\underline{\Delta}_{\,\underline{\cal
    V}}^{-}:=\underline{\Delta}_{\,\underline{\cal
    V}}^{-+}\oplus\underline{\Delta}_{\,\underline{\cal V}}^{+-}$. The
corresponding tachyon operators (\ref{mmudef}) then define
noncommutative versions of the Clifford multiplications
(\ref{CliffmultV}). Since $\dim\underline{V}_{\,|p|}=|p|$, from
(\ref{twistedspinpm}), (\ref{Deltaiadef}) and (\ref{dimkermTia}) it
follows that the index of the tachyon field (\ref{mcalTdef}) is given
by
\bea
{\rm index}\bigl(\mcalT\bigr)&=&\dim\ker\bigl(\mcalT\bigr)-
\dim\ker\bigl(\mcalT\bigr)^\dag \nonumber\\[4pt] &=&
\sum_{i=m_1^+}^{m_1}~\sum_{\a=m_2^+}^{m_2}\,|m_1-2i|\,|m_2-2\a|\,
\nonumber\\ && \times\,
\Bigl[\bigl(N_{i\,\a-m_2^--1}+N_{i-m_1^--1\,\a}\bigr)-\bigl(
N_{i-m_1^--1\,\a-m_2^--1}+N_{i\a}\bigr)\Bigr] \nonumber\\[4pt]
&=& -Q \ .
\label{indextachyon}\eea
The virtual Euler class generated by the cohomology of the complex
(\ref{mToddmap}) is the analytic K-homology class in $\Ka(\R^{2n})$ of
the configuration of D-branes represented by the quiver bundle
(\ref{bundlediag}). The formula (\ref{indextachyon}) then asserts that
the K-theory charge of the noncommutative quiver vortex configuration
constructed in Section~\ref{NCYMsols}, i.e. the virtual dimension of
this index class, coincides with the Yang-Mills instanton charge
(\ref{topchargedef})--(\ref{Qsugg}) on $\R_\theta^{2n}\time S^2\time
S^2$. The corresponding geometric worldvolume description in terms of
topological K-cycles may now also be worked out in exactly the same way as
in~\cite{PS1}. It relies crucially on the equivariant excision theorem
(\ref{excisionslc}) which asserts the equivalence of the brane
configurations on $M\time\C P^1\time\C P^1$ and on $M$.

\subsection{D-brane categories\label{ERform}}

The K-theory construction in Section~\ref{Kcharge} above of the
brane configuration corresponding to the quiver bundle
(\ref{bundlediag}) is somewhat primitive in that it only builds the
system at the level of topological charges. In particular, it relies
crucially on the equivariant excision theorem (\ref{excisionslc}). We
can get a more detailed picture of the dynamics of these D-branes, and
in particular how the original configuration folds itself into branes
and antibranes, by modelling our instanton solutions in the category
of quiver representations of
$(\quiver_{(m_1,m_2)}\,,\,\rel_{(m_1,m_2)})$. The ensuing homological
algebra of this category will then exemplify the roles of the
$\su\time\su$-modules and of the relations of the quiver in computing
the equivariant charges. Our previous approach based on intersection
pairings at the K-theory level misses certain quantitative aspects of
the brane configurations corresponding to the quiver bundle
(\ref{bundlediag}), while the category of quiver representations
provides a rigorous and complete framework for understanding these
systems~\cite{quivercat}.

Let us fix a vertex $(m_1-2i,m_2-2\a)\in\quiver^{(0)}_{(m_1,m_2)}$ of
the quiver and consider the distinguished representations
$\underline{\Pcal}_{\,i\a}$ and $\underline{\Lcal}_{\,i\a}$ introduced
in Sections~\ref{CPquiver} and~\ref{BPSsolns} respectively. Then one
has a canonical projective resolution given by the exact
sequence~\cite{Quiverbooks}
\beq
0~\longrightarrow~\underline{\Pcal}_{\,i-1\,\a-1}~\longrightarrow~
\underline{\Pcal}_{\,i-1\,\a}\oplus
\underline{\Pcal}_{\,i\,\a-1}~\longrightarrow~
\underline{\Pcal}_{\,i\a}~\longrightarrow~\underline{\Lcal}_{\,i\a}~
\longrightarrow~0 \ .
\label{Ringelres}\eeq
The first term corresponds to the independent relations of the quiver
which are indexed by $(i,\a)$ with paths starting at $(i,\a)$ and
ending at $(i-1,\a-1)$. The second sum corresponds to the arrows which
start at node $(i,\a)$. Since there are no ``relations among the
relations'', there are no further non-trivial modules to the far left
of the exact sequence (\ref{Ringelres}).

Consider now the module (\ref{calTdef}) generated by a fixed
noncommutative instanton solution. From Section~\ref{modsp} it follows
that this quiver representation specifies the loci of the D-branes in
$\R^{2n}$, and since all the moduli of our solutions come from the
noncommutative quiver solitons it will suffice to recover the appropriate
topological charge. Taking the tensor product of (\ref{Ringelres})
with the components $\ker T_{N_{i\a}}^\dag$ of $\underline{\cal T}$
and summing over all nodes $(i,\a)$ of the quiver
$\quiver_{(m_1,m_2)}$ gives the projective Ringel resolution of
$\underline{\cal T}$ as
\bea
0~\longrightarrow~\bigoplus_{i=0}^{m_1}~\bigoplus_{\a=0}^{m_2}\,
\underline{\Pcal}_{\,i-1\,\a-1}\otimes\ker T_{N_{i\a}}^\dag&
\longrightarrow&\bigoplus_{i=0}^{m_1}~
\bigoplus_{\a=0}^{m_2}\,\bigl(\,\underline{\Pcal}_{\,i-1\,\a}\oplus
\underline{\Pcal}_{\,i\,\a-1}\bigr)\otimes\ker T_{N_{i\a}}^\dag~
\longrightarrow\nonumber\\ &\longrightarrow&
\bigoplus_{i=0}^{m_1}~\bigoplus_{\a=0}^{m_2}\,
\underline{\Pcal}_{\,i\a}\otimes\ker T_{N_{i\a}}^\dag~
\longrightarrow~\underline{\cal T}~\longrightarrow~0 \ .
\label{RingelresT}\eea
Let
\beq
\underline{\cal W}\=\bigoplus_{i=0}^{m_1}~\bigoplus_{\a=0}^{m_2}\,
\underline{W}_{\,i\a}\qquad\mbox{with}\qquad\vec k_{\underline{\cal
    W}}\=\bigl(w_{i\a}\bigr)^{i=0,1,\dots,m_1}_{\a=0,1,\dots,m_2}
\label{calWany}\eeq
be any other representation of
$(\quiver_{(m_1,m_2)}\,,\,\rel_{(m_1,m_2)})$. It will be fixed below
to correctly incorporate the monopole fields at the vertices of the
quiver. Applying the contravariant functor ${\rm
  Hom}(-\,,\,\underline{\cal W}\,)$ to the projective resolution
(\ref{RingelresT}) using (\ref{HomPViso}) then induces the complex
\bea
0&\longrightarrow&\Hom\bigl(\,\underline{\cal T}\,,\,\underline{\cal
  W}\,\bigr)~\longrightarrow~\bigoplus_{i=0}^{m_1}~
\bigoplus_{\a=0}^{m_2}\,\Hom\bigl(\ker T_{N_{i\a}}^\dag\,,\,
\underline{W}_{\,i\a}\bigr)~\longrightarrow \nonumber\\
&\longrightarrow&\bigoplus_{i=0}^{m_1}~\bigoplus_{\a=0}^{m_2}\,
\Bigl(\Hom\bigl(\ker T_{N_{i\a}}^\dag\,,\,\underline{W}_{\,i-1\,\a}
\bigr)\oplus\Hom\bigl(\ker T_{N_{i\a}}^\dag\,,\,
\underline{W}_{\,i\,\a-1}\bigr)\Bigr)~\longrightarrow \nonumber\\
&\longrightarrow&\bigoplus_{i=0}^{m_1}~
\bigoplus_{\a=0}^{m_2}\,\Hom\bigl(\ker T_{N_{i\a}}^\dag\,,\,
\underline{W}_{\,i-1\,\a-1}\bigr)~\longrightarrow~
\Ext^2\bigl(\,\underline{\cal T}\,,\,\underline{\cal
  W}\,\bigr)~\longrightarrow~0 \ .
\label{gencomplex}\eea

The group $\Ext^p(\,\underline{\cal T}\,,\,\underline{\cal W}\,)$ is
defined to be the cohomology of the complex (\ref{gencomplex}) in the $p$-th
position. One has $\Ext^0(\,\underline{\cal T}\,,\,\underline{\cal
  W}\,)=\Hom(\,\underline{\cal T}\,,\,\underline{\cal W}\,)$
corresponding to the vertices of the quiver
$\quiver_{(m_1,m_2)}$. This group classifies morphisms
$\underline{ f}:\underline{\cal T}\to\underline{\cal W}$ of quiver
representations as in Section~\ref{CPquiver} and represents the
partial gauge symmetries of the combined system of D-branes and magnetic
monopoles. The group $\Ext^1(\,\underline{\cal T}\,,\,\underline{\cal
  W}\,)=\Ext(\,\underline{\cal T}\,,\,\underline{\cal W}\,)$
corresponds to the arrows of the quiver and classifies the
$\quiver_{(m_1,m_2)}$-modules $\underline{\cal U}$ which can be
defined by short exact sequences
\beq
0~\longrightarrow~\underline{\cal T}~
\stackrel{\underline{ f}}{\longrightarrow}~
\underline{\cal U}~\stackrel{\underline{ g}}{\longrightarrow}~
\underline{\cal W}~\longrightarrow~0 \ .
\label{Ext1seq}\eeq
We may regard the module $\underline{\cal U}$ as a deformation of
$\underline{\cal T}\oplus\underline{\cal W}$ which simulates the
attaching of magnetic monopoles to the D-branes to form a bound state
$\underline{\cal U}$. The arrows of (\ref{Ext1seq}) are given
by morphisms $\underline{ f}\in\Hom(\,\underline{\cal
  T}\,,\,\underline{\cal U}\,)$ and
$\underline{ g}\in\Hom(\,\underline{\cal U}\,,\,\underline{\cal
  W}\,)$, reflecting the fact that $\underline{\cal T}$ and
$\underline{\cal W}$ are constituents of $\underline{\cal U}$ arising
from partial gauge symmetries. Finally, the non-trivial $\Ext^2$ group
accounts for the relations $\rel_{(m_1,m_2)}$, while $\Ext^p=0$ for
all $p\geq3$ since there are no relations among our relations.

We now define the charge of the given configuration of noncommutative
instantons relative to the
$(\quiver_{(m_1,m_2)}\,,\,\rel_{(m_1,m_2)})$-module (\ref{calWany})
through the relative Euler character
\beq
\chi\bigl(\,\underline{\cal T}\,,\,\underline{\cal W}\,\bigr):=
\sum_{p\geq0}\,(-1)^p~\dim\,\Ext^p\bigl(\,\underline{\cal T}\,,\,
\underline{\cal W}\,\bigr) \ .
\label{relEulerdef}\eeq
This coincides with the Ringel form on the representation ring
$\rep_{\pathalg_{(m_1,m_2)}}$ of the quiver
$\quiver_{(m_1,m_2)}$. Using (\ref{gencomplex}) we may compute the
Euler form as
\bea
\chi\bigl(\,\underline{\cal T}\,,\,\underline{\cal W}\,\bigr)&=&
\dim\,\Hom\bigl(\,\underline{\cal T}\,,\,
\underline{\cal W}\,\bigr)+\dim\,\Ext^2\bigl(\,\underline{\cal T}\,,\,
\underline{\cal W}\,\bigr)-\dim\,\Ext\bigl(\,\underline{\cal T}\,,\,
\underline{\cal W}\,\bigr) \nonumber\\[4pt] &=&\sum_{i=0}^{m_1}~
\sum_{\a=0}^{m_2}\,\dim\,\Hom\bigl(\ker T_{N_{i\a}}^\dag\,,\,
\underline{W}_{\,i\a}\bigr)+\sum_{i=0}^{m_1}~\sum_{\a=0}^{m_2}\,
\dim\,\Hom\bigl(\ker T_{N_{i\a}}^\dag\,,\,\underline{W}_{\,i-1\,\a-1}
\bigr)\nonumber\\ && -\,\sum_{i=0}^{m_1}~\sum_{\a=0}^{m_2}\,\Bigl(\dim\,
\Hom\bigl(\ker T_{N_{i\a}}^\dag\,,\,\underline{W}_{\,i-1\,\a}
\bigr)+\dim\,\Hom\bigl(\ker T_{N_{i\a}}^\dag\,,\,
\underline{W}_{\,i\,\a-1}\bigr)\Bigr)\nonumber\\[4pt]
&=&\sum_{i=0}^{m_1}~\sum_{\a=0}^{m_2}\,N_{i\a}\,\bigl(w_{i\a}
+w_{i-1\,\a-1}-w_{i-1\,\a}-w_{i\,\a-1}\bigr) \ .
\label{chicomputegen}\eea

Following~\cite{PS1}, we choose the coupling representation
(\ref{calWany}) to the brane configuration of the quiver bundle
(\ref{bundlediag}) to correctly incorporate the magnetic monopole
charges through the appropriate folding of
$\su\time\su$-representations appearing in the ABS construction
(\ref{twistedspingrad})--(\ref{Deltaiadef}). We define a
non-decreasing sequence
$\underline{W}_{\,i\a}\subseteq\underline{W}_{\,j\b}$, $i\leq
j,\a\leq\b$ of representations as we move along the quiver of
constituent D-branes such that the $\su\time\su$-module
$\underline{W}_{\,i\a}$ gives an extension of the monopole fields
carried by the elementary brane state at node $(i,\a)$. Thus we take
\beq
\underline{W}_{\,i\a}=\bigoplus_{j=0}^{i-1}~\bigoplus_{\b=0}^{\a-1}\,
\underline{V}_{\,|m_1-2j|}\otimes\underline{V}_{\,|m_2-2\b|} \ .
\label{Wiadef}\eeq
As an element of the representation ring $\rep_{\pathalg_{(m_1,m_2)}}$
of the quiver $\quiver_{(m_1,m_2)}$, we view the module (\ref{Wiadef})
as a graded sum of representations with respect to the signs of the
monopole charges such that its virtual dimension is given by
\bea
w_{i\a}&=&\dim\bigl[\,\underline{W}_{\,i\a}\bigr]^{\rm vir}
\nonumber\\[4pt]&=&\sum_{j=0}^{i-1}~\sum_{\b=0}^{\a-1}\,(m_1-2j)
\,(m_2-2\b)\=i\,\a\,(m_1-i+1)\,(m_2-\a+1) \ .
\label{wiavirtual}\eea
One easily checks that the integers (\ref{wiavirtual}) obey the
inhomogeneous recursion relation
\beq
w_{i\a}+w_{i-1\,\a-1}-w_{i-1\,\a}-w_{i\,\a-1}=(m_1-2i)\,(m_2-2\a) \ .
\label{wiarecrel}\eeq
Consequently, the Euler-Ringel form (\ref{chicomputegen}) in this case
becomes
\beq
\chi\bigl(\,\underline{\cal T}\,,\,\underline{\cal W}\,\bigr)\=
\sum_{i=0}^{m_1}~\sum_{\a=0}^{m_2}\,N_{i\a}\,(m_1-2i)\,(m_2-2\a)\=Q \
,
\label{ERformQ}\eeq
reproducing again the instanton charge (\ref{topchargedef}). The
equivalence between the Euler characteristic (\ref{relEulerdef}) and
the K-theory charge of Section~\ref{Kcharge} above is a consequence of
the index theorem applied to the complex generating the cohomology
groups ${\rm H}^p(\R_\theta^{2n},\underline{\cal
  T}\otimes\underline{\cal W}^\vee\otimes\Hcal)$.

\bigskip

\section*{Acknowledgments}

\noindent
This work was supported in part by the Deutsche
Forschungsgemeinschaft~(DFG). The work of R.J.S. was supported in part
by PPARC Grant PPA/G/S/2002/00478 and by the EU-RTN Network Grant
MRTN-CT-2004-005104.

\bigskip
%\newpage

\end{document}